\documentclass[12pt]{article}
\usepackage[margin=1in]{geometry}

\usepackage[utf8]{inputenc} 
\usepackage[T1]{fontenc}    
\usepackage{url,bbm,mathrsfs,mathtools,bm,amsthm,amsmath,algorithm,algorithmic,afterpage,amssymb,setspace,amsfonts,array,verbatim,newfloat,caption,afterpage,float,rotating,subfiles,csquotes,indentfirst,cancel,subfigure,enumitem,xcolor} 
\allowdisplaybreaks
\usepackage[singlelinecheck=false]{caption}
\usepackage[backend=biber,style=numeric,citestyle=numeric-comp,sorting=none,giveninits=true,maxcitenames=1,uniquelist=false,maxbibnames=10,natbib=true]{biblatex}
\bibliography{MSMSrefs}

\makeatletter
\newcommand*\bigcdot{\mathpalette\bigcdot@{1.0}}
\newcommand*\bigcdot@[2]{\mathbin{\vcenter{\hbox{\scalebox{#2}{$\m@th#1*$}}}}}
\makeatother

\DeclarePairedDelimiter\floor{\lfloor}{\rfloor}
\usepackage{units}

\DeclareMathOperator{\im}{Im}

\DeclareMathOperator*{\argmin}{argmin}
\DeclareMathOperator*{\argmax}{argmax}
\DeclareMathOperator{\Anto}{\text{Antigen}_a}
\DeclareMathOperator{\Antt}{\text{Antigen}_b}
\DeclareMathOperator{\Entrap}{\mathcal{T}_b}
\usepackage{setspace}

\DeclareMathOperator{\E}{\mathbb{E}}
\DeclareMathOperator{\Prob}{Pr}
\newcommand{\Score}{Score}
\DeclareMathOperator{\C}{Cov}
\DeclareMathOperator{\V}{Var}

\DeclarePairedDelimiter\abs{\lvert}{\rvert}%
\DeclarePairedDelimiter\norm{\lVert}{\rVert}%
\DeclareMathOperator{\edist}{\stackrel{d}{=}}

\DeclareMathOperator{\Spfdr}{DI-fdr}
\DeclareMathOperator{\SpFDR}{DI-FDR}
\DeclareMathOperator{\RandMZ}{\mathcal{X}}
\DeclareMathOperator{\PSMfdr}{PSM-fdr}
\DeclareMathOperator{\PSMFDR}{PSM-FDR}
\DeclareMathOperator{\Targ}{\mathcal{T}}
\DeclareMathOperator{\Dec}{\mathcal{D}}
\DeclareMathOperator{\irt}{iRT}
\DeclareMathOperator{\Gen}{gen}
\DeclareMathOperator{\Int}{int}
\DeclareMathOperator{\Mass}{(m)}
\DeclareMathOperator{\IntSuper}{(y)}
\DeclareMathOperator{\logit}{logit}
\DeclareMathOperator{\expit}{expit}
\DeclareMathOperator{\Signal}{\mathcal{I}_{sig}}
\DeclareMathOperator{\Noise}{\mathcal{I}_{noi}}
\DeclareMathOperator{\signal}{sig}
\DeclareMathOperator{\noise}{noi}
\newcommand{\precmass}{pm}

\makeatletter
\let\oldabs\abs
\def\abs{\@ifstar{\oldabs}{\oldabs*}}
\let\oldnorm\norm
\def\norm{\@ifstar{\oldnorm}{\oldnorm*}}
\makeatother

\newtheorem{assumption}{\textit{Assumption}}
\newtheorem{definition}{\textit{Definition}}
\newtheorem{remark}{\textit{Remark}}
\newtheorem{myalgorithm}{\textit{Algorithm}}
\newtheorem{proposition}{\textit{Proposition}}
\newtheorem{corollary}{\textit{Corollary}}
\newtheorem{theorem}{\textit{Theorem}}
\newtheorem{lemma}{\textit{Lemma}}
\renewcommand{\theequation}{\thesection.\arabic{equation}}
\renewcommand{\thetheorem}{\thesection.\arabic{theorem}}
\renewcommand{\theremark}{\thesection.\arabic{remark}}
\renewcommand{\theproposition}{\thesection.\arabic{proposition}}
\renewcommand{\thelemma}{\thesection.\arabic{lemma}}
\renewcommand{\themyalgorithm}{\thesection.\arabic{myalgorithm}}

\title{A novel framework to quantify uncertainty in peptide-tandem mass spectrum matches with application to nanobody peptide identification}

\author{ 
  Chris McKennan$^{\ast,1}$, Zhe Sang$^2$, Yi Shi$^2$\\[4pt] 
  Department of Statistics,
University of Pittsburgh$^1$\\Department of Cell Biology, University of Pittsburgh$^2$\\
  \texttt{chm195@pitt.edu}$^{\ast}$
}

\begin{document}
\maketitle

\begin{abstract} 
Nanobodies are small antibody fragments derived from camelids that selectively bind to antigens. These proteins have marked physicochemical properties that support advanced therapeutics, including treatments for SARS-CoV-2. To realize their potential, bottom-up proteomics via liquid chromatography-tandem mass spectrometry (LC-MS/MS) has been proposed to identify antigen-specific nanobodies at the proteome scale, where a critical component of this pipeline is matching nanobody peptides to their begotten tandem mass spectra. While peptide-spectrum matching is a well-studied problem, we show the sequence similarity between nanobody peptides violates key assumptions necessary to infer nanobody peptide-spectrum matches (PSMs) with the standard target-decoy paradigm, and prove these violations beget inflated error rates. To address these issues, we then develop a novel framework and method that treats peptide-spectrum matching as a Bayesian model selection problem with an incomplete model space, which are, to our knowledge, the first to account for all sources of PSM error without relying on the aforementioned assumptions. In addition to illustrating our method's improved performance on simulated and real nanobody data, our work demonstrates how to leverage novel retention time and spectrum prediction tools to develop accurate and discriminating data-generating models, and, to our knowledge, provides the first rigorous description of MS/MS spectrum noise.
\end{abstract}
\noindent {\bf Keywords:} Bayesian model selection; Local false discovery rate; Mass spectrometry; \allowbreak Nanobody; Peptide-spectrum match; Proteomics

\section{Introduction}
\label{section:Introduction}
Nanobodies are minimal binding units derived from camelid heavy-chain antibodies \citep{Nb_overview}. Unlike conventional human IgG antibodies, these proteins are soluble, have remarkable tissue penetration, and can be bioengineered in large quantities \citep{Nb_Potential}. These characteristics, along with their strong antigen-binding affinity and sequence similarities to IgG, make them promising, cost-effective candidates to study humoral immunity and for therapeutics to diagnose and combat an array of diseases, including cancer, inflammation, and infectious diseases \citep{Nb_Potential}. Classes of nanobodies have even been shown to neutralize SARS-CoV-2 \citep{Nb_Covid}.\par
\begin{figure}
\centering
\includegraphics[width=0.85\textwidth]{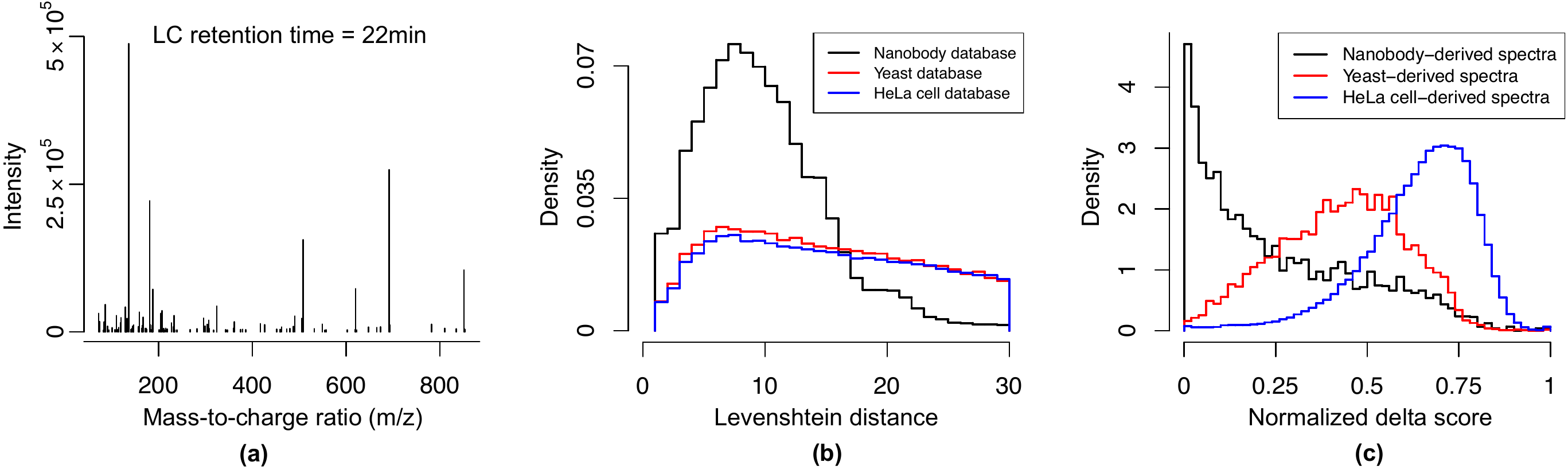}
\caption{Nanobody data and proteome complexity. (a): An example MS/MS spectrum and its LC retention time. The spectrum's retention time and peak locations (m/z's) and intensities are used to infer its generating peptide. (b): First Levenshtein distance quartiles for database peptides within 10 part per million mass bins. Smaller distances imply greater peptide sequence similarity. Histograms were truncated at 30. Yeast and HeLa cell proteomes are used to validate existing target-decoy methods. (c) Delta scores for peptide-spectrum matches significant at a 1\% target-decoy-determined false discovery rate. Normalized delta scores for each spectrum are $1-\allowbreak\frac{\text{2$^\text{nd}$ highest score}}{\text{highest score}}$; values closer to 0 indicate peptide-spectrum matches at risk of score-ordering errors.}
\label{Figure:Database}
\end{figure}

 
\indent To help effectuate their clinical potential, bottom-up proteomics has recently been proposed as a way to identify antigen-specific nanobodies at the proteome scale \citep{Nb_Yi}. Briefly, a camelid is inoculated with an antigen to generate an immune response, its blood collected, and a custom database $\Targ$ containing the constituent peptide sequences of millions of nanobody proteins that may exist in the camelid is created. Antigen-specific nanobodies are then isolated, digested with an endoprotease to yield peptides, and subjected to liquid chromatography-tandem mass spectrometry (LC-MS/MS). LC-MS/MS separates, isolates, and fragments peptides to generate MS/MS spectra whose LC retention times (the times at which the spectra were generated) and peaks can be used to identify their parent peptides in $\Targ$ (Figure~\ref{Figure:Database}(a)). Since nanobodies can then be reconstructed by their identified constituent peptides \citep{NatureNano,Nb_Yi}, correctly matching spectra to their generating peptides is an essential component of this pipeline.

\indent The goal of peptide-spectrum matching is to assign to each spectrum its most likely generating peptide, and quantify the uncertainty in said assignments \citep{PSM_overview}. While methods of varying complexity have been developed \citep{Sequest,XTandem,Andromeda,MSGFplus,Crux_Pvalues,Percolator}, nearly all rely on a target-decoy paradigm to achieve these aims. In brief, each observed spectrum $S$ is searched against a database $\Targ$ of candidate ``target'' peptides whose predicted spectra are used to define scores $\Score(P,S)$ reflecting the plausibility peptide $P \in \Targ$ generated $S$. The spectrum's inferred match is $\hat{P} = \argmax_{\substack{P \in \Targ}} \Score(P,S)$, where $\hat{P}$ may be incorrect if (i) $\Targ$ contains $S$'s true generator $P^*$ but $\Score(\hat{P},S)>\Score(P^{*},S)$, or (ii) $\Targ$ does not contain the analyte that generated $S$ \citep{MatchUncertainty}. We refer to these as score-ordering and database incompleteness errors, respectively. To quantify these uncertainties, the scoring procedure is repeated with $\Targ$ replaced with a database $\Dec$ of ``decoy'' peptides not present in $\Targ$. As decoy peptides do not generate spectra, decoy scores should ideally mirror scores from database incompleteness errors. However, to also control score-ordering errors, it is necessary to assume the maximum decoy and non-generating target peptide scores, $z^{(\Dec)}=\max_{P\in \Dec}\Score(P,S)$ and $z^{(\Targ\setminus\{P^*\})}=\max_{P\in \Targ\setminus\{P^*\}}\Score(P,S)$, are identically distributed \citep{MatchUncertainty}, where $z^{(A)}$ will typically be small if peptides in $A$ are dissimilar to $S$'s generator $P^{*}\in\Targ$ (and vice-versa). As peptides in $\Dec$ are dissimilar to target peptides $P\in\Targ$, $z^{(\Dec)}$ and $z^{(\Targ\setminus\{P^*\})}$ will only have the same distribution if non-generating target peptides do not resemble generating peptides.\par
\indent While this target-decoy paradigm performs well in relatively simple proteomes, the similarity between nanobody peptides in $\Targ$ implies non-generating peptides likely resemble generating peptides (Figure~\ref{Figure:Database}(b)), which belies the above assumption necessary to control score-ordering errors. As a result, Figure~\ref{Figure:Database}(c) suggests existing target-decoy-based methods return an abundance of score-ordering errors at nominally low false discovery rates in nanobody proteomes, which is congruent with recent experimental evidence indicating these methods significantly underestimate error rates in nanobody applications \citep{Nb_Yi}. One solution might be to use the method proposed in \citet{LikelihoodScoring}, which is, to our knowledge, the only method to avoid the above assumption. However, this ignores database incompleteness errors, and is therefore not applicable to real proteomic data \citep{MatchUncertainty}. A second option might be to use the method proposed in \citet{Nb_Yi}, which requires users have access to a second custom nanobody peptide database generated from a second camelid inoculated with a completely different antigen to filter identifications made by an existing target-decoy method. While this method significantly outperformed the target-decoy method its authors compared it with \citep{Nb_Yi}, its second database requirement is impractical in more modest experiments, it is only able to identify a small class of nanobody peptides, and, as we show in Section~\ref{subsection:CDR3}, it likely fails to suitably account for score-ordering errors.\par 
\indent In this work, we address these issues by casting peptide-spectrum matching as a model selection problem with an incomplete model space, where we use observed LC retention times and spectra to distinguish between different peptide ``models''. Our framework and method require only a single nanobody peptide database and are, to our knowledge, the first to account for both score-ordering and database incompleteness errors without assuming the former mirror decoy matches, which we prove helps avoid error rate inflation and, using real data, show facilitates uniquely accurate and powerful inference in nanobody proteomes. While our work is motivated by nanobody proteomes, it is also applicable to other proteomic data, including proteogenomic and metaproteomic data, in which peptide sequence similarity precludes the use of existing methods \citep{AutoRT}. We make several other important contributions:
\begin{enumerate}[label=(\alph*)]
\item We show how cutting edge, deep learning-based retention time and spectrum prediction tools can be used to develop discriminating scoring functions.\label{item:intro:tools}
\item We build a generative model to assess peptide model selection uncertainty that incorporates retention time, inter-spectra heterogeneity, the relationship between mass accuracy and intensity, and the dependence of observed on predicted peak intensities.\label{item:intro:model}
\item We develop a novel and mathematically rigorous model for spectrum noise.\label{item:intro:noise}
\end{enumerate}
\noindent While others have attempted to use the tools in \ref{item:intro:tools}, they only use them for downstream error rate control, ignore retention time, or do not exploit the dependence between predicted and observed peak intensities \citep{NN_pred,AutoRT,Prediction_Cox}. Our models in \ref{item:intro:model} and \ref{item:intro:noise} transcend those in \citet{LikelihoodScoring}, which, to our knowledge, is the only other work to consider peptide model selection uncertainty, and \ref{item:intro:noise} eclipses the standard, but incorrect, assumption that noise peaks occur uniformly at random \citep{UniformNoise,LikelihoodScoring}.\par 
\indent The rest of the article is organized as follows. We describe the data in Section~\ref{section:Data}, outline our assumptions and framework in Section~\ref{section:Model}, and prove existing target-decoy-based methods tend to inflate error rates in nanobody proteomes in Section~\ref{subsection:TDProof}. We then derive our data generating model in Section~\ref{section:BayesFactors}, demonstrate the efficacy of our method, MSeQUiP (\underline{Q}uantifying \underline{U}ncertainty \underline{i}n \underline{P}eptide-spectrum matches), using simulated and real data in Sections~\ref{section:Simulations} and \ref{section:Results}, and end with a discussion in Section~\ref{section:Discussion}. The proofs of all theoretical statements are given in the Supplement.

\section{Observed data and spectral libraries}
\label{section:Data}
Each datum consists of an observed MS/MS spectrum, retention time pair, where retention time is the time at which the generating analyte eluted off the LC column and reflects the analyte's hydrophobicity. We process MS/MS spectra by consolidating peaks with mass-to-charge (m/z) ratios $\leq 40$ parts per million apart to reduce the dependencies between high-resolution peaks \citep{PeakFiltering}, and deisotope and assign peak charge states by extending the method proposed in \citet{Deisotope}. We use Prosit \citep{NN_pred} to generate a predicted spectrum and indexed retention time ($\irt$) for each peptide in a given target $\Targ$ and decoy $\Dec$ database, where $\irt$ is a unitless quantity predictive of retention time. We apply the same processing procedures to predicted spectra as we do for observed spectra. For the purposes of Figure~\ref{Figure:TrainSignal} and Section~\ref{section:Simulations}, our data example in Section~\ref{section:Results} contains two groups, group 1 and group 2, of three LC-MS/MS datasets derived from two different classes of nanobodies. Sections~\ref{supp:section:Processing} and \ref{supp:subsection:ExperimentalDetails} in the Supplement contain additional processing and data details.

\section{Our statistical framework}
\label{section:Model}
We assume each spectrum's generating peptide has the same charge, and generalize to multiple charge states in Section~\ref{supp:subsection:MultipleCharge} of the Supplement. We let $\Targ$ and $\Dec$ be disjoint target and decoy peptide databases, where $\Dec$, which is created by reversing peptides in $\Targ$ and contains no generating peptides, will be used to help quantify uncertainty in Section~\ref{subsection:lfdr}. For $[m]=\{1,\ldots,m\}$, we let $S_g$ and $t_g$ be the observed spectrum and retention time for spectrum $g\in [q]$, and define the random indicator $H_g=I($spectrum $g$ was generated by some $P \in \Targ)$. Our goal is to use the observed data, $S_g$ and $t_g$, to define spectrum $g$'s most likely generator, $\hat{P}_g$, and estimate our uncertainty in $\hat{P}_g$ while allowing for the possibility that spectrum $g$'s generator is not in $\Targ$.


%
\subsection{Quantifying uncertainty in peptide-spectrum matches}
\label{subsection:lfdr}
Let $P \in \Targ$ and define $G_g(P)$ to be the event $\{\text{peptide $P$ generated spectrum $g$}\}$ and $N_g$ to be the event $\{$all peaks in spectrum $g$ are noise peaks$\}$, where we characterize noise peaks in Section \ref{subsection:GeneratingModel}. Then because $G_g(P)\subseteq \{H_g=1\}$ and assuming $\{G_g(P)\}_{P\in\Targ}$ \textit{a priori} have equal measure,
\begin{subequations}
\label{equation:Posterior} 
\begin{align}
    \Prob\{ G_g(P) \mid S_g,t_g\} &= \Prob\{G_g(P) \mid H_g=1,S_g,t_g\}\Prob(H_g=1\mid S_g,t_g)\nonumber\\
\label{equation:Posterior:Post}
    &= \frac{BF_g(P)}{\sum_{P' \in \Targ} BF_g(P') } \Prob(H_g=1\mid S_g,t_g)\\
\label{equation:Posterior:BF}
    BF_g(P) &= \Prob\{ S_g,t_g \mid G_g(P)\}/\Prob( S_g,t_g \mid N_g).
\end{align}
\end{subequations}
The Bayes factor $BF_g(P)$ assesses the evidence in favor of $G_g(P)$ as compared to the noise model $N_g$. Since $\Prob\{ G_g(P) \mid S_g,t_g\}$ and $BF_g(P)$ have the same ordering as functions of $P$, $BF_g(P)$ is the optimal scoring function in the sense that the inferred match $\argmax_{P \in \Targ} BF_g(P)$ is the Bayes optimal classifier. We use this reasoning to define 
\begin{align}
\label{equation:fdr}
    \hat{P}_g=\argmax_{P \in \Targ} BF_g(P), \, \PSMfdr_g = 1 - \Prob\{ G_g(\hat{P}_g) \mid S_g,t_g\},\, \Spfdr_g = \Prob(H_g=0 \mid S_g,t_g),
\end{align}
where the peptide-spectrum match (PSM) local false discovery rate, $\PSMfdr_g$, is our inferential target, and quantifies our uncertainty in spectrum $g$'s inferred match $\hat{P}_g$. The decomposition of $\Prob\{ G_g(P) \mid S_g,t_g\}$ in \eqref{equation:Posterior:Post} is a reflection of our treatment of peptide-spectrum matching as a model selection problem with an incomplete model space $\Targ$, and implies $\PSMfdr_g$ is an increasing function of the two measures of uncertainty discussed in Section~\ref{section:Introduction}. The first is the score-ordering error rate $1-\Prob\{G_g(\hat{P}_g) \mid H_g=1,S_g,t_g\}=1-BF_g(\hat{P}_g)/\sum_{P' \in \Targ} BF_g(P')$, which evaluates the data's capacity to differentiate between potential generators, i.e. models, in $\Targ$, and helps control score-ordering errors. The second is the database incompleteness (DI) error rate $\Spfdr_g$, which quantifies our uncertainty as to whether $\Targ$ contains spectrum $g$'s generator, and helps control database incompleteness errors.\par
\indent The above observation that $BF_g(P)$ is interpretable as a scoring function motivates using the decoy database $\Dec$, whose elements do not overlap with $\Targ$, to approximate the distribution of scores for spectra whose generators are not in $\Targ$. The following proposition formalizes this and motivates an estimator for $\Spfdr_g$.

\begin{proposition}
\label{Proposition:Pvalues}
Let $z_g^{(\Targ)}=\mathop{\max}\limits_{P \in \Targ} BF_g(P)$ and $z_g^{(\Dec)}=\mathop{\max}\limits_{P \in \Dec} BF_g(P)$ be spectrum $g$'s target and decoy score. Suppose (i) $(z_1^{(\Targ)},z_1^{(\Dec)},H_1),\ldots,(z_q^{(\Targ)},z_q^{(\Dec)},H_q)$ are independent and identically distributed and (ii) $z_g^{(\Dec)} \edist (z_g^{(\Targ)} \mid H_g=0)$. Then if $z_g^{(\Dec)}$ has continuous distribution function, $p_g = [1+\sum_{h=1}^q I\{z_g^{(\Targ)} \geq z_h^{(\Dec)}\}]/(q+1)$ satisfies $\mathop{\sup}\limits_{u \in [0,1]}\abs*{\Prob(p_g \leq u \mid H_g=0)-u} = O(q^{-1})$.
\end{proposition}

\noindent If Assumptions (i) and (ii) hold, this suggests we can determine a p-value for the null hypothesis $H_g=0$ and use the method proposed in \citet{qvalue} to estimate $\Spfdr_g$. Assumption (i) is technical and (ii) requires decoy scores mirror target scores for spectra with generators not in $\Targ$, which we show appears to hold in Section~\ref{supp:subsection:AssumProp} of the Supplement. We derive an estimator for $BF_g(P)$ in Section \ref{section:BayesFactors}, which, in conjunction with our estimator for $\Spfdr_g$, is sufficient to estimate $\PSMfdr_g$. 

\subsection{Existing target-decoy methods risk inflating PSM-fdr's}
\label{subsection:TDProof}
Our framework in Section~\ref{subsection:lfdr} explicitly accounts for score-ordering and database incompleteness errors, where we use peptide model uncertainty to determine the score-ordering error rate and, as illustrated in Proposition~\ref{Proposition:Pvalues}, only use the decoy database to control database incompleteness errors via $\Spfdr_g$. By contrast, Lemma~\ref{lemma:InflatePSMfdr} below, which provides a compendious description of the theory presented in Section~\ref{supp:subsection:OtherMethodsfdr} of the Supplement, shows that existing target-decoy methods that use the decoy database to control both score-ordering and database incompleteness errors risk inflating $\PSMfdr$'s.

\begin{lemma}
\label{lemma:InflatePSMfdr}
Suppose for some scoring function $\Score$, Assumptions (i) and (ii) of Proposition~\ref{Proposition:Pvalues} hold for $z_g^{(\Targ)}=\max_{P \in \Targ}\allowbreak\Score\{P,(S_g,t_g)\}$ and $z_g^{(\Dec)}=\max_{P \in \Dec}\Score\{P,(S_g,t_g)\}$. Then if the densities for $z_g^{(\Targ)}$ and $z_g^{(\Dec)}$ exist and are known, the estimators for $\PSMfdr_g$ implied under the target-decoy assumptions listed in \citet{MatchUncertainty} are $O(\Spfdr_g)$.
\end{lemma}
\noindent This shows existing target-decoy-based estimates for $\PSMfdr_g$ are small if and only if $\Spfdr_g$ is small, and have virtually no dependence on the score-ordering error rate. As the score-ordering error rate will likely be large in nanobody proteomes even if $\Spfdr_g$ is small, Lemma~\ref{lemma:InflatePSMfdr} implies target-decoy methods likely return excess false discoveries. We assume the densities for $z_g^{(\Targ)}$ and $z_g^{(\Dec)}$ are known because they are estimated in practice \citep{PEP_kall}, and use \citet{MatchUncertainty} to characterize the target-decoy paradigm because it and its companion \citet{OutsideDatabase} are the only to consider score-ordering and database incompleteness errors.

\section{Defining and estimating Bayes factors}
\label{section:BayesFactors}

\subsection{Data generating models}
\label{subsection:GeneratingModel}
\begin{figure}[t!]
\centering
\includegraphics[width=0.85\textwidth]{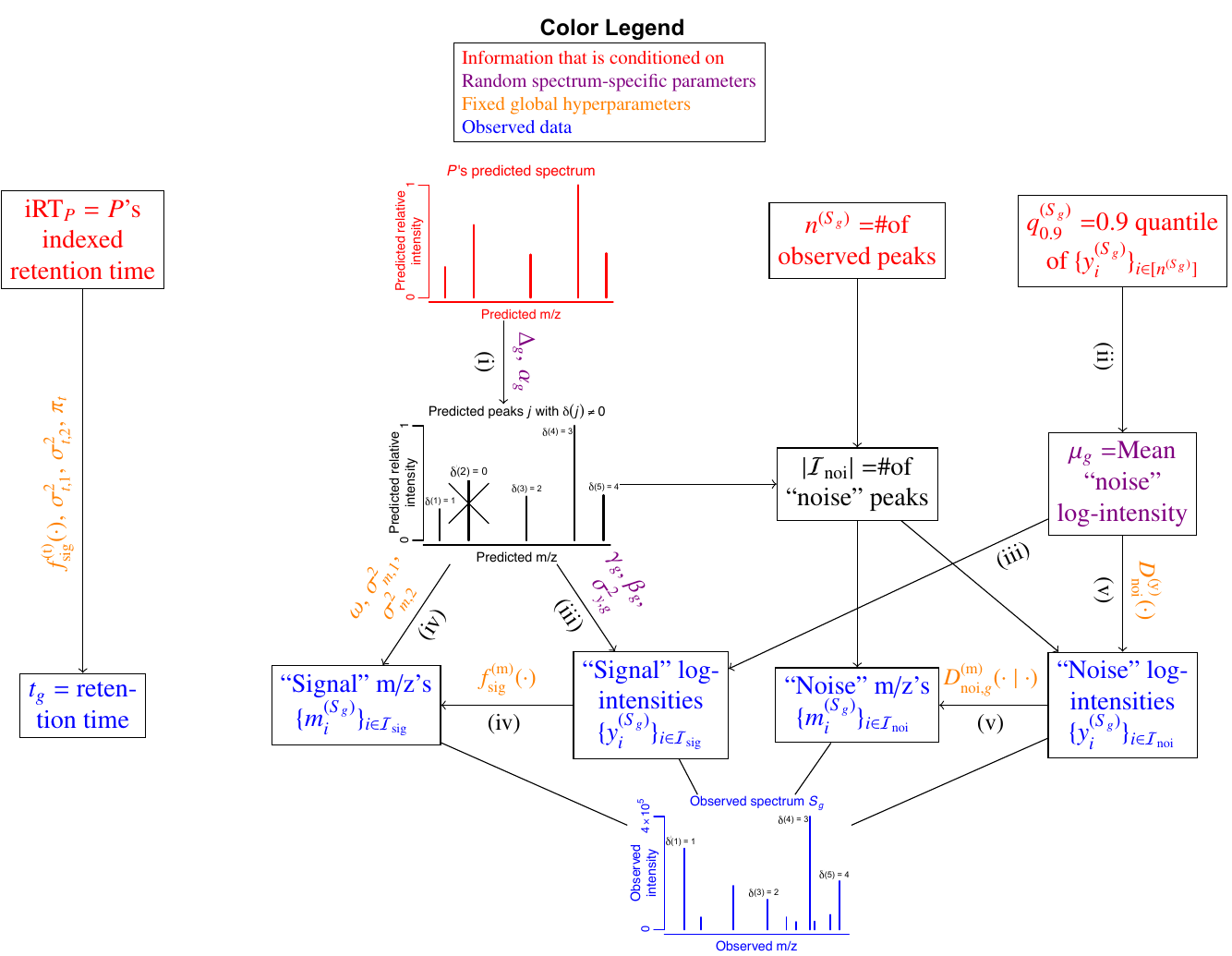}
\caption{An illustration of how $P$'s indexed retention time and predicted spectrum, which are known conditional on $G_g(P)$, generate an observed retention time $t_g$ and spectrum $S_g$. We implicitly condition on the two top rightmost nodes, and roman numerals below arrows reference steps in Algorithm~\ref{algorithm:SpectrumGen}. The map $\delta$ is defined in step (i) of Algorithm~\ref{algorithm:SpectrumGen}, where $\delta(j)=0$ if predicted peak $j$ does not generate a peak in $S_g$. If $\delta(j)\neq 0$, $\delta(j) \in \Signal$ indexes the signal peak in $S_g$ generated by predicted peak $j$. Noise peaks in $S_g$ are unlabeled.} 
\label{Figure:SpectrumGen}
\end{figure}

\begin{figure}[t!]
\centering
\includegraphics[width=0.85\textwidth]{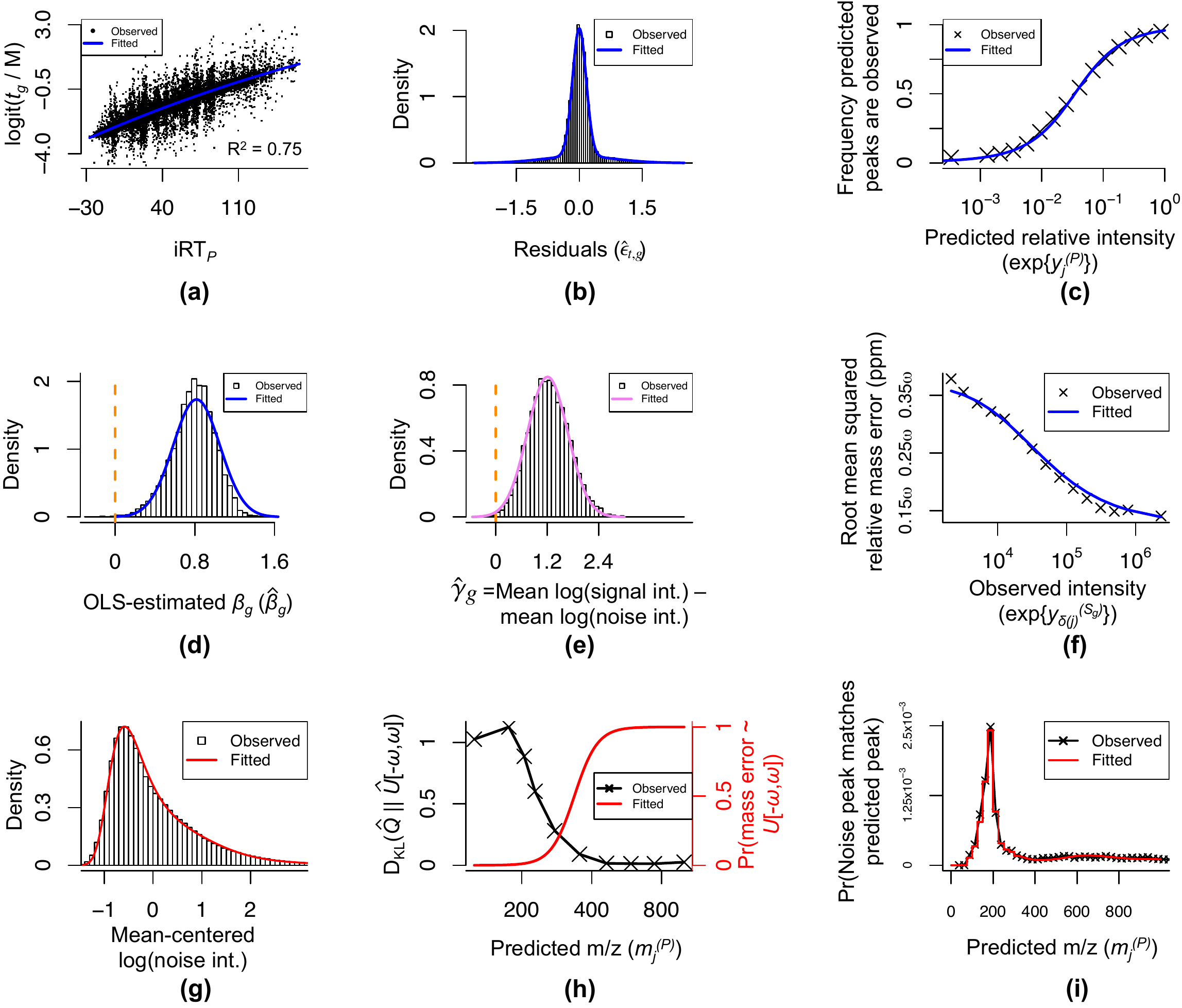}
\caption{Hyperparameter estimates. Observed points and fitted curves derive from high-confidence peptide-spectrum matches from group 2 and group 1 datasets, respectively. Parameters in (c), (d), (e), (g), and (i) are precursor charge-specific, and are fit using spectra with +2 precursors. For any parameter $\theta$ defined in Section~\ref{subsection:GeneratingModel}, $\hat{\theta}$ denotes its estimate; $\mathcal{N}(x;a,b)$ is the density at $x$ of a normal with mean $a$ and variance $b$. (a): $\hat{f}^{(\text{t})}_{\signal}(\cdot)$. (b): Resulting residuals. Blue line is $\hat{\pi}_{t}\mathcal{N}(\hat{\epsilon}_{t,g}; 0,\allowbreak \hat{\sigma}_{t,1}^2) \allowbreak + (1-\hat{\pi}_{t})\mathcal{N}(\hat{\epsilon}_{t,g}; 0,\hat{\sigma}_{t,2}^2)$. (c): Frequency predicted peaks $j$ generate observed peaks $\delta(j)$. Blue line is $h(x)=\smallint \expit(\Delta + x\alpha)\mathcal{N}\{(\Delta,\alpha)^T;\hat{\bm{\mu}}_{\Gen}, \hat{\bm{\Sigma}}_{\Gen}\}\text{d}\Delta \text{d}\alpha$ for $x=\exp\{y_j^{(P)}\}$. (d): Ordinary least squares (OLS) estimates for $\beta_g$ in \eqref{equation:ySignal}. Blue line is $\mathcal{N}(\hat{\beta}_g; \hat{\mu}_{\beta},\allowbreak \hat{\sigma}_{\beta}^2 + \bar{v}_{\beta})$; $\bar{v}_{\beta}$ is the average, over group 2 spectra $g$, OLS estimate for $\text{Var}(\hat{\beta}_g \mid \beta_g)$. (e): Estimates for $\gamma_g$. Violet line is $\mathcal{N}(\hat{\gamma}_g; \hat{\bm{\mu}}_{\text{int}_{2}},\allowbreak \hat{\bm{\Sigma}}_{\text{int}_{22}} + \bar{v}_{\gamma})$ for $\bar{v}_{\gamma}$ the analogue of $\bar{v}_{\beta}$. (f): For observed intensity $\exp(x)$, the fitted line is $[\hat{\pi}_{\signal}(x)\hat{\sigma}^2_{m,1} + \{1-\hat{\pi}_{\signal}(x)\}\hat{\sigma}^2_{m,2}]^{1/2}$. (g): Histogram of $\{y_{i}^{(S_g)}-\bar{y}_{\noise}^{(S_g)}\}_{g \in \{\text{group 2 spectra}\},i \in \Noise}$ for $\bar{y}_{\noise}^{(S_g)}=\lvert\Noise\rvert^{-1}\sum_{i \in \Noise}y_{i}^{(S_g)}$. Red curve is $\hat{D}_{\noise}^{\IntSuper}(\cdot)$. (h): Black: KL divergence between the empirical distribution of relative mass errors for matched noise peaks ($\hat{Q}$) and a discretized $U[-\omega,\omega]$ ($\hat{U}[-\omega,\omega]$). Red: $\hat{\pi}_{\noise}(\cdot)$ from \eqref{equation:dnoise:ma}. (i): Estimate for $\lambda_{\noise,g}^{\Mass}(\cdot)$.}
\label{Figure:TrainSignal}
\end{figure}

Here, we present the data generating models $\Prob\{S_g,t_g \mid G_g(P)\}$ and $\Prob(S_g,t_g \mid N_g)$ for $P \in \Targ \cup \Dec$, which are used to determine $BF_g(P)$. We assume $S_g$ and $t_g$ are independent conditional on $G_g(P)$ or $N_g$ and that $\Prob\{t_g \mid G_g(P)\}/\Prob(t_g \mid N_g) \propto \Prob\{t_g \mid G_g(P)\}$. The latter is for convenience, and plays no role in our estimators in \eqref{equation:fdr}. Therefore, we need only define $\Prob\{t_g \mid G_g(P)\}$, $\Prob\{S_g\mid G_g(P)\}$, and $\Prob(S_g\mid N_g)$. Compendious descriptions of the former two are given in the left and right connected components of Figure~\ref{Figure:SpectrumGen}, and should be referenced while reading \eqref{equation:RTModel} and Algorithm~\ref{algorithm:SpectrumGen} below. We define $\Prob(S_g\mid N_g)$ at the end of the section. First, $\Prob\{t_g \mid G_g(P)\}$, $g \in [q]$, is defined as
\begin{align}
\label{equation:RTModel}
    \logit(t_g/M) \mid \irt_P = f_{\signal}^{(\text{t})}(\irt_P) + \epsilon_{t,g},\quad \epsilon_{t,g} \sim \pi_{t} N(0,\sigma_{t,1}^2) + (1-\pi_{t}) N(0,\sigma_{t,2}^2),
\end{align}
where $M$ is the length of the LC gradient, $\irt_P$ is $P$'s indexed retention time, and $f_{\signal}^{(\text{t})}$ is a quadratic function. The $\logit$ helps remove any mean-variance relationship (Figure \ref{Figure:TrainSignal}(a)), and the mixture normal provides a sufficiently flexible model for $\epsilon_{t,g}$ (Figure \ref{Figure:TrainSignal}(b)).

\indent Our model for $\Prob\{S_g\mid G_g(P)\}$ is more complex. Suppose $P$'s predicted spectrum and $S_g$, depicted in the top and bottom of Figure~\ref{Figure:SpectrumGen}, have $n^{(P)}$ and $n^{(S_g)}$ peaks. Let $m^{(P)}_{j},y^{(P)}_{j}$ be $P$'s $j$th predicted peak's m/z and log-relative intensity, and define $m^{(S_g)}_{i},y^{(S_g)}_{i}$ to be $S_g$'s $i$th peak's observed m/z and log-intensity. The spectrum generating mechanism outlined in Algorithm~\ref{algorithm:SpectrumGen} and Figure~\ref{Figure:SpectrumGen} show how $(m^{(S_g)}_{i},y^{(S_g)}_{i})$ is assumed to be either a noisy realization of $(m^{(P)}_{j},y^{(P)}_{j})$ for some $j \in [n^{(P)}]$ or generated from some noise process, and determines $\Prob\{S_g \mid G_g(P)\}=\Prob[ \{(m^{(S_g)}_{i},y^{(S_g)}_{i})\}_{i\in[n^{(S_g)}]}\mid \{(m^{(P)}_{j},y^{(P)}_{j})\}_{j\in[n^{(P)}]} ]$. We implicitly condition on $\{n^{(S_g)},q_{0.9}^{(S_g)}\}$, where $q_{0.9}^{(S_g)}$ is the 0.9 quantile of $\{y^{(S_g)}_{i}\}_{i\in[n^{(S_g)}]}$. 
\begin{myalgorithm}[Spectrum generating mechanism]
\label{algorithm:SpectrumGen}
\textbf{Input}: Predicted spectrum $\{(m^{(P)}_{j},y^{(P)}_{j})\}_{j\in[n^{(P)}]}$ and hyperparameters $\{\bm{\mu}_{\Gen}, \bm{\Sigma}_{\Gen}\}$, $\{\bm{\mu}_{\Int}, \bm{\Sigma}_{\Int}\}$, $\{\mu_{\beta},\sigma_{\beta}^2,\phi, \nu\}$, $\{\sigma_{m_1}^2, \sigma_{m_2}^2, f_{\signal}^{\Mass}\}$, and $\{D_{\noise}^{\IntSuper}, D_{\noise,g}^{\Mass}\}$.\\
\textbf{Output}: An observed spectrum $\{(m^{(S_g)}_{i},y^{(S_g)}_{i})\}_{i\in[n^{(S_g)}]}$.
\begin{enumerate}
\item[(i)] Assume $m_1^{(P)} < \cdots < m_{n^{(P)}}^{(P)}$. For $E_{gj}^{(P)} = \{\text{predicted peak $j$ generates a peak in $S_g$}\}$,
\begin{align}
\label{equation:Obs}
\begin{aligned}
I\{E_{gj}^{(P)}\}\mid (\Delta_g,\alpha_g) &\sim \text{Ber}[\expit\{\Delta_g + y_j^{(P)}\alpha_g\}], \quad j \in [n^{(P)}]\\
(\Delta_g,\alpha_g)^T &\sim N( \bm{\mu}_{\Gen},\bm{\Sigma}_{\Gen} ).
\end{aligned}
\end{align}
Let $\delta: [n^{(P)}] \to \{0\}\cup [n^{(S_g)}]$ be $\delta(j) = I\{E_{gj}^{(P)}\}\sum_{j'=1}^j I\{E_{gj'}^{(P)}\}$. Then $\Signal=\im( \delta )\setminus\{0\}$ and $\Noise = \{\abs{\Signal}+1,\ldots,n^{(S_g)}\}$ contain observed ``signal'' and ``noise'' peak labels.
\item[(ii)] Independently of (i), let $(\mu_g,\gamma_g)^T \sim N(\bm{\mu}_{\Int} + (q_{0.9}^{(S_g)},0)^T,\bm{\Sigma}_{\Int})$. 
\item[(iii)] Let $\delta^{-1}$ be the pre-image of $\delta$ and $\bar{y}_{\signal}^{(P)} = \abs{\Signal}^{-1}\sum_{j\in \delta^{-1}(\Signal)} y_{j}^{(P)}$. For $j\in \delta^{-1}(\Signal)$,\begin{subequations}  
\label{equation:ySignal}
\begin{align}
    \label{equation:ySignal:y}
    &y_{\delta(j)}^{(S_g)} \mid (\mu_g,\gamma_g,\beta_g,\sigma_{y,g}^2,\delta) =  (\gamma_g + \mu_g) + \{y_{j}^{(P)} -  \bar{y}_{\signal}^{(P)}\}\beta_g + \epsilon_{gj}, \,\, \epsilon_{gj} \sim N(0,\sigma_{y,g}^2)\\
    \label{equation:ySignal:prior}
     &\beta_g \mid (\mu_g,\gamma_g,\sigma_{y,g}^2,\delta) \sim N(\mu_{\beta},\sigma^2_{\beta}), \quad \sigma_{y,g}^{-2} \mid (\mu_g,\gamma_g,\delta) \sim (\phi/\nu) \chi^2_\nu,\quad \phi,\nu >0.
\end{align}
\end{subequations}
\item[(iv)] Let $r_{\delta(j)} = 10^6\{m_{\delta(j)}^{(S_g)}/m_j^{(P)}-1\}$, $j \in \delta^{-1}(\Signal)$, be signal peak $\delta(j)$'s relative part per million deviation from $m_j^{(P)}$. Then $\Prob\{r_{\delta(j)}\mid y^{(S_g)}_{\delta(j)},\delta\} = d_{\signal}^{\Mass}\{r_{\delta(j)}\mid y^{(S_g)}_{\delta(j)}\}$, where for $\sigma_{m,1}^2\leq \sigma_{m,2}^2$, quadratic function $f_{\signal}^{\Mass}$, and constant $\omega>0$,
\begin{align}
\label{equation:mSignal}
\begin{aligned}
d_{\signal}^{\Mass}( r \mid y) &= \pi_{\signal}(y)\mathcal{TN}_{\pm \omega}[r;0,\sigma_{m,1}^2] + \{1-\pi_{\signal}(y)\}\mathcal{TN}_{\pm \omega}[r;0,\sigma_{m,2}^2]\\
\pi_{\signal}(y) &= \expit\{f_{\signal}^{\Mass}(y)\}.
\end{aligned}
\end{align}
$\mathcal{TN}_{\pm c}(x;a,b)$ is the density at $x$ of the truncated normal with mean $a$, variance $b$, and truncated at $\pm c$.
\item[(v)] $\{( m_i^{(S_g)},y_i^{(S_g)} )\}_{i \in \Noise}$ are independent and identically distributed given $(\delta,\mu_g)$ and
\begin{align}
\label{equation:NoiseDist}
\begin{aligned}
\Prob\{ m_i^{(S_g)},y_i^{(S_g)} \mid \delta,\mu_g \} =& \Prob\{m_i^{(S_g)}\mid y_i^{(S_g)},\delta\}\Prob\{y_i^{(S_g)}\mid \delta,\mu_g\}\\
    =& D^{\Mass}_{\noise,g}\{ m_i^{(S_g)} \mid y_i^{(S_g)} \} D^{\IntSuper}_{\noise}\{ y_i^{(S_g)}-\mu_g \}, \quad i \in \Noise
\end{aligned}
\end{align}
for $D^{\IntSuper}_{\noise}(y) = \exp(\sum\limits_{k=0}^7 b_k y^k),D^{\Mass}_{\noise,g}(m\mid y)$ densities with respect to Lebesgue measure and $\int_{-\infty}^{\infty} yD^{\IntSuper}_{\noise}(y)\text{d}y = 0$. Order $\{ (m_i^{(S_g)},y_i^{(S_g)}) \}_{i \in \Noise}$ so $m_i^{(S_g)}<m_{i+1}^{(S_g)}$ for all $i,i+1\in \Noise$.
\end{enumerate}
\end{myalgorithm}

\begin{remark}
\label{remark:omega}
The constant $\omega$ in \eqref{equation:mSignal} is known and reflects the mass accuracy of the mass spectrometer at the parts per million (ppm) scale. In our application, $\omega=20$.
\end{remark}
\begin{remark}
\label{remark:delta}
The map $\delta$ in step (i) determines which predicted peaks $j$ generate observed signal peaks $i \in \Signal$. In practice, we set $\delta(j)=i$ if $10^6\abs*{ m_i^{(S_g)}/m_j^{(P)}-1 }\leq \omega$ and $\delta(j)=0$ if no such $i$ exists, which assumes observed peaks within $\omega$ppm of $m_j^{(P)}$ were not generated by the noise process, and is standard in modern high mass accuracy data ($\omega \asymp 10$) \citep{MaxQuant}. The map $\delta$ also informs observed peak ordering, where $m_{\delta^{-1}(i)}^{(P)}<m_{\delta^{-1}(i+1)}^{(P)}$ for all $i,i+1 \in \Signal=\{1,\ldots,\abs*{\Signal}\}$. Since no ordering exists for noise peaks $i \in \Noise$, we order them in step (v) so that noise m/z's are increasing, which is equivalent to observing order statistics.
\end{remark}
\begin{remark}
The condition $\int yD^{\IntSuper}_{\noise}(y)\text{d}y = 0$ in (v) implies $\mu_g$ is the mean log-intensity of $S_g$'s noise peaks. Including $q_{0.9}^{(S_g)}$ in (ii) is akin to normalizing $y_i^{(S_g)}$ by $q_{0.9}^{(S_g)}$, and helps stabilize $\bm{\Sigma}_{\Int}$.
\end{remark}

Step (i) captures the fact that peaks predicted to be intense are more likely to generate peaks in $S_g$ (Figure \ref{Figure:TrainSignal}(c)), and $\beta_g$ in \eqref{equation:ySignal} relates observed and predicted relative intensities, where $\beta_g \approx 1$ implies observed intensities are approximately proportional to predicted relative intensities (Figure \ref{Figure:TrainSignal}(d)). The term $\gamma_g$ in step (ii) and \eqref{equation:ySignal:y} is the difference in mean signal and noise log-intensities, since for $\bar{y}^{(S_g)}_{\signal}=\abs*{\Signal}^{-1}\sum_{i\in \Signal} y_i^{(S_g)}$ and $\bar{y}^{(S_g)}_{\noise}=\abs*{\Noise}^{-1}\sum_{i\in \Noise} y_i^{(S_g)}$, $\gamma_g = \E\{\bar{y}^{(S_g)}_{\signal} - \bar{y}^{(S_g)}_{\noise} \mid \mu_g,\gamma_g,\beta_g,\sigma_{y,g}^2,\delta\}$. While an ideal prior on $\gamma_g$ would place no mass on negative values, our estimates for $\bm{\mu}_{\Int},\bm{\Sigma}_{\Int}$ suggest this is unnecessary (Figure \ref{Figure:TrainSignal}(e)). The density $d_{\signal}^{\Mass}$ in \eqref{equation:mSignal} determines a signal peak's mass accuracy, where higher density around 0 implies observed m/z's tend to be closer to their predicted m/z's, and the increasing function $f_{\signal}^{\Mass}$ reflects the fact that more intense peaks have greater mass accuracy (Figure \ref{Figure:TrainSignal}(f)) \citep{MassAccuracyInt}. Our assumption that noise log-intensities are drawn from a location family of distributions defined by $D^{\IntSuper}_{\noise}$ in \eqref{equation:NoiseDist} is driven by observations from real data (Figure \ref{Figure:TrainSignal}(g); see Section~\ref{supp:subsection:NoiseInt} in the Supplement). We let $\{(m^{(S_g)}_{i},y^{(S_g)}_{i})\}_{i \in [n^{(S_g)}]}$ be generated according to steps (ii) and (v) of Algorithm \ref{algorithm:SpectrumGen} under $N_g$, which defines $\Prob(S_g \mid N_g)$.\par 
\indent While others have developed nominal data generating models, they typically behave as scoring functions that can only rank peptides, as opposed to likelihoods capable of uncertainty quantification \citep{StatNormal,UniformNoise,DRIP,BayesNetwork}. To our knowledge, \citet{LikelihoodScoring} is the only other work to consider building a likelihood. Besides incorporating retention time and having application to high mass accuracy data, our model for $\Prob\{t_g,S_g \mid G_g(P)\}$ transcends that in \citeauthor{LikelihoodScoring} by considering the covariation between noise and signal intensities (step (ii)), the dependence of observed intensities on predicted intensities (step (iii)), the link between peak intensity and mass accuracy (step (iv)), and, as indicated by the violet parameters in Figure~\ref{Figure:SpectrumGen}, inter-spectra heterogeneity. The latter is critical, as Figure~\ref{supp:figure:Crux} and Remark~\ref{supp:remark:Heterogeneity} in the Supplement show parameters vary substantially between spectra.

\subsection{A model for noise m/z's}
\label{subsection:NoiseMZ}
Of the input hyperparameters in Algorithm~\ref{algorithm:SpectrumGen}, it remains only to define $D^{\Mass}_{\noise,g}(m \mid y)$, which is the density at m/z $m$ for a noise peak in spectrum $g$ with log-intensity $y$. One possibility is to assume $D^{\Mass}_{\noise,g}(m \mid y)$ is uniform on an interval containing $\{m^{(S_g)}_i\}_{i \in [n^{(S_g)}]}$, which is standard in low mass accuracy data \citep{UniformNoise,LikelihoodScoring}. However, we show in Section~\ref{supp:subsection:simulationDetails} of the Supplement that the probability a noise peak gets mapped to a predicted peak is $O(10^{-5})$ in our high mass accuracy data, which, as suggested in Sections~\ref{section:Simulations} and \ref{section:Results}, is an underestimate in many m/z regions and inflates Bayes factors. To design a more appropriate model, the definition of $\Prob(S_g \mid N_g)$ implies $BF_g(P) = c \prod_{i \in \Signal}[D_{\noise,g}^{\Mass}\{ m_i^{(S_g)} \mid y_i^{(S_g)} \}]^{-1}$, where $c$ does not depend on $\{m_i^{(S_g)}\}_{i \in \Noise}$. Since $i \in \Signal$ only if $10^6\abs*{m_i^{(S_g)}/m_j^{(P)} - 1} \leq \omega$ by \eqref{equation:mSignal} for some predicted m/z $m_j^{(P)}$ and $\omega$ defined in step (iv) of Algorithm~\ref{algorithm:SpectrumGen} and Remark~\ref{remark:omega}, we need only determine $D_{\noise,g}^{\Mass}( m\mid y )$ for $m$ within $\omega$ppm of predicted m/z's.\par
\indent To do so, let $(M_g,Y_g)$ be an observed noise m/z, log-intensity pair in spectrum $g$ and define $F\{m_j^{(P)}\} = \{x >0: \abs*{x/m_j^{(P)} - 1} \leq \omega\}$ to be the $\omega$ppm interval around $m_j^{(P)}$. Then if $m \in F\{m_j^{(P)}\}$, we can express $D_{\noise,g}^{\Mass}( m\mid y )$ as
\begin{align*}
&D_{\noise,g}^{\Mass}( m\mid y )= \Prob(M_g=m \mid Y_g=y) = 10^{6}\{m_j^{(P)}\}^{-1}\lambda_{\noise,g}^{\Mass}\{ m_j^{(P)} \mid y \} d_{\noise,g}^{\Mass}\{r_{m} \mid m_j^{(P)}; y \} \nonumber\\
&\lambda_{\noise,g}^{\Mass}\{ m_j^{(P)} \mid y \} = \Prob[ M_g \in F\{m_j^{(P)}\} \mid Y_g=y], \quad r_{m}=10^6\{m/m_j^{(P)}-1\}\\
& d_{\noise,g}^{\Mass}\{r \mid m_j^{(P)}; y \} = 10^{-6}m_j^{(P)}\Prob[ M_g = m_j^{(P)}(1+10^{-6}r) \mid M_g \in F\{m_j^{(P)}\}, Y_g=y ].\nonumber
\end{align*}
Here, $\lambda_{\noise,g}^{\Mass}\{ m_j^{(P)} \mid y \}$ is the probability a noise peak in the $g$th observed spectrum gets mapped to the $j$th peak in $P$'s predicted spectrum, and the density $d_{\noise,g}^{\Mass}\{r \mid m_j^{(P)}; y \}$ determines its mass accuracy, where higher density around 0 indicates the its m/z tends to be closer to $m_j^{(P)}$. We express $d_{\noise,g}^{\Mass}\{r \mid m_j^{(P)}; y \}$ in terms of relative ppm mass error $r\in[-\omega,\omega]$ to be consistent with $d_{\signal}^{\Mass}(r \mid y)$ in \eqref{equation:mSignal}. While this helps break $D_{\noise,g}^{\Mass}$ into interpretable pieces, we must account for the fact that noise peaks may be either manifestations of background noise, or genuine peptide fragments not contained in the generating peptide's predicted spectrum \citep{LikelihoodScoring}, where $d_{\noise,g}^{\Mass}$ and $\lambda_{\noise,g}^{\Mass}$ likely depend on the peak's origin. For example, $d_{\noise,g}^{\Mass}$ likely resembles $d_{\signal}^{\Mass}$ for peptide fragments, but, due to the homogeneity in background noise \citep{Background_Indep_mz2}, is likely uniform for background noise peaks. We address this by developing, to the best of our knowledge, the first mathematical description of MS/MS spectral noise in Section~\ref{supp:section:theory} of the Supplementary Material. Theorem~\ref{theorem:noise} below outlines the key assumptions and results, which helps posit models and estimators for $d_{\noise,g}^{\Mass}$ and $\lambda_{\noise,g}^{\Mass}$.
\begin{theorem}
\label{theorem:noise}  
Let $(M_g,Y_g)$ be an observed noise m/z, log-intensity pair in spectrum $g$. Suppose $M_g = z M_g^{(\text{B})} + (1-z)M_g^{(\text{PF})}$ for $z \sim \text{Ber}(\pi)$ and $\pi \in [0,1]$, where the background and peptide fragment noise m/z's $M_g^{(\text{B})}$ and $M_g^{(\text{PF})}$ satisfy:
\begin{enumerate}
\item[(a)] $M_g^{(\text{B})}$ has density $B_g(m)$ whose logarithm is continuous on a closed interval.
\item[(b)] Let $\RandMZ$ be an inhomogeneous Poisson point process with intensity $\Lambda$ so that $\RandMZ=\{$peptide fragment m/z's that may produce noise peaks$\}$. Then $M_g^{(\text{PF})}=X_g(1+10^{-6}R_g)$, where $R_g \mid Y_g=y$ has density $d_{\signal}^{\Mass}(r \mid y)$, $X_g \mid \RandMZ \sim \text{Categorical}(\RandMZ,p_g)$, and $p_g$ is a probability mass function on $\RandMZ$.
\end{enumerate}
Then if $\log\{p_g(m)\}$ and $\log\{\Lambda(m)\}$ are continuous on compact intervals and the assumptions in Section~\ref{supp:subsection:assumptions} hold, the following are true:
\begin{enumerate}
\item[(i)] Theorems~\ref{supp:theorem:CondSmall}\textendash\ref{supp:theorem:Condlarge} and Corollary~\ref{supp:corollary:dModAvg}: $d_{\noise,g}^{\Mass}\{r\mid m_{j}^{(P)}; y\}$ can be approximated as\\$\pi_{\noise,g}\{m_{j}^{(P)}\}\mathcal{U}_{\pm \omega}(r) + [1-\pi_{\noise,g}\{m_{j}^{(P)}\}] d_{\signal}^{\Mass}(r \mid y)$, where $\logit[\pi_{\noise,g}\{m_{j}^{(P)}\}]$ is continuous in $m_{j}^{(P)}$ and $\mathcal{U}_{\pm \omega}(r)$ is the density of $U[-\omega,\omega]$ at $r$.
\item[(ii)] Theorem~\ref{supp:theorem:lambda} and Corollary~\ref{supp:corollary:lambda}: $\lambda_{\noise,g}^{\Mass}\{m_{j}^{(P)}\mid y\}$ can be approximated function that is continuous in $m_j^{(P)}$ and does not depend on $y$.
\end{enumerate}
\end{theorem}
\begin{remark}
\label{remark:theorem:assumptions}
Consistent with the above discussion, $M_g$ is either a background, $M_g^{(\text{B})}$, or peptide fragment, $M_g^{(\text{PF})}$, noise m/z. Assumption (a) is more general than the usual assumption that background noise is uniform on a closed interval \citep{UniformNoise}, and Assumption (b) describes the generative model for peptide fragment noise peaks. First, a fragment m/z $X_g$ is chosen from $\RandMZ$, and then, like peptide fragment signal m/z's in step (iv) of Algorithm \ref{algorithm:SpectrumGen}, its relative ppm mass error $R_g$ is drawn according to $d_{\signal}^{\Mass}(r \mid y)$.
\end{remark}  
\begin{remark}
\label{remark:theorem:poisson}
We let $\RandMZ$ be random to account for our uncertainty in the exact position of all potential peptide fragment noise m/z's. The continuity of $\log\{\Lambda(m)\}$ assumes the density of potential fragment noise m/z's does not change too abruptly, and the continuity of $\log\{p_g(m)\}$ implies the probability a fragment noise peak is generated from two neighboring regions is approximately proportional to the number of points in each region. We further justify these assumptions, as well as $\RandMZ$'s Poisson assumption, in Section~\ref{supp:subsection:assumptions} using observed data and implicit assumptions from current best practices in the literature. In practice, and consistent with previous work \citep{UniformNoise2}, we assume $\RandMZ$ is equivalent to $\{x>0:x$ is the m/z of a peptide fragment from a permuted peptide in the target database $\Targ\}$.
\end{remark} 
\noindent In addition to Remark \ref{supp:remark:Dnoiseg} arguing that $\pi_{\noise,g}\{m_{j}^{(P)}\}$ has little dependence on $g$, (i) and (ii) in Theorem~\ref{theorem:noise} imply $d_{\noise,g}^{\Mass}$ and $\lambda_{\noise,g}^{\Mass}$ can be modeled as
\begin{align}
\label{equation:dnoise:ma}
\begin{aligned}
d_{\noise,g}^{\Mass}\{r\mid m_{j}^{(P)}; y\}&=d_{\noise}^{\Mass}\{r\mid m_{j}^{(P)}; y\} = \pi_{\noise}\{ m_{j}^{(P)} \}\mathcal{U}_{\pm \omega}(r) + [1-\pi_{\noise}\{ m_{j}^{(P)} \}] d_{\signal}^{\Mass}(r \mid y)\\
\logit[\pi_{\noise}\{ m_{j}^{(P)} \}]&=f_{\noise}^{\Mass}\{ m_{j}^{(P)} \}, \quad \lambda_{\noise,g}^{\Mass}\{m_j^{(P)} \mid y\} = \lambda_{\noise,g}^{\Mass}\{m_j^{(P)}\}
\end{aligned}
\end{align}
for continuous functions $f_{\noise}^{\Mass}$ and $\lambda_{\noise,g}^{\Mass}$, where $\mathcal{U}_{\pm \omega}(r)$ is the density at $r$ of $U[-\omega,\omega]$. Since $d_{\signal}^{\Mass}$ was defined in \eqref{equation:mSignal}, $f_{\noise}^{\Mass}$ and $\lambda_{\noise,g}^{\Mass}$ are the only unknowns in \eqref{equation:dnoise:ma}. Here, $1-\pi_{\noise}\{ m_{j}^{(P)} \}$ is the probability the noise peak arose from a peptide fragment with predicted m/z $m_{j}^{(P)}$, and $\pi_{\noise}\{ m_{j}^{(P)} \}$ is the probability background noise or a different peptide fragment begat the noise peak. Informally, the latter set of noise peaks are uniformly distributed around $m_{j}^{(P)}$ because the combinatorial explosion of, and therefore our uncertainty in, possible noise peptide fragment m/z's suggests they, and therefore their begotten noise peaks, are equally likely to appear at any point in a small interval surrounding $m_{j}^{(P)}$. The function $f_{\noise}^{\Mass}\{ m_{j}^{(P)} \}$ is likely increasing, since the number of potential noise-generating fragment m/z's surrounding $m_{j}^{(P)}$, and consequently the likelihood the noise peak neighboring $m_{j}^{(P)}$ was generated by an m/z other than $m_{j}^{(P)}$, increases with $m_{j}^{(P)}$ (Figure~\ref{supp:Figure:Proof}). 

\subsection{Estimating Bayes factors}
\label{subsection:GlobalHyp}
Having defined all generating models, we can now compute $BF_g(P)$. For $\delta$ and $r_{\delta(j)}$ defined in steps (i) and (iv) of Algorithm \ref{algorithm:SpectrumGen}, $BF_g(P)$ is the product of the below four terms:
\begin{align*}
    &BF_g^{(\text{rt})}(P) \propto \Prob\{ \logit(t_g/M) \mid \Theta^{(\text{rt})},\irt_P \}, \quad \Theta^{(\text{rt})}=\{ f_{\signal}^{(\text{t})}, \sigma_{t,1}^2,\sigma_{t,2}^2,\pi_{t} \}\\
    &BF_g^{(\text{gen})}(P) = \frac{ \Prob[ \delta \mid \Theta^{(\text{gen})},\{ y_j^{(P)} \}_{j \in [n^{(P)}]} ] }{ \{n^{(S_g)}\}!/\{n^{(S_g)}-\abs*{\Signal}\}! \prod_{j: \delta(j)>0} \lambda_{\noise,g}^{\Mass}\{m_j^{(P)}\} }, \quad \Theta^{(\text{gen})} = \{\bm{\mu}_{\Gen},\bm{\Sigma}_{\Gen},\lambda_{\noise,g}^{\Mass}\}\\
    &BF_g^{(\text{int})}(P) = \frac{ \Prob[ \{ y_{\delta(j)}^{(S_g)} \}_{j: \delta(j)>0}, \{ y_i^{(S_g)} \}_{i \in \Noise} \mid \delta, \Theta^{(\text{int})},\{ y_j^{(P)} \}_{j \in [n^{(P)}]} ] }{ \int \prod_{i\in[n^{(S_g)}]} D_{\noise}^{\IntSuper}\{ y^{(S_g)}_i-\mu_g \} \Prob(\mu_g \mid \bm{\mu}_{\Int},\bm{\Sigma}_{\Int})\text{d}\mu_g }, \,\, \Theta^{(\text{int})}= \left\{\hspace{-2.5mm}\begin{array}{l}
         \mu_{\beta},\sigma_{\beta}^2,\phi,\nu,\\[1pt]
         \bm{\mu}_{\Int},\bm{\Sigma}_{\Int},D_{\noise}^{\IntSuper}
    \end{array}\hspace{-2.7mm}\right\}\\
    &BF_g^{(\text{ma})}(P) = \prod_{j: \delta(j)>0}\frac{  d_{\signal}^{\Mass}\{ r_{\delta(j)} \mid y_{\delta(j)}^{(S_g)} \} }{ d_{\noise}^{\Mass}\{ r_{\delta(j)} \mid m_{j}^{(P)};y_{\delta(j)}^{(S_g)} \} }, \quad \Theta^{(\text{ma})}=\{f_{\signal}^{\Mass},\,\sigma_{m,1}^2,\,\sigma_{m,2}^2, \,f_{\noise}^{\Mass} \}.
\end{align*}
The sets of hyperparameters $\Theta^{(\text{rt})}, \Theta^{(\text{gen})}, \Theta^{(\text{int})}$, and $\Theta^{(\text{ma})}$ are defined in \eqref{equation:RTModel}, \eqref{equation:Obs}, \eqref{equation:ySignal:prior} and steps (ii) and (v) of Algorithm \ref{algorithm:SpectrumGen}, and \eqref{equation:mSignal} and \eqref{equation:dnoise:ma}, respectively. Larger values of any of the above likelihood ratios provide evidence supporting $G_g(P)$, the hypothesis that spectrum $g$ was generated by $P$, and are therefore interpretable as ``scores'' for retention time (rt), peak generation (gen), peak intensity (int), and mass-accuracy (ma). The second follows because $BF_g^{(\text{gen})}(P)$ is interpretable as the ratio of the probabilities that we observe $\abs*{\Signal}$ peaks within $\omega$ppm of $\{m_j^{(P)}\}_{j: \delta(j)>0}$ under $G_g(P)$ and the noise model $N_g$. We show how we calculate the above four terms in Section~\ref{supp:subsection:BF:Calc} of the Supplement.\par
\indent Similar to previous work \citep{LikelihoodScoring,MSGFplus,NN_pred}, we use training data to estimate the hyperparameters, and derive three sets of estimators for spectra with precursor charges $+2$, $+3$ and $\geq +4$. Briefly, we use an existing search engine to curate a set of high confidence peptide-spectrum matches, use the procedure outlined in Remark \ref{remark:delta} to map predicted to observed peaks, and use standard methods to estimate all parameters except $f_{\noise}\subset \Theta^{(\text{ma})}$ and $\lambda_{\noise,g}^{\Mass} \subset \Theta^{(\text{gen})}$. Sections~\ref{supp:subsection:SignalHyp} and \ref{supp:subsection:NoiseHyp} of the Supplement contain the details.\par
\indent Recall $f_{\noise}^{\Mass}\{m_j^{(P)}\}$ and $\lambda_{\noise,g}^{\Mass}\{m_j^{(P)}\}$ characterize the distribution of noise peaks surrounding predicted m/z's $m_j^{(P)}$. To avoid confusing notation, we outline their estimators below, and provide exact algorithms in Section~\ref{supp:subsection:NoiseHyp} of the Supplement. In brief, we use high quality training spectra with their mapped signal peaks removed to estimate these functions. For each training spectrum $g$, we randomly draw a permuted peptide from the target database $\Targ$ with similar mass to spectrum $g$'s precursor mass, fragment it \textit{in silico}, and use the mapping procedure outlined in Remark~\ref{remark:delta} to map fragment to noise m/z's. These randomly generated fragments are samples from $\RandMZ$ defined in Remark~\ref{remark:theorem:poisson}, and represent fragments that could generate noise peaks. Having already estimated $\{f_{\signal}^{\Mass},\sigma_{m,1}^2,\sigma_{m,2}^2\}$ above, which defines the estimate for $d_{\signal}^{\Mass}$ in \eqref{equation:dnoise:ma}, we use the set of matched fragment-noise m/z pairs to estimate $f_{\noise}^{\Mass}$ via maximum likelihood. We find letting $f_{\noise}^{\Mass}\{m_j^{(P)}\}$ be linear in $\log\{m_j^{(P)}\}$ to be an appropriate choice for $f_{\noise}^{\Mass}$, where, consistent with our discussion in Section~\ref{subsection:NoiseMZ}, we estimate $f_{\noise}^{\Mass}$ to be increasing (Figure~\ref{Figure:TrainSignal}(h)).\par 
\indent For $\lambda_{\noise,g}^{\Mass}\{m_j^{(P)}\}$, we find it depends most on spectrum $g$'s precursor mass (Figure~\ref{supp:figure:NoiseObsOther}), and therefore partition spectra into precursor mass bins and assume $\lambda_{\noise,g}=\lambda_{\noise,h}$ if $g$ and $h$ are in the same bin. Using Theorem~\ref{theorem:noise}, which shows $\lambda_{\noise,g}$ can be approximated with a continuous function, we estimate $\lambda_{\noise,g}$ nonparametrically as a piecewise-constant function using the above randomly sampled peptide fragment m/z's, and smooth the estimator with B-splines. Our estimator for $+2$ charged spectra is given in Figure~\ref{Figure:TrainSignal}(i), where we only need one precursor mass bin for these spectra. Figures \ref{Figure:TrainSignal}(h) and \ref{Figure:TrainSignal}(i) indicate $1-\pi_{\noise}\{m_j^{(P)}\}$ and $\lambda_{\noise,g}^{\Mass}\{m_j^{(P)}\}$ are large for small $m_j^{(P)}$, and imply small peptide fragments tend to beget noise peaks. This has the effect of reducing the influence of low mass, non-specific fragment ions on Bayes factors. We illustrate the importance of this phenomenon is Section~\ref{section:Results}.

\section{Simulations}
\label{section:Simulations} 
To assess the fidelity of our method, we developed a novel simulation technique that uses real nanobody data to generate simulated data, and contains three important features. First, these simulated data consisted of two sets of spectra to mirror real data prone to both database incompleteness and score-ordering errors. Second, all simulated spectra contained real noise peaks, which allowed us to assess the veracity of our new model for spectral noise outlined in Section~\ref{subsection:NoiseMZ}. Lastly, we used hyperparameters derived from real group 1 and group 2 spectra, defined in Section~\ref{section:Data}, to simulate and analyze the data, respectively. As these groups of spectra were generated by different classes of nanobodies, this enabled evaluation of the inter-dataset generalizability of model hyperparameters defined in Section~\ref{subsection:GlobalHyp}.\par 
\indent The first set of spectra were subject to database incompleteness errors. To simulate them, we note that in practice we apply a common peptide screening procedure for each spectrum $S$ that discards a peptide if its predicted m/z lies outside a window of length $L_S$ surrounding $S$'s generator's observed m/z, as peptides outside this window could not have generated $S$. We therefore followed previous work \citep{TargetDecoy} and simulated these spectra by shifting generator m/z's of randomly chosen real group 1 spectra by 10m/z, which is substantially larger than $L_S$ for all spectra $S$. As such, these spectra were not generated by any screened peptides in the target database $\Targ$, and were therefore prone to database incompleteness errors.\par 
\indent Spectra in the second set, which were prone to score-ordering errors, were generated by peptides in $\Targ$ and consisted of real noise and simulated signal peaks. Briefly, we used existing software \citep{ProteomeDiscoverer} to identify group 1 spectra $S_g^{(1)}$, $g=1,\ldots,q^{(1)}=$\#group 1 spectra, whose generating peptides were likely in $\Targ$, but whose inferred matches may have been ambiguous. We removed matching signal peaks from each spectrum so the resulting spectra $S_{\noise,g}^{(1)}$, $g=1,\ldots, q^{(1)}$, contained only noise peaks. We then added simulated signal peaks to $S_{\noise,g}^{(1)}$ by choosing each simulated spectrum's generating peptide uniformly at random from $S_g^{(1)}$'s identified potential generators, and then simulating retention time and signal peaks according to \eqref{equation:RTModel} and steps (i)-(iv) of Algorithm~\ref{algorithm:SpectrumGen}, where the signal hyperparameters $\Theta_{\signal}=\{ \Theta^{\text{(rt)}}$,\allowbreak$\{\bm{\mu}_{\Gen}, \bm{\Sigma}_{\Gen}\}$,\allowbreak$\{\bm{\mu}_{\Int}, \bm{\Sigma}_{\Int}\}$,\allowbreak$\{\mu_{\beta},\sigma_{\beta}^2,\phi,v\}$,\allowbreak$\{f_{\signal}^{\Mass}, \sigma_{m,1}^2, \sigma_{m,2}^2\}\}$ were estimated using group 1 data. To evaluate our method's sensitivity to distribution assumptions, we substituted the normal distribution in the expression for $\epsilon_{gj}$ in \eqref{equation:ySignal:y} with a t-distribution with four degrees of freedom. Each dataset contained 7382 such spectra.\par
\begin{figure}
\centering
\includegraphics[width=0.75\textwidth]{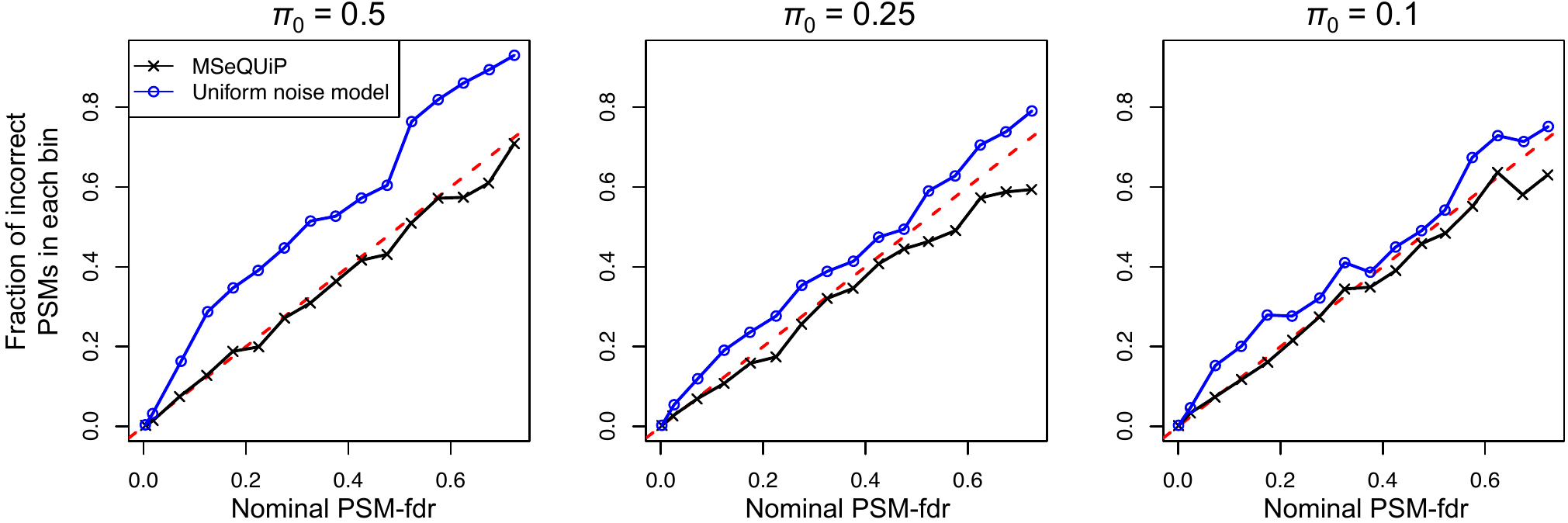}
\caption{Simulation results for different $\pi_0$'s, defined as the fraction of simulated spectra generated by an analyte outside of $\Targ$. The y-axis is the fraction of all simulated spectra in that $\PSMfdr$ bin whose inferred generator did not match its true generator. The dashed red line is $y=x$.}
\label{Figure:Simulation}
\end{figure}

\indent We analyzed each simulated spectrum using hyperparameter estimates for $\Theta_{\signal}$, $D_{\noise}^{\IntSuper}$, $\lambda_{\noise,g}^{\Mass}$, and $d_{\noise}^{\Mass}$ derived from the group 2 data, and used \eqref{equation:fdr} to infer each spectrum's generator $\hat{P}_g$ and $\PSMfdr_g$. To demonstrate the importance of using our novel model for noise m/z's defined by $\lambda_{\noise,g}^{\Mass}$ and $d_{\noise}^{\Mass}$ in Section~\ref{subsection:NoiseMZ}, we repeated this procedure by estimating Bayes factors using the standard assumption that noise m/z's are uniform on the smallest interval containing the spectrum's observed m/z's \citep{UniformNoise,LikelihoodScoring}. The results for $3 \times 30$ simulated datasets are given in Figure~\ref{Figure:Simulation}, and shows that our method, MSeQUiP, is well-calibrated despite the data being simulated and analyzed using data derived from orthogonal nanobody classes. The inflation exhibited by the uniform noise model suggests noise m/z's are non-uniform, where, unlike our model, the uniform model underestimates the probability noise peaks match signal peaks, which inflates Bayes factors and overstates the confidence in inferred generators. Section~\ref{supp:section:simulations} contains additional results evaluating our estimates for $\pi_0$ (see Figure~\ref{Figure:Simulation}) and comparing our Bayes factor score to other scoring functions. Since existing methods use different spectral features not considered by our data generating model to perform inference, it is unfair to use these simulations to compare MSeQUiP with existing methods. Instead, we postpone comparisons to our real data analysis below.


\section{Real data application} 
\label{section:Results}
\subsection{Data description}
\label{subsection:NanobodyData}
We demonstrate the efficacy of our novel framework and method by analyzing real mass spectrometry data from \citet{Nb_Yi}, whose goal was to identify nanobodies with antigen-specific binding affinities. The authors inoculated a single llama (\textit{Lama glama}) with a specific antigen (denoted $\Anto$), and created a custom target nanobody and decoy peptide databases $\Targ_{\text{a}}$ and $\Dec_{\text{a}}$. The entries of $\Targ_{\text{a}}$ contain peptides derived from nanobody proteins that could exist in the llama, which include both non-specific and $\Anto$-specific nanobodies. The authors then used two different experimental protocols to isolate $\Anto$-specific nanobodies from the llama's blood plasma, and digested nanobodies to create two protocol-determined groups of nanobody peptides. For purposes of calibration in Section~\ref{subsection:Calibration}, we also had access to a second database, $\Entrap$, generated from a second llama inoculated with a different antigen (denoted $\Antt$), and defined its decoy database $\Dec_{\text{b}}$ accordingly.\par 
\indent \citeauthor{Nb_Yi} found the second protocol was more capable of isolating drug-quality nanobodies. We therefore focus our analysis on the resulting second group of three LC-MS/MS datasets, denoted D1, D2, and D3, and use the first group of datasets to estimate hyperparameters. We used the pipeline outlined in Section~\ref{section:Data} to process all LC-MS/MS data and generate a library of predicted spectra and indexed retention times for peptides in $\Targ_{\text{a}} \cup \Dec_{\text{a}}$ and, for the purposes of Section~\ref{subsection:Calibration}, $\Entrap \cup \Dec_{\text{b}}$. Sections~\ref{supp:section:RealData} and \ref{supp:section:Analysis} in the Supplement contain additional data details and results for the first group of datasets.

\subsection{Assessment of our statistical framework}
\label{subsection:Sensitivity}
Here, we demonstrate the importance of our statistical framework outlined in Section~\ref{subsection:lfdr}, whereby we cast peptide-spectrum matching as a model selection problem with an incomplete candidate model space. We define a peptide-spectrum match (PSM) to be an inferred peptide-spectrum pair, where $\PSMfdr_g$ indicates our uncertainty in the $g$th PSM. Recall from Section~\ref{subsection:lfdr} that $\PSMfdr_g$ is an increasing function of the database incompleteness error rate, $\Spfdr_g$, and the score-ordering error rate, where a large value of the latter implies a peptide in $\Targ_{\text{a}}$ other than the inferred match could have generated the spectrum and, due to the similarity in their peptide sequences, is likely in nanobody proteomes. As small changes in a nanobody's sequence can have drastic effects on its binding properties \citep{IL_Antibody}, controlling the score-ordering error rate is essential for identifying antigen-specific nanobodies. We were therefore interested in validating the conclusions of Lemma~\ref{lemma:InflatePSMfdr} that existing methods underestimate score-ordering error rates, and consequently, $\PSMfdr$'s, and whether our framework can circumvent this.\par 
\indent To do so, we compared our proposed method MSeQUiP to three popular analysis pipelines, which first score PSMs with one of Crux \citep{Crux_Pvalues}, MS-GFplus \citep{MSGFplus}, or X!Tandem \citep{XTandem}, and subsequently use the post-processing software Percolator \citep{Percolator} to compute $\PSMfdr$'s. To assess MSeQUiP's capacity to prune PSMs prone to score-ordering error, we compared its reported $\Spfdr$'s and $\PSMfdr$'s, where a small $\Spfdr_g$ but large $\PSMfdr_g$ implies that while the spectrum was likely generated by a peptide in $\Targ_{\text{a}}$, there is ambiguity in the inferred match due to its high score-ordering error rate. Since existing methods only report $\PSMfdr$'s, we assessed whether they understate score-ordering errors by considering each significant PSM's delta score, defined as the difference in scores between the spectrum's highest and second highest scoring peptide. A small delta score is suggestive of a score-ordering error, since it indicates another peptide in $\Targ_{\text{a}}$ besides the highest scoring inferred match could have generated the spectrum. To gauge if there is an excess of such ambiguous PSMs, and because existing pipelines are validated using simpler yeast proteomes \citep{MSGFplus,Crux_Pvalues,Percolator}, we used delta scores derived from a yeast digest to determine each method's expected ambiguity.\par
\begin{figure}
    \centering
    \includegraphics[width=0.95\textwidth]{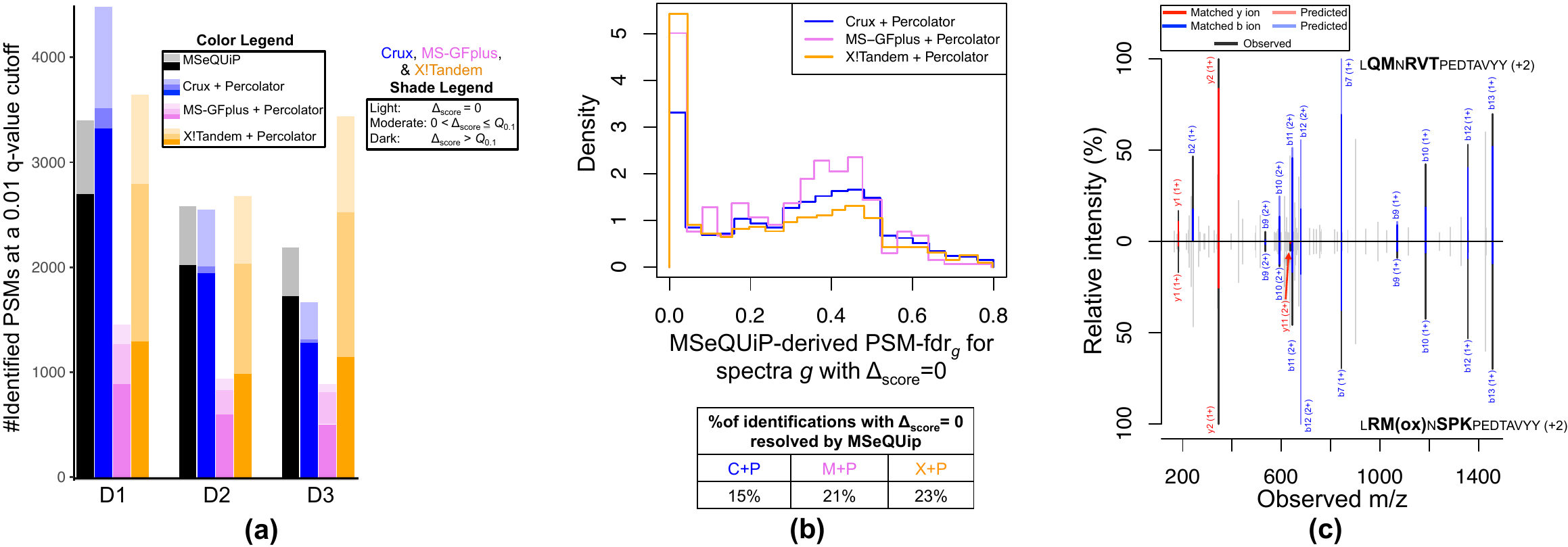}
    \caption{Overview of the nanobody peptide search results. (a): The height of MSeQUiP's black and black+grey bars are the number of PSMs using $\PSMfdr_g$ and $\Spfdr_g$, respectively, to compute q-values. PSM delta score ($\Delta_{\text{score}}$) and the 0.1 delta score quantile for yeast PSMs with q-value $\leq 0.01$ ($Q_{0.1}$) are utilized in the shade legend. PSMs with $\Delta_{\text{score}} \leq Q_{0.1}$ are defined to be ambiguous. (b): Histograms are truncated at 0.8, and resolved PSMs have MSeQUiP-derived $\PSMfdr$'s significant at the 0.01 q-value cutoff. (c): An example of a PSM identified by Crux with $\Delta_{\text{score}}=0$ that is resolved by MSeQUiP. Both peptides matched 12 out of 27 predicted peaks, which are indicated in red and blue. The top peptide, whose predicted peak intensities better match observed intensities, was MSeQUiP's inferred generator with $\PSMfdr=0.006$.}
    \label{Figure:AllResults} 
\end{figure}

  
\indent Figure~\ref{Figure:AllResults}(a) contains the results at a 1\% global false discovery rate. As suspected, existing methods return PSMs prone to score-ordering error, where Crux, MS-GFplus, and X!Tandem return over 2.5, four, and six times as many ambiguous nanobody PSMs compared to yeast. On the other hand, the grey bar, whose height gives the difference between the number of MSeQUiP-derived PSMs identified using $\Spfdr_g$ and $\PSMfdr_g$, reflects MSeQUiP's capacity to prune PSMs prone to score-ordering error. The fact that some methods ostensibly identify more PSMs than MSeQUiP is inconsequential, since, as suggested by their ambiguous identifications and shown in Section~\ref{subsection:Calibration}, these methods inflate error rates.

\indent We postulated that because Crux, MS-GFplus, and X!Tandem do not model predicted ion fragments' relative intensities, many of their ambiguities could be due to degeneracies in their predicted spectra, which could be resolved using MSeQUiP's relative intensity model. For example, predicted spectra, and therefore scores, for peptides that differ by amino acid isomers will be identical in existing methods, but will vary in MSeQUiP. This is tested in Figure~\ref{Figure:AllResults}(b), which suggests MSeQUiP can resolve several ambiguous, but nominally significant, Crux-, MS-GFplus-, or X!Tandem-derived PSMs, where Figure~\ref{Figure:AllResults}(c) gives an example. As expected, Figure~\ref{Figure:AllResults}(b) also indicates that MSeQUiP's $\PSMfdr$'s are large, i.e. $\gtrsim 0.5$, for many of these PSMs, implying MSeQUiP, unlike existing methods, is able to prune these ambiguous PSMs.

\subsection{CDR3 peptides}
\label{subsection:CDR3}
\begin{figure}
    \centering
\includegraphics[width=0.95\textwidth]{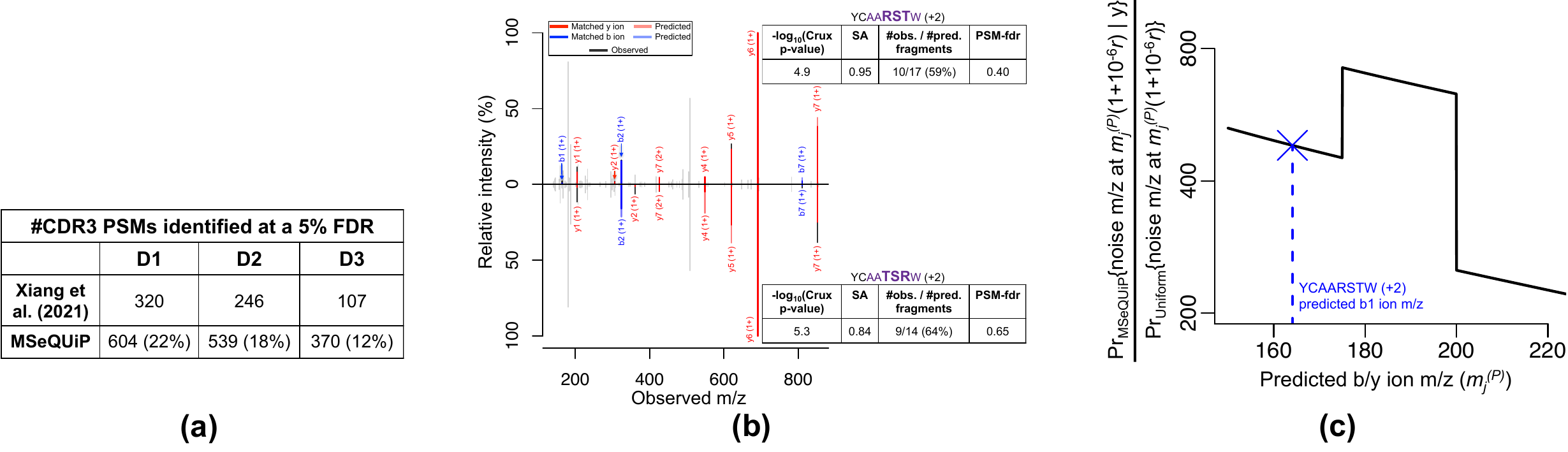}
    \caption{(a): Numbers in parentheses are the fraction of the CDR3 PSMs identified in \citet{Nb_Yi} that are not identified by MSeQUiP. (b): An example of a spectrum confidently matched to a CDR3 peptide in \citet{Nb_Yi} (bottom), but whose MSeQUiP-derived match (top) is ambiguous ($\PSMfdr = 0.40$). Violet letters are amino acids in the CDR3 region. The spectral angle (SA) is used to train the model that generates predicted spectra. An SA closer to 1 suggests a better match. (c): $r$ and $y$ are the relative ppm mass error and intensity of the observed b1 ion in (b); y-axis is the ratio between the likelihood a noise peak with intensity $y$ was generated at $m_j^{(P)}(1+10^{-6}r)$ under MSeQUiP's and the uniform noise m/z model.} 
    \label{Figure:CDR3Sensitivity}
\end{figure}

We next compare MSeQUiP to the method proposed in \citet{Nb_Yi}, which uses ad hoc criteria and the auxiliary database $\Entrap$ derived from the second llama inoculated with $\Antt$, and defined in Section~\ref{subsection:NanobodyData}, to filter PSMs identified by existing software by treating $\Entrap$ as an additional decoy database. As this method can only identify nanobody peptides in $\Targ_{\text{a}}$ that are specific to $\Anto$, they restrict their attention to peptides overlapping the complementarity determining region 3 (CDR3) sequence, where the CDR3 region is a contiguous, nanobody-specific subsequence of amino acids that nearly completely determines a nanobody's binding properties \citep{xu2000diversity}. Figure~\ref{Figure:CDR3Sensitivity}(a) shows that this method is underpowered in comparison to MSeQUiP, where we consider a 5\% global false discovery rate to be consistent with \citet{Nb_Yi}. Interestingly, $18\%$ of the CDR3 PSMs identified by \citeauthor{Nb_Yi} were not identified by MSeQUiP, which could be because \citeauthor{Nb_Yi} estimated 14\% of their identified nanobodies were false discoveries. Although the aforementioned filtering criteria likely alleviates some of the PSM ambiguity issues discussed in Section~\ref{subsection:Sensitivity}, their relatively large estimated false discovery proportion and failure to account for score-ordering error suggest they still return an excess of uncertain CDR3 PSMs, as CDR3 sequences typically only differ by a few amino acids \citep{Nb_Yi}. We give one such PSM in Figure~\ref{Figure:CDR3Sensitivity}(b), which was significant in \citeauthor{Nb_Yi} but, due to the similarity in fragment ion matches and predicted intensities, identified as ambiguous using MSeQUiP. This example also demonstrates the importance of our non-uniform noise m/z model described in Section~\ref{subsection:NoiseMZ}. Unlike MSeQUiP, the probability of observing a noise m/z adjacent to the b1 ion's low predicted m/z is small under the uniform model, and nominally differentiates the top peptide from the bottom by increasing its Bayes Factor and decreasing its $\PSMfdr$ by a factor of 500 (Figure \ref{Figure:CDR3Sensitivity}(c)).

\subsection{Fidelity of MSeQUiP's PSM-fdr}
\label{subsection:Calibration}
We lastly assessed the fidelity of MSeQUiP's $\PSMfdr$ to ensure it accurately reflects our uncertainty in PSMs. For example, if a set of PSMs have $\PSMfdr$ equal to 0.1, then about 10\% of them should be incorrect. Typical approaches to do so involve appending to $\Targ_{\text{a}}$ a set of randomly generated ``entrapment'' peptides known to be absent from the biological sample, where a PSM whose peptide belongs to the entrapment set is incorrect \citep{Entrapment1}. However, this entrapment set is inappropriate, as differences between entrapment and nanobody peptides are much larger than those between incorrect inferred and true nanobody peptide generators. Instead, we took inspiration from \citet{Nb_Yi} and designed an entrapment procedure using the nanobody peptides in $\Entrap$ defined in Section~\ref{subsection:NanobodyData}. In brief, we re-analyzed spectra using the concatenated target and decoy databases $\Targ_{\text{a}} \cup \Entrap$ and $\Dec_{\text{a}} \cup \Dec_{\text{b}}$. Since spectra were generated by peptides whose parent nanobodies are $\Anto$-specific, any inferred match from $\Entrap$, which contains $\Antt$-related nanobody peptides, would ideally be incorrect. However, due to the incompleteness of $\Targ_{\text{a}}$ and cross-antigen degeneracies in non-CDR3 regions \citep{xu2000diversity}, $\Entrap$ may contain legitimate generators absent from $\Targ_{\text{a}}$. To avoid overstating the number of false PSMs, we focused on CDR3 peptides, where, as discussed above, the antigen specificity of and variation in CDR3 sequences imply matches to CDR3 peptides in $\Entrap$ are incorrect \citep{Nb_Yi}. We estimated the true false discovery rate in a nominal $\PSMfdr$ window to be the fraction of CDR3 PSMs whose peptides lie in $\Entrap$, and include results for all PSMs in Section \ref{supp:subsection:Calibration} of the Supplement. While some PSMs with peptides in $\Targ_{\text{a}}$ may also be incorrect, we argue in Section \ref{supp:subsection:EntrapmentEst} of the Supplement that their contribution is likely minor. In addition to assessing $\PSMfdr$'s, this procedure also helps judge the veracity of MSeQUiP's CDR3 PSMs from Section~\ref{subsection:CDR3}, which currently form the basis for antigen-specific nanobody inference \citep{Nb_Yi}.\par
\begin{figure}
    \centering
    \includegraphics[width=0.95\textwidth]{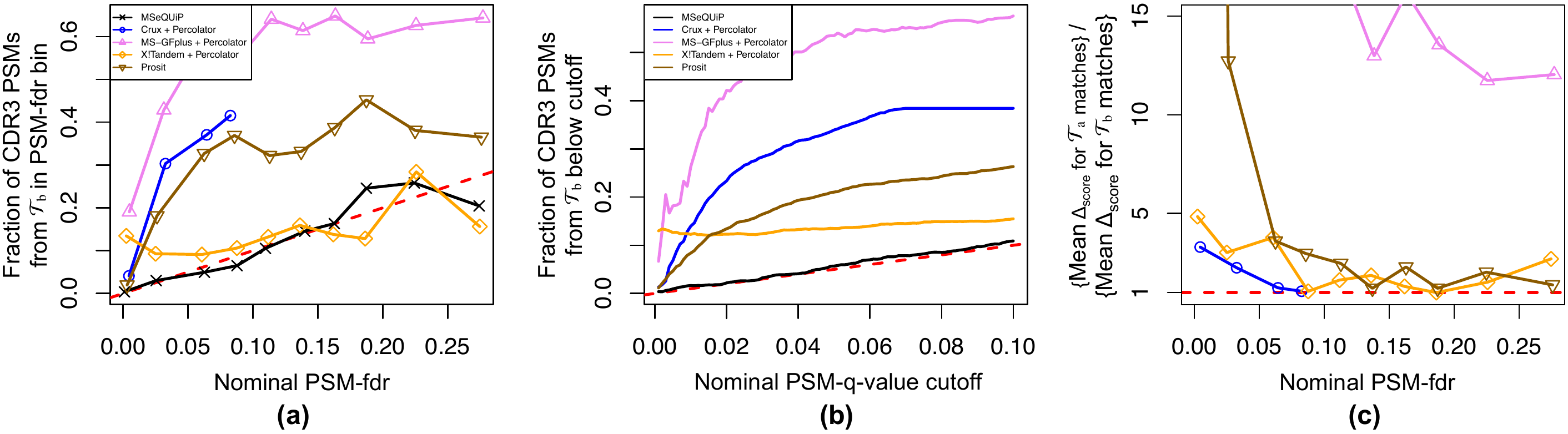}
    \caption{$\PSMfdr$ calibration, where results were pooled across datasets D1, D2, and D3 to increase precision. The black lines give results for MSeQUiP in (a) and (b), and dashed red lines are $y=x$ in (a) and (b), and $y=1$ in (c). MSeQUiP is not plotted in (c).} 
    \label{Figure:CDR3_FDR}
\end{figure}

 
\indent In addition to MSeQUiP and the three methods considered in Section~\ref{subsection:Sensitivity}, we used this entrapment procedure to assess the fidelity of $\PSMfdr$'s reported by Prosit, the method proposed in \citet{NN_pred} to re-score MaxQuant \citep{MaxQuant} PSMs. Notably, Prosit and MSeQUiP use the same predicted spectra to score PSMs. However, Prosit uses a black box algorithm that ignores score-ordering error to compute $\PSMfdr$'s. We could not use this entrapment procedure to evaluate the method proposed in \citet{Nb_Yi}, as their reported error rates are defined by it. Figures~\ref{Figure:CDR3_FDR}(a) and \ref{Figure:CDR3_FDR}(b) show that MSeQUiP is the only method to accurately estimate local and global PSM false discovery rates, where MSeQUiP's superb calibration is further verified in Figure \ref{supp:Figure:HighAndLowPEP} in the Supplement, which gives results for the first group of datasets discussed in Section~\ref{subsection:NanobodyData}. Since the primary difference between MSeQUiP and existing methods is the latter's failure to model score-ordering error, we sought to determine if an abundance of score-ordering errors could account for their inflated error rates. To do so, we compared delta scores between PSMs whose peptide belonged to one of $\Targ_{\text{a}}$ or $\Entrap$, where the latter set of delta scores typify those of false PSMs. Since, as discussed in Section~\ref{subsection:Sensitivity}, small delta scores are indicative of potential score-ordering errors, Figure~\ref{Figure:CDR3_FDR}(c) suggests errant score-ordering is responsible for many of the existing methods' incorrect PSMs. 

\section{Discussion}
\label{section:Discussion}
We have developed a novel framework and method to infer peptide-spectrum matches (PSMs) in tandem mass spectrometry data that casts peptide-spectrum matching as a model selection problem with an incomplete model space. To our knowledge, our work is the first to account for both score-ordering and database incompleteness errors without relying on the assumption that the former mirror decoy peptide matches. We demonstrated the improved performance of our method using simulated and real nanobody data.\par
\indent Our Bayes factors $BF_g(P)$ utilize recently developed deep learning-based spectrum and indexed retention time prediction tools. While we are not the first to leverage these tools, available software merely uses them to estimate error rates for previously identified PSMs \citep{NN_pred}. Instead, our data generating model in Algorithm~\ref{algorithm:SpectrumGen}, which we use to define $BF_g(P)$, illustrates how these tools can better identify PSMs by discriminating between a spectrum's potential peptide generators with similar amino acid sequences. It also implies spectrum prediction can be improved. For example, rather than predicting all peaks' relative intensities by optimizing their Euclidean distance from observed relative intensities \citep{NN_pred}, steps (i) and (iii) in Algorithm~\ref{algorithm:SpectrumGen} suggest separately predicting whether peaks are observed and their log-intensities could improve discriminatory power.\par 
\indent In addition to spectrum prediction, there are several important areas of future research. These include experimentally validating our results from Section~\ref{section:Results}, using CDR3 and other PSMs to quantify uncertainty in inferred nanobody proteins, using our framework to analyze proteogenomic, metaproteomic, and other notoriously challenging proteomic data, as well as designing integrated software to predict spectra, estimate hyperparameters from Algorithm~\ref{algorithm:SpectrumGen}, and infer PSMs.

\printbibliography

@article{MatchUncertainty,
	author = {Keich, Uri and Noble, William Stafford},
	journal = {Journal of the American Statistical Association},
	month = {07},
	number = {523},
	pages = {973--982},
	title = {Controlling the {FDR} in Imperfect Matches to an Incomplete Database},
	volume = {113},
	year = {2018}
}

@article{CruxSoftware,
	author = {Diament, Benjamin J. and Noble, William Stafford},
	journal = {Journal of Proteome Research},
	month = {09},
	number = {9},
	pages = {3871--3879},
	title = {Faster {SEQUEST} Searching for Peptide Identification from Tandem Mass Spectra},
	volume = {10},
	year = {2011}
}

@article{Sequest,
	author = {Eng, Jimmy K. and McCormack, Ashley L. and Yates, John R.},
	journal = {Journal of the American Society for Mass Spectrometry},
	number = {11},
	pages = {976--989},
	title = {An approach to correlate tandem mass spectral data of peptides with amino acid sequences in a protein database},
	volume = {5},
	year = {1994}
}

@article{XTandem,
	author = {Craig, Robertson and Beavis, Ronald C.},
	journal = {Rapid Communications in Mass Spectrometry},
	month = {09},
	number = {20},
	pages = {2310--2316},
	title = {A method for reducing the time required to match protein sequences with tandem mass spectra},
	volume = {17},
	year = {2003}
}

@Manual{qvalueSoftware,
    title = {qvalue: {Q}-value estimation for false discovery rate control},
    author = {John D. Storey and Andrew J. Bass and Alan Dabney and David Robinson},
    year = {2019},
    note = {R package version 2.18.0},
    url = {http://github.com/jdstorey/qvalue},
  }

@article{qvalue,
	author = {Storey, John D. },
	journal = {Journal of the Royal Statistical Society: Series B (Statistical Methodology)},
	month = {08},
	number = {3},
	pages = {479--498},
	title = {A direct approach to false discovery rates},
	volume = {64},
	year = {2002}
}

@article{MaxQuant,
	author = {Tyanova, Stefka and Temu, Tikira and Cox, Juergen},
	journal = {Nature Protocols},
	number = {12},
	pages = {2301--2319},
	title = {The {MaxQuant} computational platform for mass spectrometry-based shotgun proteomics},
	volume = {11},
	year = {2016}
}

@article{Prediction_Cox,
	author = {Tiwary, Shivani and Levy, Roie and Gutenbrunner, Petra and Salinas Soto, Favio and Palaniappan, Krishnan K. and Deming, Laura and Berndl, Marc and Brant, Arthur and Cimermancic, Peter and Cox, J{\"u}rgen},
	journal = {Nature Methods},
	number = {6},
	pages = {519--525},
	title = {High-quality {MS/MS} spectrum prediction for data-dependent and data-independent acquisition data analysis},
	volume = {16},
	year = {2019}
}

@article{PoiDens,
	author = {Bradley Efron and Robert Tibshirani},
	journal = {The Annals of Statistics},
	month = {12},
	number = {6},
	pages = {2431--2461 },
	title = {Using specially designed exponential families for density estimation},
	volume = {24},
	year = {1996}
}

@article{MassAccuracyInt,
  title={Best Practice Guide: {M}ethodology for Accurate Mass Measurement of Small Molecules},
  author={Webb, K and Bristow, ATW and Sargent, M and Stein, BK},
  journal={London: LGC Ltd},
  year={2004}
}

@article{UniformNoise,
	author = {Wan, Yunhu and Yang, Austin and Chen, Ting},
	journal = {Analytical Chemistry},
	month = {01},
	number = {2},
	pages = {432--437},
	title = {{PepHMM}:  A Hidden Markov Model Based Scoring Function for Mass Spectrometry Database Search},
	volume = {78},
	year = {2006}
}

@article{Background_Indep_mz2,
	author = {Gallia, Jason and Lavrich, Katelyn and Tan-Wilson, Anna and Madden, Patrick H.},
	journal = {BMC Genomics},
	number = {7},
	pages = {S2},
	title = {Filtering of MS/MS data for peptide identification},
	volume = {14},
	year = {2013}
}

@article{ContinuousNoise,
	author = {Awan, Muaaz Gul and Saeed, Fahad},
	journal = {09},
	month = {09},
	number = {20},
	pages = {1800206},
	title = {{MaSS}-Simulator: A Highly Configurable Simulator for Generating {MS/MS} Datasets for Benchmarking of Proteomics Algorithms},
	volume = {18},
	year = {2018}
}

@article{CRMs,
	author = {John Frank Charles Kingman},
	journal = {Pacific Journal of Mathematics},
	month = {1},
	number = {1},
	pages = {59--78 },
	title = {Completely random measures},
	volume = {21},
	year = {1967}
}

@article{StatNormal,
	author = {Ryu, Soyoung and Goodlett, David R and Noble, William S and Minin, Vladimir N},
	journal = {Proceedings. IEEE International Conference on Bioinformatics and Biomedicine},
	month = {10},
	pages = {648--653},
	title = {A statistical approach to peptide identification from clustered tandem mass spectrometry data},
	year = {2012}
}

@article{TukeyGandH,
  doi = {10.1111/j.1740-9713.2019.01273.x},
  year = {2019},
  month = may,
  publisher = {Wiley},
  volume = {16},
  number = {3},
  pages = {12--13},
  author = {Yuan Yan and Marc G. Genton},
  title = {The Tukey g-and-h distribution},
  journal = {Significance}
}

@article{QLS_Tukey,
	author = {Xu, Yihuan and Iglewicz, Boris and Chervoneva, Inna},
	journal = {Computational statistics \& data analysis},
	month = {07},
	pages = {66--80},
	title = {Robust Estimation of the Parameters of g-and-h Distributions, with Applications to Outlier Detection},
	volume = {75},
	year = {2014}
}

@Manual{Ashr,
    title = {ashr: {Methods} for Adaptive Shrinkage, using Empirical {B}ayes},
    author = {Matthew Stephens and Peter Carbonetto and David Gerard and Mengyin Lu and Lei Sun and Jason Willwerscheid and Nan Xiao},
    year = {2020},
    note = {R package version 2.2-50},
    url = {https://github.com/stephens999/ashr}
}

@article{Percolator,
	Author = {The, Matthew and MacCoss, Michael J and Noble, William S and K{\"a}ll, Lukas},
	Journal = {Journal of the American Society for Mass Spectrometry},
	Month = {11},
	Number = {11},
	Pages = {1719--1727},
	Title = {Fast and Accurate Protein False Discovery Rates on Large-Scale Proteomics Data Sets with {P}ercolator 3.0},
	Volume = {27},
	Year = {2016}
}

@article{MSGFplus,
	Author = {Kim, Sangtae and Pevzner, Pavel A.},
	Journal = {Nature Communications},
	Number = {1},
	Pages = {5277},
	Title = {{MS-GF+} makes progress towards a universal database search tool for proteomics},
	Volume = {5},
	Year = {2014}
}

@article{NN_pred,
	Author = {Gessulat, Siegfried and Schmidt, Tobias and Zolg, Daniel Paul and Samaras, Patroklos and Schnatbaum, Karsten and Zerweck, Johannes and Knaute, Tobias and Rechenberger, Julia and Delanghe, Bernard and Huhmer, Andreas and Reimer, Ulf and Ehrlich, Hans-Christian and Aiche, Stephan and Kuster, Bernhard and Wilhelm, Mathias},
	Journal = {Nature Methods},
	Number = {6},
	Pages = {509--518},
	Title = {Prosit: proteome-wide prediction of peptide tandem mass spectra by deep learning},
	Volume = {16},
	Year = {2019}
}

@article{AutoRT,
	author = {Wen, Bo and Li, Kai and Zhang, Yun and Zhang, Bing},
	journal = {Nature Communications},
	number = {1},
	pages = {1759},
	title = {Cancer neoantigen prioritization through sensitive and reliable proteogenomics analysis},
	volume = {11},
	year = {2020}
}

@article{BayesNetwork,
	author = {Klammer, Aaron A and Reynolds, Sheila M and Bilmes, Jeff A and MacCoss, Michael J and Noble, William Stafford},
	journal = {Bioinformatics (Oxford, England)},
	month = {07},
	number = {13},
	pages = {i348--i356},
	title = {Modeling peptide fragmentation with dynamic Bayesian networks for peptide identification},
	volume = {24},
	year = {2008}
}

@article{Nb_overview,
  doi = {10.1038/363446a0},
  year = {1993},
  month = jun,
  publisher = {Springer Science and Business Media {LLC}},
  volume = {363},
  number = {6428},
  pages = {446--448},
  author = {C. Hamers-Casterman and T. Atarhouch and S. Muyldermans and G. Robinson and C. Hammers and E. Bajyana Songa and N. Bendahman and R. Hammers},
  title = {Naturally occurring antibodies devoid of light chains},
  journal = {Nature}
}

@article{DRIP,
	author = {Halloran, John T and Bilmes, Jeff A and Noble, William S},
	journal = {Uncertainty in artificial intelligence},
	pages = {320--329},
	title = {Learning Peptide-Spectrum Alignment Models for Tandem Mass Spectrometry},
	volume = {30},
	year = {2014}
}

@article{LikelihoodScoring,
author = "Li, Qunhua and Eng, Jimmy K. and Stephens, Matthew",
doi = "10.1214/12-AOAS568",
journal = "The Annals of Applied Statistics",
month = "12",
number = "4",
pages = "1775--1794",
publisher = "The Institute of Mathematical Statistics",
title = "A likelihood-based scoring method for peptide identification using mass spectrometry",
volume = "6",
year = "2012"
}

@article{Nb_Potential,
	author = {Jov{\v c}evska, Ivana and Muyldermans, Serge},
	journal = {BioDrugs},
	number = {1},
	pages = {11--26},
	title = {The Therapeutic Potential of Nanobodies},
	volume = {34},
	year = {2020}
}

@article {Nb_Covid,
	author = {Xiang, Yufei and Nambulli, Sham and Xiao, Zhengyun and Liu, Heng and Sang, Zhe and Duprex, W. Paul and Schneidman-Duhovny, Dina and Zhang, Cheng and Shi, Yi},
	title = {Versatile and multivalent nanobodies efficiently neutralize {SARS-CoV-2}},
	volume = {370},
	number = {6523},
	pages = {1479--1484},
	year = {2020},
	doi = {10.1126/science.abe4747},
	issn = {0036-8075},
	journal = {Science}
}

@article{ProteomeDiscoverer,
  doi = {10.3390/proteomes9010015},
  year = {2021},
  month = mar,
  volume = {9},
  number = {1},
  pages = {15},
  author = {Benjamin C. Orsburn},
  title = {Proteome Discoverer{\textemdash}A Community Enhanced Data Processing Suite for Protein Informatics},
  journal = {Proteomes}
}

@article{UniformNoise2,
	author = {Cannon, William R. and Jarman, Kristin H. and Webb-Robertson, Bobbie-Jo M. and Baxter, Douglas J. and Oehmen, Christopher S. and Jarman, Kenneth D. and Heredia-Langner, Alejandro and Auberry, Kenneth J. and Anderson, Gordon A.},
	journal = {Journal of Proteome Research},
	month = {10},
	number = {5},
	pages = {1687--1698},
	title = {Comparison of Probability and Likelihood Models for Peptide Identification from Tandem Mass Spectrometry Data},
	volume = {4},
	year = {2005}
}

@article{IL_Antibody,
	author = {Bagal, Dhanashri and Kast, Eddie and Cao, Ping},
	journal = {Analytical Chemistry},
	month = {01},
	number = {1},
	pages = {720--727},
	title = {Rapid Distinction of Leucine and Isoleucine in Monoclonal Antibodies Using Nanoflow {LCMS$^{\text{n}}$}},
	volume = {89},
	year = {2017}
}

@article{Nb_Yi,
title = {Integrative proteomics identifies thousands of distinct, multi-epitope, and high-affinity nanobodies},
journal = {Cell Systems},
year = {2021},
issn = {2405-4712},
volume = {12},
number = {3},
pages = {220--234.e9},
doi = {https://doi.org/10.1016/j.cels.2021.01.003},
author = {Yufei Xiang and Zhe Sang and Lirane Bitton and Jianquan Xu and Yang Liu and Dina Schneidman-Duhovny and Yi Shi}
}

@article{Andromeda,
	author = {Cox, J{\"u}rgen and Neuhauser, Nadin and Michalski, Annette and Scheltema, Richard A. and Olsen, Jesper V. and Mann, Matthias},
	journal = {Journal of Proteome Research},
	month = {04},
	number = {4},
	pages = {1794--1805},
	title = {Andromeda: A Peptide Search Engine Integrated into the {MaxQuant} Environment},
	volume = {10},
	year = {2011}
}

@article{NatureNano,
	author = {Fridy, Peter C and Li, Yinyin and Keegan, Sarah and Thompson, Mary K and Nudelman, Ilona and Scheid, Johannes F and Oeffinger, Marlene and Nussenzweig, Michel C and Feny{\"o}, David and Chait, Brian T and Rout, Michael P},
	journal = {Nature Methods},
	number = {12},
	pages = {1253--1260},
	title = {A robust pipeline for rapid production of versatile nanobody repertoires},
	volume = {11},
	year = {2014}
}

@article{TargetDecoy,
	author = {Elias, Joshua E and Gygi, Steven P},
	journal = {Nature Methods},
	number = {3},
	pages = {207--214},
	title = {Target-decoy search strategy for increased confidence in large-scale protein identifications by mass spectrometry},
	volume = {4},
	year = {2007}
}

@article{Crux_Pvalues,
	author = {Howbert, J. Jeffry and Noble, William Stafford},
	journal = {Molecular \& Cellular Proteomics},
	number = {9},
	pages = {2467--2479},
	title = {Computing Exact p-values for a Cross-correlation Shotgun Proteomics Score Function},
	volume = {13},
	year = {2014}
}

@article{PSM_overview,
	author = {K{\"a}ll, Lukas and Vitek, Olga},
	journal = {PLOS Computational Biology},
	month = {12},
	number = {12},
	pages = {e1002277--},
	title = {Computational Mass Spectrometry--Based Proteomics},
	volume = {7},
	year = {2011}
}

@article{PEP_kall,
    author = {K{\"a}ll, Lukas and Storey, John D. and Noble, William Stafford},
    title = "{Non-parametric estimation of posterior error probabilities associated with peptides identified by tandem mass spectrometry}",
    journal = {Bioinformatics},
    volume = {24},
    number = {16},
    pages = {i42-i48},
    year = {2008},
    month = {08},
    issn = {1367-4803},
    doi = {10.1093/bioinformatics/btn294}
}

@article{OutsideDatabase,
	author = {Keich, Uri and Kertesz-Farkas, Attila and Noble, William Stafford},
	journal = {Journal of Proteome Research},
	month = {08},
	number = {8},
	pages = {3148--3161},
	title = {Improved False Discovery Rate Estimation Procedure for Shotgun Proteomics},
	volume = {14},
	year = {2015}
}

@article{StoreyControl,
	author = {Storey, John D. and Taylor, Jonathan E. and Siegmund, David},
	journal = {Journal of the Royal Statistical Society: Series B (Statistical Methodology)},
	month = {12},
	number = {1},
	pages = {187--205},
	title = {Strong control, conservative point estimation and simultaneous conservative consistency of false discovery rates: a unified approach},
	volume = {66},
	year = {2004}
}

@article{SimScore,
	author = {Sun, Shaojun and Meyer-Arendt, Karen and Eichelberger, Brian and Brown, Robert and Yen, Chia-Yu and Old, William M. and Pierce, Kevin and Cios, Krzysztof J. and Ahn, Natalie G. and Resing, Katheryn A.},
	journal = {Molecular \& Cellular Proteomics},
	number = {1},
	pages = {1--17},
	title = {Improved Validation of Peptide {MS/MS} Assignments Using Spectral Intensity Prediction},
	volume = {6},
	year = {2007}
}

@article{PeakFiltering,
	author = {Carvalho, Paulo C and Xu, Tao and Han, Xuemei and Cociorva, Daniel and Barbosa, Valmir C and Yates, John R, 3rd},
	journal = {Bioinformatics (Oxford, England)},
	month = {10},
	number = {20},
	pages = {2734--2736},
	title = {{YADA}: a tool for taking the most out of high-resolution spectra},
	volume = {25},
	year = {2009}
}

@article{Deisotope,
	author = {Goldfarb, Dennis and Lafferty, Michael J. and Herring, Laura E. and Wang, Wei and Major, Michael B.},
	journal = {ACS Omega},
	month = {09},
	number = {9},
	pages = {11383--11391},
	title = {Approximating Isotope Distributions of Biomolecule Fragments},
	volume = {3},
	year = {2018}
}

@article{xu2000diversity,
  title={Diversity in the {CDR3} region of {V(H)} is sufficient for most antibody specificities},
  author={Xu, John L and Davis, Mark M},
  journal={Immunity},
  volume={13},
  number={1},
  pages={37--45},
  year={2000}
}

@article{Entrapment1,
	author = {Feng, Xiao-Dong and Li, Li-Wei and Zhang, Jian-Hong and Zhu, Yun-Ping and Chang, Cheng and Shu, Kun-Xian and Ma, Jie},
	journal = {BMC genomics},
	month = {03},
	number = {Suppl 2},
	pages = {143--143},
	title = {Using the entrapment sequence method as a standard to evaluate key steps of proteomics data analysis process},
	volume = {18},
	year = {2017}
}

\newpage
\begin{center}
    {\Large Supplementary material for ``A novel framework to quantify uncertainty in peptide-tandem mass spectrum matches with application to nanobody peptide identification''}
\end{center}
\vspace{4mm}

\setcounter{equation}{0}
\setcounter{theorem}{0}
\setcounter{assumption}{0}
\setcounter{myalgorithm}{0}
\setcounter{lemma}{0}
\setcounter{corollary}{0}
\setcounter{figure}{0}
\setcounter{table}{0}
\setcounter{section}{0}
\setcounter{remark}{0}
\renewcommand{\thefigure}{S\arabic{figure}}
\renewcommand{\thesubfigure}{(\alph{subfigure})}
\renewcommand{\theequation}{\thesection.\arabic{equation}}
\renewcommand{\thesection}{S\arabic{section}}
\renewcommand{\thetheorem}{\thesection.\arabic{theorem}}
\renewcommand{\thedefinition}{\thesection.\arabic{definition}}
\renewcommand{\theassumption}{\thesection.\arabic{assumption}}
\renewcommand{\thelemma}{\thesection.\arabic{lemma}}
\renewcommand{\thecorollary}{\thesection.\arabic{corollary}}
\renewcommand{\theproposition}{\thesection.\arabic{proposition}}
\renewcommand{\theremark}{\thesection.\arabic{remark}}
\renewcommand{\themyalgorithm}{\thesection.\arabic{myalgorithm}}

\section{LC-MS/MS data processing}
\label{supp:section:Processing}
\subsection{Fragment de-isotoping and charge assignment}
\label{supp:subsection:qc}
We converted Thermo RAW files to mzML format and de-isotoped MS/MS fragments and inferred their charge using an in-house method that extends \citet{Deisotope}. Our method scores observed and expected isotopic profile pairs, and, provided the score is large enough, replaces the observed profile's peak insensities and m/z's with its summed intensities and putative monoisotopic m/z. To describe the method, consider an MS/MS spectrum with precursor monoisotopic mass $M_p$, charge $z_p \geq 2$, and precursor isolation window $\Delta_p$, and, for $f \in \{1,\ldots,\text{\#observed fragments in the MS/MS spectrum}\}$, let $m_{f_0}$ be an observed fragment's m/z. For posited charge $z_f \in \{1,\ldots,z_p\}$, let $\mathcal{G} = \{m_{f_0},z_f,M_p,z_p,\Delta_p\}$, $j_{\max}=\floor{\Delta_p z_p}$, and
\begin{align*}
    \mathcal{F}_j(M) &= \{\text{we sample a fragment's $A_j$ peak with monoisotopic mass $M$}\}, \quad j \in \{0,1,2,\ldots\}\\
    \mathcal{P}_j(M) &= \{\text{we sample a precursor's $A_j$ peak with monoisotopic mass $M$}\}, \quad j \in \{0,1,2,\ldots\}.
\end{align*}
We assume that (i) isotopic peaks outside the precursor's isolation window could not have generated peaks in its corresponding MS/MS spectrum, and (ii) that the number of neutrons in each of the precursor's atoms is independent across atoms. Let
\begin{align*}
    R_j(\mathcal{G})=& \E(\text{Relative intensity of the fragment's $A_j$ peak} \mid \mathcal{G}), \quad j \in \{0,1,2,\ldots\}\\
    \hat{g}_j(M)=& \widehat{\Prob}\{\mathcal{F}_j(M)\} = \sum_{k=0}^{r_j}\hat{\beta}_{j,k}M^k, \quad j \in \{0,1,2,\ldots\},
\end{align*}
where the estimates $\hat{\beta}_{j,k}$ were obtained by first calculating $\Prob\{\mathcal{F}_j(M)\}$ for a randomly chosen subset of nanobody peptide b and y ion fragments, and subsequently regressing them onto $M,\ldots,M^{r_j}$. Then assuming for simplicity that the precursor isolation window is centered at $M_p/z_p$,
\begin{align*}
     R_j(\mathcal{G})=& \Prob\{\mathcal{F}_j(m_{f_0} z_f) \mid \mathcal{P}_0(M_p) \cup\cdots\cup \mathcal{P}_{j_{\max}}(M_p),\mathcal{G}\}= \frac{\sum_{s=j}^{j_{\max}}\Prob\{\mathcal{F}_j(m_{f_0} z_f)\cap \mathcal{P}_s(M_p) \mid \mathcal{G}\}}{\sum_{s=0}^{j_{\max}} \Prob\{\mathcal{P}_s(M_p) \mid \mathcal{G}\}}\\
     =& \frac{\sum_{s=j}^{j_{\max}}\Prob\{\mathcal{F}_j(m_{f_0} z_f)\} \Prob\{\mathcal{F}_{s-j}(M_p-m_{f_0} z_f)\}}{\sum_{s=0}^{j_{\max}} \Prob\{\mathcal{F}_s(M_p)\}}, \quad j\in \{0,1,\ldots,j_{\max}\},
\end{align*}
which suggests the following estimator for $R_j(\mathcal{G})$:
\begin{align*}
    \hat{R}_j(\mathcal{G}) = \begin{cases}
    \frac{\sum_{s=j}^{j_{\max}} \hat{g}_{j}(m_{f_0} z_f)\hat{g}_{s-j}(M_p-m_{f_0} z_f)}{\sum_{s=0}^{j_{\max}}\hat{g}_s(M_p)} & \text{if $j \in \{0,1,\ldots,j_{\max}\}$}\\
    0 & \text{if $j > j_{\max}$}
    \end{cases}.
\end{align*}
For observed fragment m/z's $m_{f_0},m_{f_1},\ldots,m_{f_{j_{\max}}}$ and intensities $O_{f_0},O_{f_1},\ldots,O_{f_{j_{\max}}}$, isotopic profiles were scored using a standardized chi-squared statistic
\begin{align*}
    t = \frac{\sum_{j=0}^{j_{\max}} \{ \hat{R}_j(\mathcal{G})-O_{f_j}/K \}^2}{(j_{\max}+1)\sum_{j=0}^{j_{\max}}\hat{R}_j(\mathcal{G})} = \frac{\sum_{j=0}^{j_{\max}} \{ \hat{R}_j(\mathcal{G})-O_{f_j}/K \}^2}{j_{\max}+1}, \quad K = \sum_{j=0}^{j_{\max}} O_{f_j}
\end{align*}
We defined $\{(m_{f_0},O_{f_0}),(m_{f_{1}},O_{f_1}),\ldots,(m_{f_{j_{\max}}},O_{f_{j_{\max}}})\}$ as an isotopic profile with monoisotopic m/z $m_{f_0}$, intensity $K$, and charge $z_f$ if $m_{f_{j}}-m_{f_0} \in j\eta/z_f(1 \pm m_{f_0} \omega_I 10^{-6})$ for $j=1,\ldots,j_{\max}$ and $t \leq t_{\max}$, where $\omega_I=15$, $\eta=1.00284$ is the average mass of a neutron, and $t_{\max}$ is a user-defined quantity. We set $t_{\max}$ in our application so that when we randomly chose $j_{\max}+1$ peak intensities from a training MS/MS spectrum, $t \geq t_{\max}$ no more than 10\% of the time. This threshold was large enough to recover nearly 95\% of the matched b and y ion isotopic profiles from our training data. While most methods use a dot product to score potential isotopic profiles, our modified chi-squared statistic provided the greatest discriminating power. This method, which is computationally efficient due to $\hat{g}_j$ being a simple polynomial function, is available as a stand-alone \texttt{R} function.

\subsection{Transforming observed and predicted relative fragment intensities}
\label{supp:subsection:FragmentNormal}
Suppose peptide $P$ is spectrum $S$'s generating peptide, and let $y_{j}^{(P)} \in (0,1]$ and $y_{\delta(j)}^{(S)} \in (0,1]$, $j=1,\ldots,n$, be predicted fragment $j$'s predicted and observed relative intensity, where $\delta(j)$ indexes the matched fragment in spectrum $S$. We used the Box Cox family of transformations to transform $y_{j}^{(P)}$ and $y_{\delta(j)}^{(S)}$ to ensure their joint distribution was approximately normal. That is, for $\lambda \in \mathbb{R}$, we considered transformations of the form
\begin{align}
\label{supp:equation:BoxCox}
    \tilde{\bm{y}}_{\lambda,j}= \begin{cases}
    ([\{y_{j}^{(P)}\}^{\lambda}-1]/\lambda, [\{y_{\delta(j)}^{(S)}\}^{\lambda}-1]/\lambda)^T & \text{if $\lambda \neq 0$}\\
    (\log\{y_{j}^{(P)}\}, \log\{y_{\delta(j)}^{(S)}\})^T & \text{if $\lambda = 0$}
    \end{cases},
\end{align}
and for a given pair ($P$,$S$), maximized the profile likelihood
\begin{align*}
    \ell^{(P,S)}(\lambda) =  \abs{\bm{J}_{\lambda}}^{-1}\argmax_{\substack{\bm{\mu} \in \mathbb{R}^2\\ \bm{\Sigma} \in \mathbb{R}^{2 \times 2}}} \prod_{j=1}^n \mathcal{N}_2(\tilde{\bm{y}}_{\lambda,j}; \bm{\mu}, \bm{\Sigma}),
\end{align*}
where $\bm{J}_{\lambda}$ is the Jacobian of the transformation in \eqref{supp:equation:BoxCox} and $\mathcal{N}_2(\bm{x}; \bm{\mu}, \bm{\Sigma})$ is the density at $\bm{x}$ of a bivariate normal with mean $\bm{\mu}$ and variance $\bm{\Sigma}$. We found that 70\% of all 95\% confidence intervals for $\lambda$ computed using the ($P$,$S$) pairs in our training data contained 0, suggesting that $\log(\cdot)$ is an appropriate transformation for predicted and observed relative intensities.

\section{Real data example experimental and database search details}
\label{supp:section:RealData}
\subsection{Experimental details}
\label{supp:subsection:ExperimentalDetails}
The below method for extracting antigen-specific nanobodies was adapted from \citet{Nb_Yi}.
\begin{enumerate}[label=(\alph*)]
\item \textbf{Llama inoculation}: A male llama was immunized with the antigen glutathione S-transferase (GST; $\Anto$) at a primary dose of 1mg, followed by three consecutive boosts of 0.5mg every 3 weeks.\label{supp:item:GST}
\item \textbf{Constructing the target database}: The blood from the animal was collected 10 days after the final boost. The mRNA from peripheral blood mononuclear cells was isolated and reverse-transcribed into cDNA, and VHH genes were PCR amplified. Next-generation sequencing (NGS) of the VHH repertoire was then performed, and sequences were translated into proteins and subsequently digested with chymotrypsin \textit{in silico} to create the target database $\Targ_{\text{a}}$.\label{supp:item:Target}
\item \textbf{Isolating antigen-specific nanobodies}: VHH antibodies from the plasma of the llama were isolated using by a two-step purification protocol using protein G and protein A sepharose beads. Antigen-specific VHH antibodies were isolated by incubating the VHH antibodies with antigen-conjugated CNBr resin, and were subsequently washed. VHH antibodies were then released from the resin by using one of the following elution conditions: low pH (group 1; 0.1M glycine, pH 1, 2, and 3) or high pH (group 2; 1-100mM NaOH, pH 11, 12, and 13). The nanobodies released from the resin under low and high pH conditions contained nanobodies with predominantly low and high affinities for the antigen, respectively.
\item \textbf{Mass spectrometry analysis}: Released VHH antibodies were reduced, alkylated, and in-solution digested using chymotrypsin. The resulting nanobody peptides were analyzed with a nano-LC 1200 that was coupled online with a Q Exactive$^{\text{TM}}$ HF-X Hybrid Quadrupole Orbitrap$^{\text{TM}}$ mass spectrometer (Thermo Fisher).
\item \textbf{$\Antt$ and $\Entrap$}: The target database $\Entrap$ was created by first inoculating a second llama with human serum albumin ($\Antt$). The remaining steps were analogous to those given in \ref{supp:item:GST} and \ref{supp:item:Target} above.
\end{enumerate}

\subsection{MSeQUiP and other database search details}
\label{supp:subsection:search}
All searches were performed using chymotrypsin (cleavage C-terminal to F, W, Y, or L, but not N-terminal to P) to digest nanobody databases \textit{in silico}, as well as a fixed carbamidomethylation modification on C and at most two variable oxidized M's. Other search engine-specific parameters are given below.
\begin{itemize}
\item \textbf{MSeQUiP}: We processed MS/MS spectra using the pipeline described in Section~\ref{supp:section:Processing}, and searched each spectrum using a 10ppm precursor and $\omega=20$ppm MS/MS mass tolerance. Spectra with fewer than 20 peaks were assumed to be uninformative and were not searched. After computing Bayes factors, we used Proposition~\ref{Proposition:Pvalues} to define spectrum-specific p-values and the R package \texttt{qvalue} \citep{qvalueSoftware} to compute $\Spfdr$'s.
\item \textbf{Crux, MS-GFplus, or X!Tandem $+$ Percolator}: We used Crux v3.2 \citep{CruxSoftware}, MS-GFplus v20200805 \citep{MSGFplus}, or X!Tandem v20170201 \citep{XTandem} to score peptide-spectrum matches (PSMs), and Percolator v3.05.0 to compute $\PSMfdr$'s. All searches were performed with a 10ppm precursor mass tolerance. With the exception of Crux, which discretizes MS/MS m/z space into bins of length 0.02m/z, MS/MS mass tolerance was set to 20ppm. Besides the enzyme and modification parameters given above, all other parameters were set to their software defaults. We note that by default, X!Tandem includes a number of additional post-translational modifications not considered by other search engines, including N-terminal acetylation, N-terminal deamination of Q to pyroglutamic acid, and a water loss for N-terminal E's.
\item \textbf{Prosit}: We used MaxQuant v1.6.7.0 with a 10ppm precursor and 20ppm MS/MS mass tolerance to score PSMs, where besides the modifications listed above, all other parameters were set to their recommended defaults. We then used Prosit \citep{NN_pred} to compute $\PSMfdr$'s via their web server https://www.proteomicsdb.org/prosit/.
\end{itemize}



\section{Additional details regarding error-rate control}
\label{supp:section:Error}
\subsection{Generalizing error-rate control to multiple charge states}
\label{supp:subsection:MultipleCharge}
We partition spectra by their precursor charge into three groups: $+2$, $+3$, and $\geq +4$, where the dependence of error-rate control on charge state has two sources. The first comes from estimates for $\Spfdr_g$, which is defined in \eqref{equation:fdr}. In practice, we estimate $\Spfdr_g$ by applying the method outlined in \citet{qvalue} to each of the three charge-dependent sets of p-values $p_g$ defined in Proposition~\ref{Proposition:Pvalues}. The second source arises from the hyperparameters used to define $BF_g(P)$, where some hyperparameters are assumed to depend on precursor charge. We detail this dependence in Section~\ref{supp:section:BF}. 

\subsection{Estimates for the PSM-fdr in other methods}
\label{supp:subsection:OtherMethodsfdr}
Here we study how estimates for the $\PSMfdr$ behave in existing methods. To do so, we use the framework presented in Section 2 of \citet{MatchUncertainty}, whose theoretical set up applies to all existing methods that assume decoy peptide-spectrum matches (PSMs) are representative of incorrect matches \citep{MatchUncertainty}. Let $\Score(P,S) \in \mathbb{R}$ be a scoring function that represents the plausability that peptide $P \in \Targ$ generated spectrum $S$, and let $H_g$ be as defined in Section~\ref{section:Model}. If $H_g=1$, let $P_g \in \Targ$ be spectrum $g$'s generating peptide, and define the following random variables:
\begin{align*}
    W_g =& \max_{P \in \Targ} \Score(P,S_g), \quad X_g=\begin{cases}
    -\infty & \text{if $H_g=0$}\\
    \Score(P_g,S_g) & \text{if $H_g=1$}
    \end{cases}\\
    Y_g =& \begin{cases}
    W_g & \text{if $H_g=0$}\\
    \mathop{\max}\limits_{P \in \Targ\setminus \{P_g\}}\Score(P,S_g) & \text{if $H_g=1$}
    \end{cases}.
\end{align*}
Under this set-up, $W_g = \max(X_g,Y_g)$ is observable, but $X_g$ and $Y_g$ are not because $H_g$ and $P_g$ are unknown. Lemma \ref{supp:lemma:OtherPSM} and Remarks~\ref{supp:remark:OtherFDRs} and \ref{supp:remark:fdrToFDR} below shows that under the assumptions of \citet{MatchUncertainty}, estimates for $\PSMfdr_g$ that do not employ target-decoy competition are $O(\Spfdr_g)$.
\begin{lemma}
\label{supp:lemma:OtherPSM}
Let $W_g,X_g,Y_g,H_g$ be as defined and above, $J_g = I($the $g$th PSM is correct$)$, and let (i), (ii), and (iii) below be the set of assumptions outlined in Section 2 of \citet{MatchUncertainty}, which may not necessarily hold in our data:
\begin{enumerate}[label=(\roman*)]
\item $X_g$ and $Y_g$ are independent conditional on $H_g=1$.\label{supp:item:XYInd}
\item The observable random variable $Z_g = \max_{P \in \Dec}\Score(P,S_g)$ satisfies
\begin{align*}
    Z_g \edist (Y_g \mid H_g=0), \quad Z_g \edist (Y_g \mid H_g=1).
\end{align*}\label{supp:item:DecoyAssump}
\item $\{(X_g,Y_g,W_g,Z_g)\}_{g \in [q]}$ are independent and identically distributed.\label{supp:item:IID}
\end{enumerate}
Assume the following hold:
\begin{enumerate}[label=(\alph*)]
\item $\{(Z_g,W_g)\}_{g \in [q]}$ are independent and identically distributed and $Z_g$ and $W_g$ have known densities with respect to Lebesgue measure.\label{supp:item:Density}
\item $Z_g \edist (Y_g \mid H_g=0)$\label{supp:item:ZDecoy0}
\item $\Prob(H_g=0)=\pi_0$ for all $g \in [q]$.\label{supp:item:pi0}
\end{enumerate}
Define $\PSMfdr_g=\Prob(J_g=0 \mid W_g=w)$ and $\widehat{\PSMfdr}_g$ to be the estimate for $\PSMfdr_g$ implied by conditions \ref{supp:item:XYInd}, \ref{supp:item:DecoyAssump}, and \ref{supp:item:IID} above that depends only on the known distributions of the observable random variables $Z_g$ and $W_g$. Then
\begin{align*}
    \widehat{\PSMfdr}_g= O\{\Prob(H_g=0 \mid W_g)\} = O(\Spfdr_g).
\end{align*}
\end{lemma}

\begin{proof}
Under conditions \ref{supp:item:XYInd}, \ref{supp:item:DecoyAssump}, and \ref{supp:item:IID}, $\PSMfdr_g$ can be expressed as
\begin{align*}
    \PSMfdr_g =& \Prob(J_g=0 \mid W_g=w)\\
    =& \frac{\Prob(Y_g=w \mid H_g=0)\pi_0 + \Prob(W_g=w, J_g=0 \mid H_g=1)(1-\pi_0)}{\Prob(W_g=w)}\\
    \underbrace{=}_{\ref{supp:item:DecoyAssump}}& \frac{\Prob(Z_g=w)\pi_0 + \Prob(X_g < Y_g=w \mid H_g=1)(1-\pi_0)}{\Prob(W_g=w)}\\
    \underbrace{=}_{\ref{supp:item:XYInd}}& \frac{\Prob(Z_g=w)\pi_0 + \Prob(Y_g=w \mid H_g=1)\Prob(X_g < w \mid H_g=1)(1-\pi_0)}{\Prob(W_g=w)}\\
    \underbrace{=}_{\ref{supp:item:DecoyAssump}}& \frac{\Prob(Z_g=w)\pi_0 + \Prob(Z_g=w)\Prob(X_g < w \mid H_g=1)(1-\pi_0)}{\Prob(W_g=w)}
\end{align*}
where $\Prob(Z_g=w)$ and $\Prob(W_g=w)$ are the densities of $Z_g$ and $W_g$ evaluated at $w$. Next, under \ref{supp:item:XYInd} and \ref{supp:item:DecoyAssump},
\begin{align*}
    \Prob(W_g < w) =& \Prob(Y_g < w \mid H_g=0)\pi_0 + \Prob(W_g < w \mid H_g=1)(1-\pi_0)\\
    \underbrace{=}_{\ref{supp:item:DecoyAssump}}& \Prob(Z_g < w)\pi_0+ \Prob(W_g < w \mid H_g=1)(1-\pi_0)\\
    \Prob(W_g < w \mid H_g=1)=& \Prob\{\max(X_g,Y_g) < w \mid H_g=1\}\\
    \underbrace{=}_{\ref{supp:item:XYInd}}&  \Prob(X_g < w \mid H_g=1)\Prob(Y_g < w \mid H_g=1)\\
    \underbrace{=}_{\ref{supp:item:DecoyAssump}}& \Prob(X_g < w \mid H_g=1)\Prob(Z_g < w).
\end{align*}
Therefore, since $\Prob(Z_g=w)$ and $\Prob(W_g=w)$ are known, $\Prob(X_g < w \mid H_g=1)$ is also known under \ref{supp:item:XYInd} and \ref{supp:item:DecoyAssump}, and can be expressed as
\begin{align*}
    \Prob(X_g < w \mid H_g=1) = \frac{\Prob(W_g < w)}{(1-\pi_0)\Prob(Z_g< w)} - \frac{\pi_0}{1-\pi_0}.
\end{align*}
Then since the implied value of $\Prob(X_g < w \mid H_g=1)$ is trivially bounded above by 1, we see that
\begin{align*}
    \widehat{\PSMfdr}_g =& \frac{\Prob(Z_g=w)\pi_0 + \Prob(Z_g=w)\Prob(X_g < w \mid H_g=1)(1-\pi_0)}{\Prob(W_g=w)}\\
     \leq & \frac{\Prob(Z_g=w)}{\Prob(W_g=w)} \underbrace{=}_{\ref{supp:item:ZDecoy0}} \frac{\Prob(W_g=w \mid H_g=0)}{\Prob(W_g=w)} = \frac{\Prob(H_g=0 \mid W_g=w)}{\pi_0}\\
     =& \frac{\Spfdr_g}{\pi_0},
\end{align*}
which completes the proof.
\end{proof}

\begin{remark}
\label{supp:remark:DecoyMatch}
It is assumed that $\Dec$ is a randomly drawn decoy database in \citet{MatchUncertainty}. In existing works that treat $\Dec$ as random, peptide sequences in $\Dec$ are typically generated by independently drawing each amino acid, with amino acid frequencies proportional to the number of times they appear in $\Targ$ \citep{MSGFplus,Crux_Pvalues}.
\end{remark}

\begin{remark}
\label{supp:remark:Density}
Assumption \ref{supp:item:Density} requiring $Z_g$ and $W_g$ have densities with respect to Lebesgue measure is a trivial assumption, and is satisfied by nearly all scoring functions. Assumption \ref{supp:item:ZDecoy0} is equivalent to Assumption (ii) in Proposition \ref{Proposition:Pvalues} in the main text.
\end{remark}

\begin{remark}
\label{supp:remark:MixMax}
The expression for $\widehat{\PSMfdr}_g$ in the proof of Lemma~\ref{supp:lemma:OtherPSM} is exactly what one would derive using the mixture-maximum procedure outlined in Section 3.3.2 of \citet{MatchUncertainty}.
\end{remark}

\begin{remark}
\label{supp:remark:OtherFDRs}
It is easy to see that the Benjamini-Hochberg-like estimator $\Prob(Z_g=w)/\Prob(W_g=w)$ outlined in Section 3.2.1 of \citet{MatchUncertainty} and the estimator $\pi_0\Prob(Z_g=w)/\Prob (\allowbreak W_g=w)$ proposed in \citet{PEP_kall} (which is exactly the Adaptive Benjamini-Hochberg estimator outlined in Section 3.2.2 of \citet{MatchUncertainty}) are both bounded above by $\Spfdr_g/\pi_0$.
\end{remark}

\begin{remark}
\label{supp:remark:fdrToFDR}
Lemma~\ref{supp:lemma:OtherPSM} and Remark~\ref{supp:remark:OtherFDRs} prove results for local false discovery rates. It is easy to extend these to global false discovery rates (FDRs) by noting that the global PSM and database incompleteness FDRs $\PSMFDR_g(w) = \Prob(J_g=0 \mid W_g \geq w)$, $\SpFDR_g(w) = \Prob(H_g=0 \mid W_g \geq w)$, satisfy $\PSMFDR_g(w) = \E\{ \PSMfdr_g(W_g) \mid W_g \geq w \}$ and $\SpFDR_g(w) = \E\{ \Spfdr_g(W_g) \mid W_g \geq w \}$ for $\PSMfdr_g(x)=\Prob(J_g=0 \mid W_g=x)$ and $\Spfdr_g(x)=\Prob(H_g=0 \mid W_g=x)$.
\end{remark}

Lemma~\ref{supp:lemma:OtherPSM} and Remark~\ref{supp:remark:OtherFDRs} show that if conditions \ref{supp:item:XYInd}, \ref{supp:item:DecoyAssump}, and \ref{supp:item:IID} are assumed, as they are in existing target-decoy methods \citep{MatchUncertainty}, existing estimators for $\PSMfdr_g$ that do not employ target-decoy competition are $O(\Spfdr_g)$. Lemma~\ref{supp:lemma:TDC} below shows a similar result holds for those that do employ target-decoy competition.

\begin{lemma}
\label{supp:lemma:TDC}
In addition to Assumptions \ref{supp:item:Density}, \ref{supp:item:ZDecoy0}, and \ref{supp:item:pi0} from Lemma~\ref{supp:lemma:OtherPSM}, assume the following holds:
\begin{enumerate}[label=(\alph*)]
\setcounter{enumi}{3}
\item $Z_g$ is independent of $(Y_g,H_g)$.\label{supp:item:ZIndep}
\end{enumerate}
Then the target-decoy competition-like estimator for the global false discovery rate, $\frac{ \Prob(W_g < Z_g \cap Z_g \geq w) }{ \Prob(W_g > Z_g \cap W_g \geq w) }$, is $O\{\SpFDR_g(w)\}$, where $\SpFDR_g(w)$ is defined to be
\begin{align*}
    \SpFDR_g(w) =& \Prob(H_g=0\mid W_g \geq w) =  \E\{ \Spfdr_g(W_g) \mid W_g \geq w \},
\end{align*}
where $\Spfdr_g(w) = \Prob( H_g=0 \mid W_g=w )$.
\end{lemma}

\begin{proof}
Let $\widehat{\PSMFDR}_g^{(TDC)}(w)=\frac{ \Prob(W_g < Z_g \cap Z_g \geq w) }{ \Prob(W_g > Z_g \cap W_g \geq w) }$. Then
\begin{align*}
    \widehat{\PSMFDR}_g^{(TDC)}(w) \leq \frac{ \Prob(Z_g \geq w) }{\Prob(W_g \geq w)\Prob(Z_g < W_g \mid W_g \geq w)},
\end{align*}
where
\begin{align*}
    \Prob(Z_g < W_g \mid W_g \geq w) \geq & \Prob(Z_g < W_g) \geq \pi_0 \Prob(Z_g < W_g \mid H_g=0)\\
    =& \Prob(Z_g < Y_g \mid H_g=0) \underbrace{=}_{\ref{supp:item:ZDecoy0},\ref{supp:item:ZIndep}} \pi_0/2.
\end{align*}
Therefore,
\begin{align*}
    \widehat{\PSMFDR}_g^{(TDC)}(w) \leq \frac{2}{\pi_0}\frac{ \Prob(Z_g \geq w) }{\Prob(W_g \geq w)} \underbrace{=}_{\ref{supp:item:ZDecoy0}} \frac{2}{\pi_0^2} \SpFDR_g(w) = O\{\SpFDR_g(w)\}.
\end{align*}
\end{proof}

\begin{remark}
\label{supp:remark:TDC}
As shown in \citet{MatchUncertainty}, the usual target-decoy competition estimator for the global false discovery rate at some score threshold $w$ is
\begin{align*}
    TDC(w)=\frac{\sum_{g=1}^q I(W_g<Z_g)I(Z_g \vee W_g \geq w)}{\sum_{g=1}^q I(W_g>Z_g)I(Z_g \vee W_g \geq w)},
\end{align*}
where the denominator is the total number of discoveries and the numerator is an estimate for the number of false discoveries. As decoy scores are meant to mirror those from all incorrect target matches, this estimator implicitly assumes \ref{supp:item:DecoyAssump} in the statement of Lemma~\ref{supp:lemma:OtherPSM} holds. If $q$ is large, Assumption~\ref{supp:item:Density} in the statements of Lemmas~\ref{supp:lemma:OtherPSM} and \ref{supp:lemma:TDC}, along with other regularity conditions \citep{StoreyControl}, can be used to guarantee $TDC(w)$ approximates $\widehat{\PSMFDR_g}^{(TDC)}(w)$ defined in the proof of Lemma~\ref{supp:lemma:TDC}.
\end{remark}

\begin{remark}
\label{supp:remark:Condd}
Assumption~\ref{supp:item:ZIndep} in the statement of Lemma~\ref{supp:lemma:TDC} is a standard assumption, and is among the target-decoy assumptions listed in \citet{MatchUncertainty}.
\end{remark}

As \ref{supp:item:XYInd} and  \ref{supp:item:DecoyAssump} in the statement of Lemma~\ref{supp:lemma:OtherPSM} are responsible for the behavior of the $\PSMfdr$ in existing methods, we discuss why they are invalid in nanobody data and the how they cause existing methods to underestimate score-ordering errors.
\begin{itemize}
\item Condition \ref{supp:item:XYInd}: $X_g$ is independent of $Y_g$ conditional on $H_g=1$. This is invalid in nanobody proteomes. If $H_g=1$ and due to the sequence similarity in nanobody peptide sequences, the generating peptide $P_g$ will almost surely resemble another peptide $P_g' \in \Targ$. In this case, $X_g=\Score(P_g,S_g) \approx \Score(P_g',S_g)\leq Y_g$, implying $Y_g$ is dependent on $X_g$.
\item Condition \ref{supp:item:DecoyAssump}: $Z_g \edist (Y_g \mid H_g=1)$. This is invalid in nanobody proteomes. To see this, suppose $H_g=1$. Since $Z_g$ is the top scoring decoy peptide, $Z_g$ will typically be much smaller than $X_g$, since peptides in $\Dec$ bear no resemblance to nanobody peptides. However, as discussed above, $X_g$ will likely be similar to scores for other non-generating nanobody peptides, implying $X_g \approx Y_g$. Therefore, $Z_g$ is stochastically smaller than $Y_g$, which causes existing methods to underestimate the number of score-ordering errors (i.e. the number of times $Y_g > X_g$ when $H_g=1$).
\end{itemize} 

\subsection{Assumption (ii) in Proposition~\ref{Proposition:Pvalues}}
\label{supp:subsection:AssumProp}
\begin{figure}
    \centering
    \includegraphics[width=0.75\textwidth]{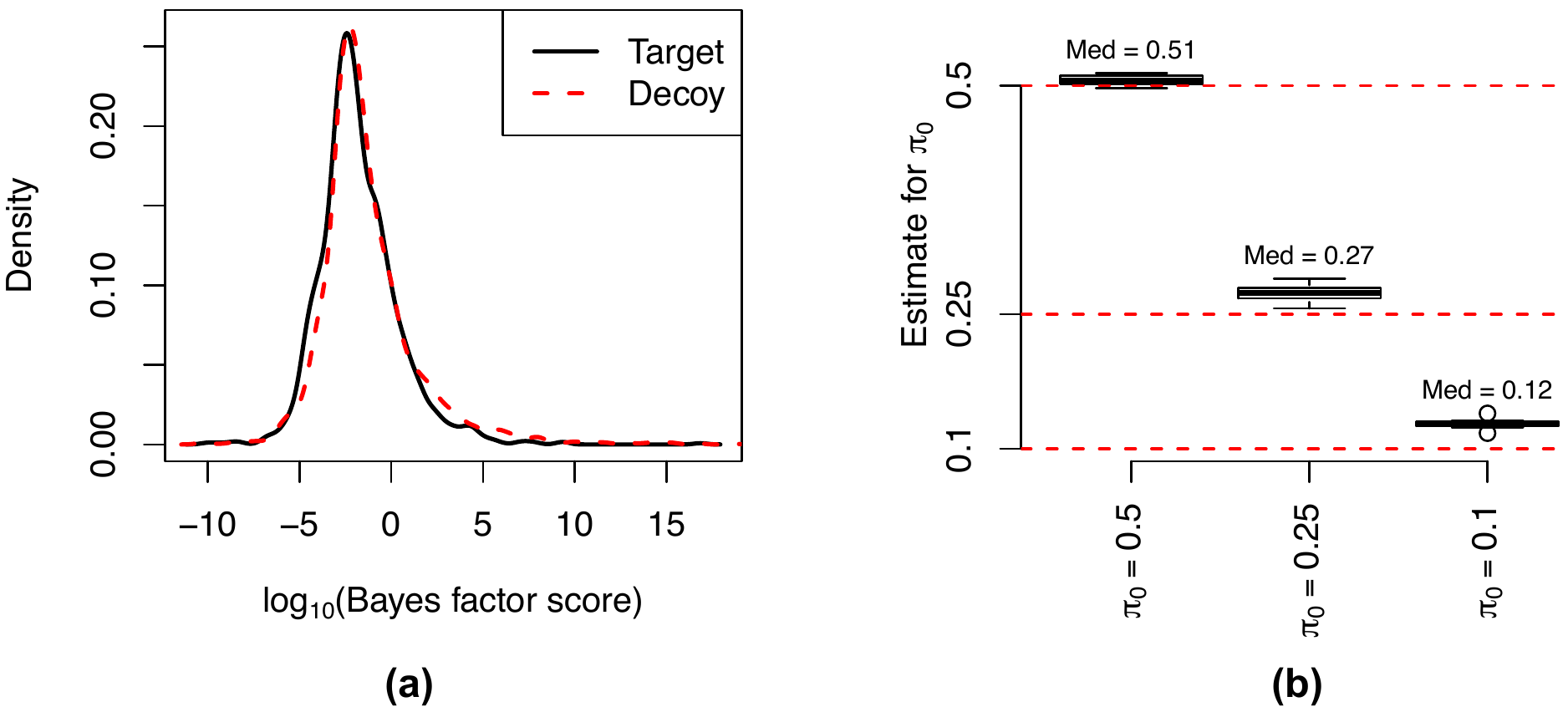}
    \caption{Simulation results from Section~\ref{section:Simulations} suggesting the Assumptions in Proposition~\ref{Proposition:Pvalues} are valid. (a): Density plots of Bayes factor scores from one simulation for spectra whose precursors were randomly shifted $\pm 10$m/z units, and therefore were generated outside the target database $\Targ$. (b) Estimates for $\pi_0$, defined as the fraction of spectra generated outside the database. Each point is a simulation, where ``Med'' is the median estimate across simulations.}\label{supp:figure:DecoyAssump}
\end{figure}

Assumption (ii) in Proposition~\ref{Proposition:Pvalues} states that Bayes factor scores $z_g^{(\Targ)}$ have the same distribution as $z_g^{(\Dec)}$ when spectrum $g$'s generator lies outside the target database $\Targ$. Since p-values $p_g$, defined in Proposition~\ref{Proposition:Pvalues}, are only valid if Assumption (ii) holds, this assumption is critical to controlling $\Spfdr_g$, and consequently $\PSMfdr_g$. We justify this assumption using the precursor mass shift technique outlined in \citet{TargetDecoy}, whereby we randomly increased or decreased a spectrum's assigned precursor monoisotopic m/z by 10 m/z units. Given the high precursor mass accuracy of these data, and because it is assumed $P \in \Targ$ is a candidate generator only if its expected monoisotopic mass is within 10 parts per million of a spectrum's assigned monoisotopic mass, none of the spectra with an artificial m/z shift could have been generated by a peptide in $\Targ$. Therefore, if Assumption (ii) were to hold, the distributions of the Bayes factor scores $z_g^{(\Targ)}$ and $z_g^{(\Dec)}$ should be the same. Further, our estimates for $\pi_0$ in our simulation studies in Section~\ref{section:Simulations}, which gives the fraction of all spectra not generated by peptides in $\Targ$, should be accurate. Figure~\ref{supp:figure:DecoyAssump} shows this both of these appear to be true.

\section{Theory justifying our model for noise m/z's}
\label{supp:section:theory}
\subsection{Assumptions on the noise model}
\label{supp:subsection:assumptions}
\begin{figure}[t!]
\centering
\includegraphics[width=0.6\textwidth]{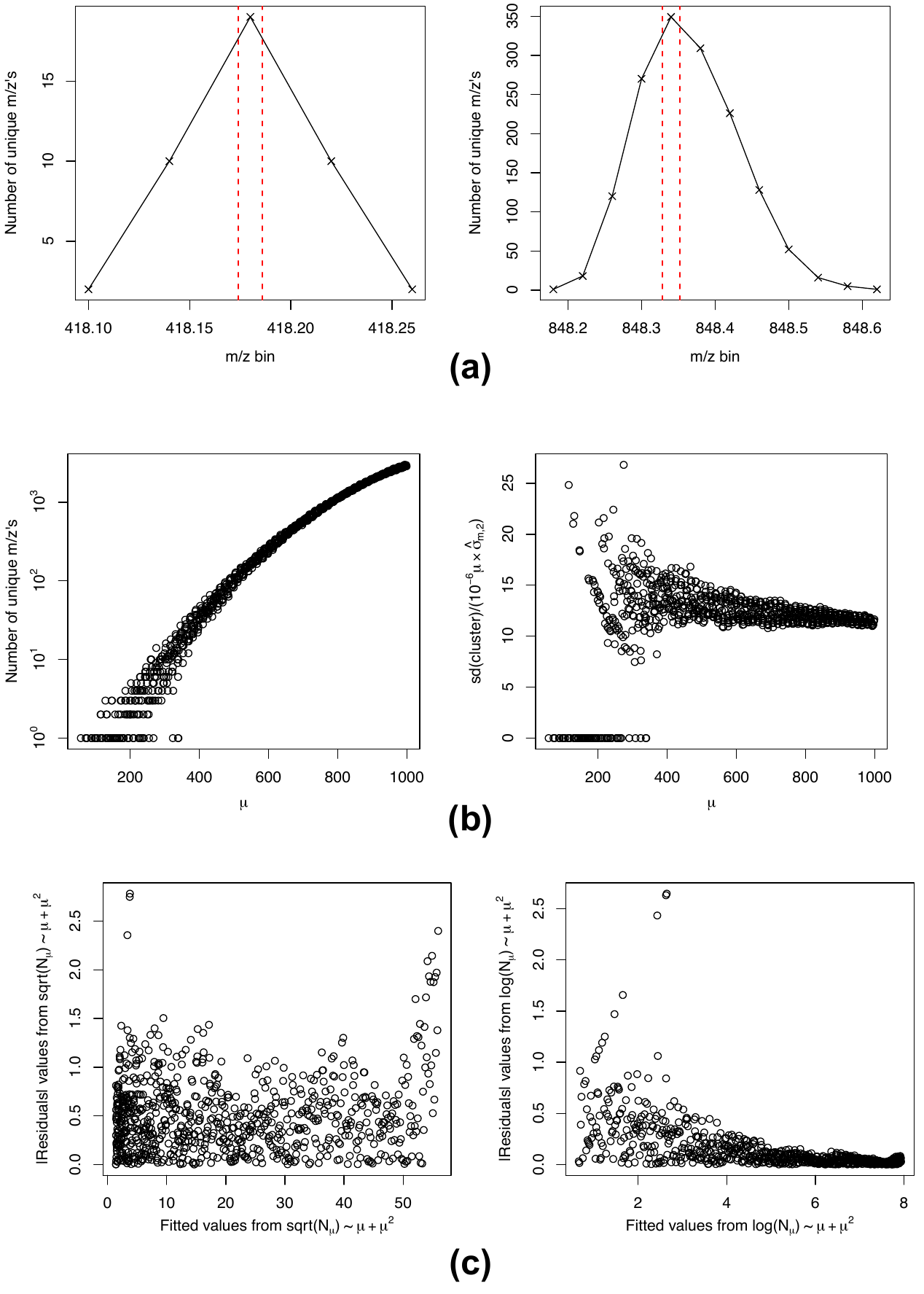}
\caption{The distribution of all unique theoretical $+1$ b and y ion m/z's with m/z $\leq 10^3$ derived from peptides with nine or fewer amino acids. (\textbf{a}): An example of clusters of m/z's in low (left) and high (right) m/z regions. The red lines are $\pm 10^{-6}\omega \mu_{\max}$ around the cluster maximum, $\mu_{\max}$. (\textbf{b}): Cluster characteristics as a function of cluster mean $\mu$, where a cluster was defined as a contiguous sequence of m/z bins with at least one theoretical b ion. ``sd(cluster)'' was the standard deviation of points comprising a cluster assuming each point had probability proportional to the number of unique m/z's in that bin. $\hat{\sigma}_{m,2}$ is our estimate for $\sigma_{m,2}$ defined in \eqref{equation:mSignal}. (\textbf{c}): Regressions involving functions of the number of unique m/z's in each cluster, $N_{\mu}$, for $\mu>200$m/z. These suggest $\sqrt{\cdot}$ is $N_{\mu}$'s variance-stabilizing transformation.}\label{supp:Figure:Proof}
\end{figure}

Let $\precmass^{(P)},\precmass^{(S_g)}$ be peptide $P$'s and $S_g$'s predicted and observed precursor mass, respectively. Define $\RandMZ_g = \{m_j^{(P)}: \abs*{\precmass^{(S_g)}- \precmass^{(P)}}\leq \Delta, j \in [n^{(P)}]\}$ to be the set of signal m/z's that could appear in $S_g$, which include peaks not included in $S_g$'s generator's predicted spectrum, as well as peaks from co-fragmented peptides. The constant $\Delta>0$ is determined by the mass spectrometer's precursor isolation window. For the remainder of Section \ref{supp:section:theory}, we make use of the following definition:
\begin{definition}
\label{supp:definition:holder}
Let $\alpha \in (0,1]$. A function $f: R \to (0,\infty)$ is \underline{log $\alpha$-H\"{o}lder continuous} if, for some constant $c>0$, $\abs*{\log\{f(x_1)\}-\log\{f(x_2)\}} \leq c\abs{x_1-x_2}^{\alpha}$ for all $x_1,x_2 \in R$.
\end{definition}
\noindent We begin with an assumption on $\RandMZ_g$.
\begin{assumption}
\label{supp:assumption:Mg}
Let $\alpha \in (0,1]$ and $0<a < m_* < b$ be constants, $A:(0,\infty) \to (0,\infty)$ a non-decreasing function, $\RandMZ_0 \subset [a,m_*)$ be a set with a finite number of elements, and $\RandMZ \subset [m_*,b]$ a realization of a inhomogeneous Poisson process with intensity $\lambda(x)$. Then for $\tilde{\RandMZ}_g = \{ m \in \RandMZ: m \leq A\{\precmass^{(S_g)}\} \}$, $\RandMZ_g =\RandMZ_0 \cup \tilde{\RandMZ}_g$ and $\lambda$ is log $\alpha$-H\"{o}lder continuous.
\end{assumption}
\begin{remark}
\label{supp:remark:AssumpMg}
The set $\RandMZ_0$ is meant to contain all unmodeled small m/z's, which represent peptide fragments with a small number of (1 or 2) amino acids. We differentiate these from $\tilde{\RandMZ}_g$ because, unlike the elements in $\tilde{\RandMZ}_g$, we are certain of their location. Due to the combinatorial explosion of moderate-to-large peptide fragments, we assume $\tilde{\RandMZ}_g$ is a realization of a Poisson process to capture our uncertainty about the location of unmodeled moderate-to-large peptide fragments. Figures \ref{supp:Figure:Proof}(b) and \ref{supp:Figure:Proof}(c) show that the Poisson approximation appears to be a reasonable approximation for the distribution of peptide fragment m/z's.
\end{remark}

\begin{remark}
The assumptions on $\RandMZ_g$ ensure $\RandMZ_g \subseteq \RandMZ_h$ if $\precmass^{(S_g)} \leq \precmass^{(S_h)}$, and captures the fact that fragments of a small peptide $P_1$ will also be fragments of a larger peptide $P_2$ if $P_1$ is a sub-sequence of $P_2$. It suffices to assume $a=57$, since this is the mass of the smallest $+1$ b-ion.
\end{remark}

\begin{remark}
\label{supp:remark:ContLambda}
We abuse notation by defining $\RandMZ$'s intensity to be $\lambda$, which was $\Lambda$ in the statement of Theorem~\ref{theorem:noise} and is unrelated to $\lambda_{\noise,g}^{\Mass}$ defined in the main text. Here, $\lambda$ reflects the distribution of moderate-to-large peptide fragments (see Remark \ref{supp:remark:AssumpMg}). The assumption that $\lambda$ is log $\alpha$-H\"{o}lder continuous appears to be a reasonable assumption, where Figure~\ref{supp:Figure:Proof} shows that unmodeled peptide fragments with moderate-to-large m/z's likely cluster in regions whose local distributions appear Gaussian with standard deviation substantially larger than the scale $\text{m/z}\times 10^{-6}\omega$ used to map observed m/z's to modeled m/z's described in Remark \ref{remark:delta}. 
\end{remark}

We next place assumptions on $D_{\noise,g}^{\Mass}( m \mid y )$.
\begin{assumption}
\label{supp:assumption:dnoise}
Let $A(\cdot),a,b$ be as defined in Assumption \ref{supp:assumption:Mg}. The density $D_{\noise,g}^{\Mass}( m \mid y) = \{\pi_0 \alpha_g B(m) + (1-\pi_0)\Gamma_g(m\mid y)\}I\{m \in [a-1,A\{\precmass^{(S_g)}]\}$ for $\pi_0 \in [0,1]$, where $B(m)$, $\alpha_g$, and $\Gamma_g(m\mid y)$ satisfy the following:
\begin{enumerate}[label=(\alph*)]
\item $B:[a-1,b+1] \to (0,\infty)$ is log $\alpha$-H\"{o}lder continuous, $\smallint_{a-1}^{b+1} B(u)\text{d}u = 1$, and $\alpha_g=[\smallint_{a-1}^{A\{\precmass^{(S_g)}\}} B(u)\text{d}u]^{-1}$.\label{supp:item:C}
\item Let $p: [a,b] \to (0,1]$ be a log $\alpha$-H\"{o}lder continuous function with $\max_{x \in [a,b]}p(x)=1$, and for $D_g = \sum_{m \in \RandMZ_g}p(m)$, let $\tilde{p}_g(x)=p(x)/D_g$ be a probability mass function on $\RandMZ_g$. Then $\Gamma_g(m\mid y)$ is the density of the random variable $M_g=m_g(10^{-6}R_g+1)$, conditional on $Y_g=y$ and $\RandMZ_g$, such that $R_g\mid Y_g=y$ has density $d_{\signal}^{\Mass}(r \mid y)$ and $m_g \mid \RandMZ_g \sim \text{Categorical}(\RandMZ_g,\tilde{p}_g)$, where $R_g$ and $m_g$ are independent conditional on $Y_g,\RandMZ_g$.\label{supp:item:p}
\end{enumerate}
\end{assumption}

\begin{remark}
\label{supp:remark:Cprop}
As defined, $B_g(m)=\alpha_g B(m)$ is the density of spectrum $g$'s contaminating background noise m/z's, which we assume is truncated at $A\{\precmass^{(S_g)}\}$. It assumes that the relative frequency of contaminating peaks in different m/z regions is consistent across spectra. That is, if, for spectra $g$ and $h$, $\precmass_g^{(S_g)} <\precmass_h^{(S_g)}$ and $m_1,m_2 \subset [a-1,A\{\precmass_g^{(S_g)}\}]$, then $B_g(m_1)/B_g(m_2) = B_h(m_1)/B_h(m_2)$. This is consistent with existing noise models, which do not differentiate between contaminating peaks and unmodeled peptide fragments, and assume $D_{\noise,g}^{\Mass}(m\mid y) \propto 1$ \citep{LikelihoodScoring,UniformNoise}.
\end{remark}

\begin{remark}
\label{supp:remark:Dprop}
$\Gamma_g(m\mid y)$ is the density of unmodeled peptide fragments in spectrum $g$. These include fragments from co-isolated peptides, as well as fragments from spectrum $g$'s generating peptide that are not included in its predicted spectrum. Just as with $B(m)$ in Remark \ref{supp:remark:Cprop} above, the assumption that $\tilde{p}_g(m) \propto \tilde{p}_h(m)$ for $m \in \RandMZ_g \cap \RandMZ_h$ implies the relative frequency of unmodeled peptide fragments in different m/z regions is consistent across spectra, which is consistent with our observations in real data (data not shown). The probability mass function $\tilde{p}_g$ can also be interpreted as a completely random probability measure \citep{CRMs}, which is ubiquitous class of random measures generated by normalizing non-negative functions of inhomogeneous Poisson process.
\end{remark}

\begin{remark}
\label{supp:remark:Contp}
The assumption that $p$ is continuous is akin to assuming it is locally constant, and implies we are more likely to observe m/z's from regions with dense clusters of potential signal m/z's. Using as an example the data in Figure~\ref{supp:Figure:Proof}(a), this suggests we are more likely to observe noise m/z's around 931.4 than we are around 931.2. Further, the continuity of $B$, $p$, and $\lambda$ helps ensure $D_{\noise,g}^{\Mass}( m \mid y )$ is continuous in $m$, and is justified by the observation that continuous, non-uniform noise densities have previously been used to simulate realistic high mass accuracy MS/MS spectra \citep{ContinuousNoise}.
\end{remark}

\subsection{Theoretical statements justifying our noise model}
\label{supp:subsection:NoiseTheory}
In the remainder of this section, state four theorems that use Assumptions \ref{supp:assumption:Mg} and \ref{supp:assumption:dnoise} to justify the following:
\begin{enumerate}[label=(\alph*)]
\item The model for $d_{\noise,g}^{\Mass}(r \mid y; m) = d_{\noise}^{\Mass}(r \mid y; m)$ does not depend on spectrum $g$.\label{supp:item:dg}
\item Model \eqref{equation:dnoise:ma} is an appropriate model for $d_{\noise}^{\Mass}(r\mid y; m)$.\label{supp:item:dcorrect}
\item $\lambda_{\noise,g}^{\Mass}(m\mid  y) = \lambda_{\noise,g}^{\Mass}(m)$ does not depend on log-intensity $y$.\label{supp:item:lambdav}
\item $\lambda_{\noise,g}^{\Mass}(m)$ can be approximated with a continuous function.\label{supp:item:lambdacont}
\end{enumerate}
We justify \ref{supp:item:dg} and \ref{supp:item:dcorrect} in Theorems \ref{supp:theorem:CondSmall}, \ref{supp:theorem:dModerate}, and \ref{supp:theorem:Condlarge}. Theorem \ref{supp:theorem:CondSmall} shows that $d_{\noise,g}^{\Mass}(r \mid y; m)$ is approximately a mixture of $d_{\signal}^{\Mass}(r \mid y)$ and a uniform distribution for small $m$, and Theorem \ref{supp:theorem:Condlarge} uses the fact that $\text{\#m/z's in $F_m$} = \sum_{m' \in \RandMZ_g}I(m' \in F_m)$ is typically large for large $m$ (Figure \ref{supp:Figure:Proof}(b)) to prove $d_{\noise,g}^{\Mass}(r \mid y; m)$ can be approximated with a uniform distribution for large $m$.\par 
\indent Studying the behavior of $d_{\noise,g}^{\Mass}(r \mid y; m)$ for moderate $m$ is more complex because of our uncertainty in the exact position of unmodeled peaks with moderate signal m/z's $m$, where $F_m$ typically contains relatively few (but $>0$) unmodeled m/z's besides $m$. To account for this uncertainty, note that 
\begin{align*}
    d_{\noise,g}^{\Mass}(r \mid y; m) = 10^{-6}m \Prob\{M_g = m(10^{-6} r+1) \mid Y_g=y, M_g \in F_m, m \in \RandMZ_g, \RandMZ_g\}.
\end{align*}
First, Theorem \ref{supp:theorem:dModerate} considers $10^{-6}m\Prob\{M_g=m(10^{-6} r+1) \mid Y_g=y, M_g \in F_m, m \in \RandMZ_g\}$, which captures the average behavior of $d_{\noise,g}^{\Mass}(r \mid y; m)$ over possible instantiations of $\RandMZ_g$, and shows this is a mixture of $d_{\signal}^{\Mass}$ and a uniform distribution. We condition on $m \in \RandMZ_g$, where $m=m_j^{(P)}$ for some peptide $P$ and predicted ion $j$, because when we score $P$ we observe $m$. We then show in Corollary \ref{supp:corollary:dModAvg} that the model for $d_{\noise,g}^{\Mass}(r \mid y; m)$ in \eqref{equation:dnoise:ma} reflects the average behavior of $d_{\noise,g}^{\Mass}(r \mid y; m)$ over neighboring intervals $F_m$. We lastly justify \ref{supp:item:dg} in Remark \ref{supp:remark:Dnoiseg}.\par
\indent We then use Theorem \ref{supp:theorem:lambda} and Corollary \ref{supp:corollary:lambda} to show \ref{supp:item:lambdav} and \ref{supp:item:lambdacont}, and discuss when it is appropriate to assume $\lambda_{\noise,g}^{\Mass}(m) = \lambda_{\noise}^{\Mass}(m)$ does not depend on $g$ in Remark \ref{supp:remark:lambdag}. We let $\eta=10^{-6}\omega$ throughout, and prove all theoretical results in Section \ref{supp:subsection:ProofTheorems}.

\begin{theorem}
\label{supp:theorem:CondSmall}
Suppose Assumptions \ref{supp:assumption:Mg} and \ref{supp:assumption:dnoise} hold, let $m \in \RandMZ_0$ such that $m < m_*/(1+\eta)$, and define $F_m = [m(1-\eta),m(1+\eta)]$. Then for $L(F_m)$ the Lebesgue measure of $F_m$,
\begin{align*}
    &\sup_{\substack{g \in [q]\\r \in [-\omega,\omega]\\y \in \mathbb{R}}} \abs*{ d_{\noise,g}^{\Mass}(r \mid y; m) - [ \tilde{\pi}_g(m)(2\omega)^{-1}I(\abs*{r}\leq \omega) + \{1-\tilde{\pi}_g(m)\}d_{\signal}^{\Mass}(r \mid y) ] } = O[ \{L(F_m)\}^{\alpha} ]\\
    &\tilde{\pi}_g(m) = \frac{\pi_0 \alpha_g B(m)}{\pi_0 \alpha_g B(m) + (1-\pi_0)\E\{\tilde{p}_g(m)\}/L(F_m)}, \quad \tilde{p}_g(m) = \frac{p(m)}{p(m) + \sum\limits_{m' \in \RandMZ_g\setminus\{m\}}p(m')},
\end{align*}
where $\log\{\tilde{\pi}_g(m)\}$ is $\alpha$-H\"{o}lder continuous as a function of $m$.
\end{theorem}


\begin{remark}
\label{supp:remark:dMsmall}
Figure~\ref{Figure:TrainSignal}(h) suggests $\tilde{\pi}_g(m)$ is small for small $m$, which suggests that $\pi_0$ is also small.
\end{remark}

According to Figure~\ref{supp:Figure:Proof}(b), the $m \in \RandMZ_0$ for $m$ corresponding to one or two amino acid-long fragment ions. We next study how $\Prob(u \mid u \in F_m, v, m \in \RandMZ_g)$ for moderate to large values of $m$.

\begin{theorem}
\label{supp:theorem:dModerate}
Suppose Assumptions \ref{supp:assumption:Mg} and \ref{supp:assumption:dnoise} hold and let $m \in \RandMZ_g\cap (m_*/(1-\eta),\infty)$. Define $F_m=[m(1-\eta),m(1+\eta)]$ and $G_m = [m/(1+\eta),m/(1-\eta)]$. Then for $L(\cdot)$ Lebesgue measure,
\begin{align}
    &\sup_{\substack{g \in [q]\\r\in[-\omega,\omega]\\y \in \mathbb{R}}} \lvert 10^{-6}m\Prob\{M_g=m(10^{-6} r+1) \mid Y_g=y, M_g \in F_m, m \in \RandMZ_g\} - [\tilde{\pi}_g(m)(2\omega)^{-1}I(\abs{r}\leq \omega)\nonumber\\
    &+\{1 - \tilde{\pi}_g(m)\}d_{\signal}^{\Mass}(r \mid y)] \rvert = O[\{L(F_m)\}^{\alpha}]\nonumber\\
    \label{supp:equation:dmoderate}
    &\tilde{\pi}_g(m) = \frac{\pi_0\alpha_g B(m) + (1-\pi_0)\E\{\tilde{N}_m \tilde{p}_{g}(m)\}/L(F_m)}{\pi_0\alpha_g B(m) + (1-\pi_0)\E\{(\tilde{N}_m +1)\tilde{p}_{g}(m)\}/L(F_m)}\\
    &\tilde{p}_{g}(m)= \frac{p(m)}{p(m) + \sum\limits_{m' \in \RandMZ_g\setminus\{m\}}p(m')}, \quad \tilde{N}_m = L(F_m)[\{L(G_m)\}^{-1}\sum_{m' \in \RandMZ \setminus \{m\}} I(m' \in G_m)]\nonumber,
\end{align}

where $\log\{\tilde{\pi}_g(m)\}$ is $\alpha$-H\"{o}lder continuous as a function of $m$.
\end{theorem}

\begin{remark}
\label{supp:remark:dAvg}
Theorem \ref{supp:theorem:dModerate} shows that Model \eqref{equation:dnoise:ma} captures the average behavior of $d_{\noise,g}^{\Mass}(r \mid y; m)$ over possible instantiations of $\RandMZ_g$, which reflects our uncertainty regarding the exact positions of unmodeled signal m/z's.
\end{remark}

\begin{remark}
\label{supp:remark:dMid}
Since $\{L(G_m)\}^{-1}\sum_{m' \in \RandMZ_g\setminus \{m\}} I(m' \in G_m)$ in Theorem \ref{supp:theorem:dModerate} is an estimate for $\lambda(m)$, $\tilde{N}(m)$ is an extrapolation estimate for the number of m/z's in $F_m$. The expression for $\tilde{\pi}_g$ in Theorem \ref{supp:theorem:dModerate} matches that in Theorem \ref{supp:theorem:CondSmall} when $\tilde{N}_m=0$, and as shown in Corollary \ref{supp:corollary:dModerate} below, $\tilde{\pi}_g(m)$ approaches 1 when the number of unmodeled signal peaks in $F_m$ increases. Since $\pi_0$ is likely close to 0 (see Remark \ref{supp:remark:dMsmall}) and the normalizing term $p(m) + \sum_{m' \in \RandMZ_g\setminus\{m\}}p(m')$, under trivial assumptions (see Corollary~\ref{supp:corollary:dModAvg}), will concentrate around its expectation, $\tilde{\pi}_g(m)\approx \E\{\tilde{N}_m\tilde{p}_g(m)\}/\E\{(\tilde{N}_m+1)\tilde{p}_g(m)\}\approx \E(\tilde{N}_m)/\E(\tilde{N}_m+1)$ has little dependence on $g$.
\end{remark}

We next state two corollaries of Theorem \ref{supp:theorem:dModerate}.

\begin{corollary}
\label{supp:corollary:dModAvg}
In addition to the assumptions of Theorem \ref{supp:theorem:dModerate}, suppose the following hold for $D_g = \sum_{m' \in \RandMZ_g}p(m')$ and $E_g = \E(D_g)$:
\begin{enumerate}[label=(\roman*)]
\item For $CV_g = \{\V(D_g)\}^{1/2}/E_g$ the coefficient of variation of $D_g$, $CV_g \leq c_1 E_g^{-\gamma}$ for some constants $c_1,\gamma>0$.\label{supp:item:AssumCV}
\item $m_1< \cdots < m_K \in \RandMZ_g \cap (m_*/(1-\eta),\infty]$ such that $G_{m_1},\ldots, G_{m_K}$ are non-overlapping, $p(m_k)\geq c_2$ for all $k \in [K]$, and $K \leq c_3 E_g^{\delta}$ for constants $c_2,c_3>0$ and $\delta \in (0,\gamma/2)$.\label{supp:item:AssumK}
\end{enumerate}
Define $R(u,m) = 10^6(u-m)/m$ to be the relative mass error. Then for all $r \in [-\omega,\omega]$ and $E_g$ large enough,
\begin{align*}
    &\sup_{\substack{g \in [q]\\ r\in[-\omega,\omega] \\y \in \mathbb{R}}}\abs*{ K^{-1}\sum\limits_{k=1}^K d_{\noise,g}^{\Mass}(r \mid y; m_k) - h(r) }= O_P[K^{-1/2} + E_g^{-\gamma/2+\delta} +\{L(F_{m_k})\}^{\alpha}]\\
    &h(r)= \bar{\pi}_g (2\omega)^{-1}I(\abs{r}\leq \omega) + (1-\bar{\pi}_g) d_{\signal}^{\Mass}(r \mid y), \quad \bar{\pi}_g = K^{-1}\sum\limits_{k=1}^K \tilde{\pi}_g(m_k),
\end{align*}
where $\abs*{\log(\bar{\pi}_g)-\log\{ \tilde{\pi}_g(K^{-1}\sum_{k=1}^K m_k) \}} = O(\abs*{m_1-m_K}^{\alpha})$.
\end{corollary}

\begin{remark}
\label{supp:remark:AvgDmod}
Corollary \ref{supp:corollary:dModAvg} provides a second interpretation of Theorem \ref{supp:theorem:dModerate}, and shows that \eqref{equation:dnoise:ma} reflects the average behavior of $d_{\noise,g}^{\Mass}(r \mid y; m_k)$ across neighboring intervals $F_{m_k}$.
\end{remark}

\begin{remark}
\label{supp:remark:AvgDmod_Assump}
The random variable $D_g$ is used to define $\tilde{p}_g(m)=p(m)/D_g$, where the assumption that $CV_g = O[\{E(D_g)\}^{-\gamma}]$ in \ref{supp:item:AssumCV} of Corollary \ref{supp:corollary:dModAvg} simply assumes that the function $\tilde{p}_g$ concentrates around the non-random function $p(m)/\E(D_g)$. This assumption is quite general, and holds whenever $p(m)$, $m \in R \subset \mathbb{R}$, is bounded from below for a sufficiently large set $R$. For example, if $R$ is the range of $\RandMZ_g$, then $\gamma=1/2$. We derive expressions for $\E(D_g)$ and $\V(D_g)$, as well as concentration results for $D_g$, in Lemma \ref{supp:lemma:subexp} in Section \ref{supp:subsection:ProofTheorems}. Assumption \ref{supp:item:AssumK} is technical and simply assumes that $K$ (the number of m/z's $m_k\in \RandMZ_g$ we are considering) is small with respect to $\E(D_g)$, which is very large.
\end{remark}

\begin{corollary}
\label{supp:corollary:dModerate}
Let $N_m = \E\{\sum_{m' \in \RandMZ}I(m' \in F_m) \mid m \in \RandMZ\} = 1 + \smallint I(x \in F_m) \lambda(x)dx$. Then under the assumptions of Theorem \ref{supp:theorem:dModerate}, $\tilde{\pi}_g(m) = 1 - O(N_m^{-1})$. 
\end{corollary}

\begin{remark}
Corollary \ref{supp:corollary:dModerate} shows that $10^{-6}m\Prob\{M_g=m(10^{-6} r+1) \mid Y_g=y, M_g \in F_m, m \in \RandMZ_g\}$ is approximately uniform for large $N_m$, which as suggested in Figure \ref{supp:Figure:Proof}(b), holds for nearly all moderate to large m/z's $m$. 
\end{remark}

We next consider how $\tilde{d}_{\noise,g}^{\Mass}(u \mid u \in F_m, v)$ behaves for large values of $m$.

\begin{theorem}
\label{supp:theorem:Condlarge}
Suppose Assumptions \ref{supp:assumption:Mg} and \ref{supp:assumption:dnoise} hold and let $F_m = [m(1-\eta),m(1+\eta)]$ for some $m \in \RandMZ_g \cap [m_*,\infty)$. Define $N_m = 1+ \smallint I(x \in F_m)\lambda(x)\text{d}x$ to be the expected number of elements in $\RandMZ_g \cap F_m$ given $m \in \RandMZ_g$. Then
\begin{align*}
    \sup_{\substack{g \in [q]\\r \in [-\omega,\omega]\\y \in \mathbb{R}}} \abs*{d_{\noise,g}^{\Mass}(r \mid y) - (2\omega)^{-1}I(\abs{r}\leq \omega)} = O_P[\{L(F_m)\}^{\alpha} + N_m^{-1/3}].
\end{align*}
\end{theorem}

Theorem \ref{supp:theorem:Condlarge} shows that $d_{\noise,g}^{\Mass}(r \mid y)$ is approximately uniform when $N_m$ is large. As shown in Figure \ref{supp:Figure:Proof}(b), $N_m$ is large for large values of $m$.

\begin{remark}
\label{supp:remark:Dnoiseg}
Theorems \ref{supp:theorem:CondSmall}, \ref{supp:theorem:dModerate}, \ref{supp:theorem:Condlarge} and Corollary \ref{supp:corollary:dModAvg} show that $d_{\noise,g}^{\Mass}$ only depends on $g$ through $\tilde{\pi}_g$ (and $\bar{\pi}_g$ in Corollary \ref{supp:corollary:dModAvg}). It is easy to see that if $\pi_0$ is small and $D_g$, defined in Corollary \ref{supp:corollary:dModAvg}, concentrates around its expectation, then $\tilde{\pi}_g$ has little dependence on $g$. Remark~\ref{supp:remark:dMsmall} argues that the former is true, and we justify the latter in Remark \ref{supp:remark:AvgDmod_Assump}.
\end{remark}

\begin{theorem}
\label{supp:theorem:lambda}
Suppose Assumptions \ref{supp:assumption:Mg} and \ref{supp:assumption:dnoise} hold and let $F_m=[m(1-\eta),m(1+\eta)]$, $G_m=[m/(1+\eta),m/(1-\eta)]$ for $m \in \RandMZ_g$ and $\tilde{N}_m,N_m$ be as defined in the statements of Theorem \ref{supp:theorem:dModerate} and Corollary \ref{supp:corollary:dModerate}. Then the following hold:
\begin{enumerate}[label=(\roman*)]
\item $\lambda_{\noise,g}^{\Mass}(m \mid y) = \pi_0\alpha_g B(m)L(F_m)( 1+O[\{L(F_m)\}^{\alpha}] ) + (1-\pi_0)\tilde{p}_g(m)$ if $m \in \RandMZ_0$ and $m < m_*/(1+\eta)$.\label{supp:item:lambdaOnem}
\item $\lambda_{\noise,g}^{\Mass}(m \mid y) = \pi_0\alpha_g B(m)L(F_m)( 1+O[\{L(F_m)\}^{\alpha}] ) + (1-\pi_0)\tilde{p}_g(m)$ if $m \in \RandMZ_g \cap (m_*/(1+\eta),\infty)$ and on the event $G_m \cap \RandMZ_g = \{m\}$.\label{supp:item:lambdaOnemMg}
\item $\Prob(M_g \in F_m \mid Y_g=y, m \in \RandMZ_g)= \pi_0 \alpha_g B(m)L(F_m)( 1+O[\{L(F_m)\}^{\alpha}] ) + [(1-\pi_0)\E\{(\tilde{N}_m+1)\tilde{p}_g(m)\}]( 1+O[\{L(F_m)\}^{\alpha}] )$ if $m \in \RandMZ_g \cap (m_*/(1-\eta),\infty)$.\label{supp:item:lambdaNmGen}
\item $\lambda_{\noise,g}^{\Mass}(m \mid y) = \pi_0\alpha_g B(m)L(F_m)( 1+O[\{L(F_m)\}^{\alpha}] ) + (1-\pi_0) \tilde{p}_g(m) N_m( 1+O[\{L(F_m)\}^{\alpha}] + O_P(N_m^{-1/2}) )$ if $m \in \RandMZ_g \cap (m_*/(1-\eta),\infty)$,\label{supp:item:lambdaNm}
\end{enumerate}
where the errors $O(\cdot)$ do not depend on $g$, $y$, or $m$. Further, the below four functions of $m$ are log $\alpha$-H\"{o}lder continuous:
\begin{align}
\label{supp:equation:ContFunctions}
    B(m)L(F_m), \quad \tilde{p}_g(m), \quad \E\{(\tilde{N}_m+1)\tilde{p}_g(m)\}, \quad \tilde{p}_g(m)N_m.
\end{align}
\end{theorem}

\begin{remark}
\label{supp:remark:Deltam}
The continuity of the functions in \eqref{supp:equation:ContFunctions} implies the expressions in \ref{supp:item:lambdaOnem}, \ref{supp:item:lambdaOnemMg}, \ref{supp:item:lambdaNmGen}, and \ref{supp:item:lambdaNm} are log $\alpha$-H\"{o}lder continuous up to $O[\{ L(F_m) \}^{\alpha}], O(N_m^{-1/2})$ errors. Note that $\tilde{p}_g(m)N_m = \tilde{p}_g(m)$ if $\RandMZ_g \cap G_m= \{m\}$, as in \ref{supp:item:lambdaOnem}, and $\tilde{N}_m = N_m\{1+O_P(N_m^{-1/2})\}$.
\end{remark}

\begin{corollary}
\label{supp:corollary:lambda}
In addition to the assumptions in Theorem \ref{supp:theorem:lambda}, suppose Assumptions \ref{supp:item:AssumCV} and \ref{supp:item:AssumK} in the statement of Corollary \ref{supp:corollary:dModAvg} hold. Then for $\bar{m} = K^{-1}\sum_{k=1}^K m_k$,
\begin{align}
    &\sup_{\substack{g\in[q]\\y \in \mathbb{R}}}\abs*{K^{-1}\sum\limits_{k=1}^K \lambda_{\noise,g}^{\Mass}(m_k\mid y)/\Lambda_g(\bar{m}) - 1} = O_P(K^{-1/2} + E_g^{-\gamma/2+\delta})\nonumber\\
    \label{supp:equation:LAMBDA}
    &\Lambda_g(\bar{m}) = \left[\pi_0 \alpha_g B(\bar{m})L(F_{\bar{m}}) + (1-\pi_0)\E\{ (\tilde{N}_{\bar{m}}+1)\tilde{p}_g(\bar{m}) \}\right]\{1 + O( \abs{m_1-m_K}^{\alpha} )\}
\end{align}
where $\tilde{N}_m$ and $\tilde{p}_g(m)$ are as defined in \eqref{supp:equation:dmoderate}.
\end{corollary}

\begin{figure}
    \centering
    \includegraphics[width=0.5\textwidth]{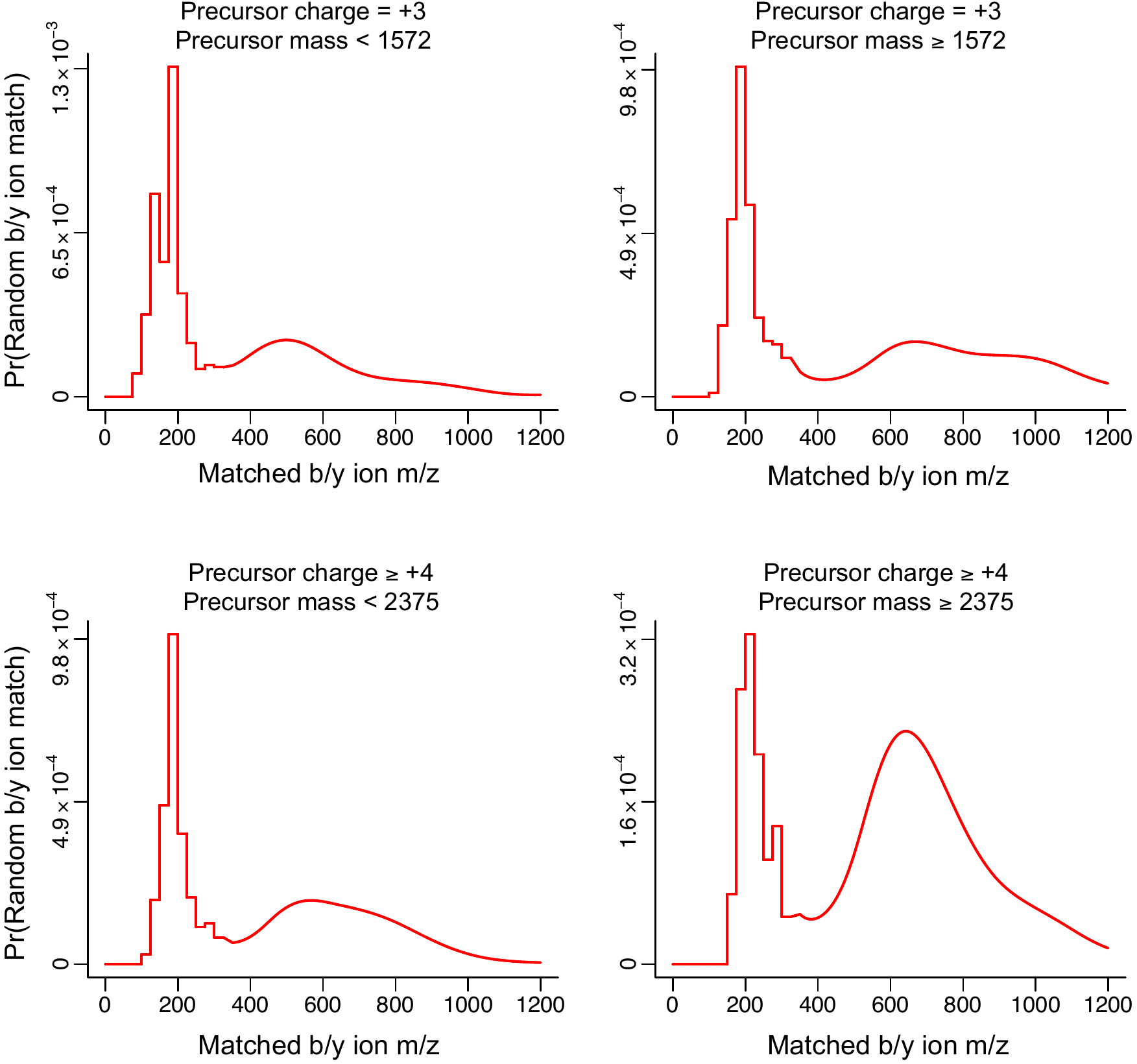}
    \caption{Estimates for $\lambda_{\noise,g}^{\Mass}(m)$ obtained from group 1 datasets for different charge state and precursor mass bins. These are the bins used to analyze our real data, where the precursor mass bin partitions are the median precursor mass for training spectra with that precursor charge.}\label{supp:figure:NoiseObsOther}
\end{figure}

\begin{remark}
\label{supp:remark:lambdag}
The expressions in \ref{supp:item:lambdaOnem}-\ref{supp:item:lambdaNm} in Theorem \ref{supp:theorem:lambda}, as well as \eqref{supp:equation:LAMBDA} in Corollary \ref{supp:corollary:lambda}, have no dependence on $y$. Further, they will be similar across spectra $g$ if $\pi_0$ is small and the majority of the mass of $\tilde{p}_g(m)N_m$ concentrates around relatively small m/z's $m$. The former, as indicated in Remark \ref{supp:remark:dMid}, appears to be true. The latter follows from the fact that if $\tilde{p}_g(m)N_m$ is largest for small $m$, the normalizing constant $\sum_{m \in \RandMZ_g}p(m)$ will be similar for all $g$, implying $\tilde{p}_g(m)\approx \tilde{p}_h(m)$ provided $m \in \RandMZ_g\cap \RandMZ_h$. The large probability mass for small $m/z$ in Figures \ref{Figure:TrainSignal}(i) and \ref{supp:figure:NoiseObsOther} suggest this is the case for spectra with $+2$ precursor charge, and is approximately true for $+3$ and $\geq +4$ precursors.
\end{remark}

\begin{remark}
\label{supp:remark:LambdaAvg}
Corollary \ref{supp:corollary:lambda} is analogous to Corollary \ref{supp:corollary:dModAvg}, where \eqref{supp:equation:ContFunctions} implies that the average $\lambda^{\Mass}_{\noise,g}(m \mid y)$ across neighboring intervals $F_m$ can be approximated with a continuous function. This, along with Remark \ref{supp:remark:lambdag}, justifies our estimator in \eqref{equation:dnoise:ma}.
\end{remark}

\section{Estimating Bayes factors $BF_g(P)$}
\label{supp:section:BF}
\subsection{Calculating $BF_g(P)$ given hyperparameters}
\label{supp:subsection:BF:Calc}
As stated in the beginning of Section~\ref{subsection:GlobalHyp},
\begin{align*}
    BF_g(P) = BF_g^{(\text{rt})}(P) \times BF_g^{(\text{gen})}(P)\times BF_g^{(\text{int})}(P)\times BF_g^{(\text{ma})}(P).
\end{align*}
The expressions for and computing $BF_g^{(\text{rt})}(P)$, $BF_g^{(\text{ma})}(P)$, and the denominator of $BF_g^{(\text{gen})}(P)$ are straightforward. Let $x^{(\text{gen})}$, $x^{(\text{int})}$, and $z^{(\text{int})}$ be the numerator of $BF_g^{(\text{gen})}(P)$, numerator of $BF_g^{(\text{int})}(P)$, and denominator of $BF_g^{(\text{int})}(P)$, respectively. First,
\begin{align*}
    x^{(\text{gen})} =& \int \prod_{j=1}^{n^{(P)}} \Prob\{ \delta(j) \mid y_j^{(P)},\alpha_g,\Delta_g \}\Prob\{ (\alpha_g,\Delta_g)^T \}\text{d}\alpha_g \text{d}\Delta_g\\
    =& \int \left[\prod_{j: \delta(j)>0} \expit\{ \Delta_g + \alpha_g y_j^{(P)} \}\right] \left(\prod_{j: \delta(j)=0}\left[ 1- \expit\{ \Delta_g + \alpha_g y_j^{(P)} \}\right]\right)\\
    &\times \mathcal{N}\{ (\alpha_g,\Delta_g)^T; \bm{\mu}_{\Gen}, \bm{\Sigma}_{\Gen} \}\text{d}\alpha_g \text{d}\Delta_g,
\end{align*}
which we approximate using a Laplace approximation. Next,
\begin{align*}
    x^{(\text{int})} =& \int \left[\int \prod_{j: \delta(j)>0}\Prob\{ y_{\delta(j)}^{(S_g)} \mid \delta, y_{j}^{(P)}, \gamma_g,\beta_g, \sigma_{y,g}^2, \mu_g \} \Prob(\gamma_g \mid \mu_g) \Prob(\beta_g, \sigma_{y,g}^2) \text{d}\gamma_g \text{d}\beta_g \text{d}\sigma_{y,g}^2 \right]\\
    &\times \prod_{i \in \Noise} D_{\noise}^{\IntSuper}\{ y_{i}^{(S_g)}-\mu_g \}\Prob(\mu_g)\text{d}\mu_g = \int h_{\signal}(\mu_g)h_{\noise}(\mu_g) \mathcal{N}(\mu_g; \bm{\mu}_{\text{int}_1}, \bm{\Sigma}_{\text{int}_{11}})\text{d}\mu_g\\
    h_{\signal}(\mu_g) =& \int \left[\prod_{j: \delta(j)>0} \mathcal{N}\{ y_{\delta(j)}^{(S_g)}; (\gamma_g+\mu_g) + \tilde{y}_{j}^{(P)}\beta_g, \phi_g^{-1} \}\right] \mathcal{N}\{ \gamma_g; a(\mu_g),b \} \mathcal{N}(\beta_g; \mu_{\beta},\sigma_{\beta}^2)\\
    &\times \{v/\phi \mathcal{X}(\phi_g v/\phi ; v)\} \text{d}\gamma_g \text{d}\beta_g\text{d}\phi_g\\
    h_{\noise}(\mu_g) =& \prod_{i \in \Noise} \exp\left[\sum\limits_{k=0}^7 b_k\{ y_{i}^{(S_g)}-\mu_g \}^k \right]\\
    \tilde{y}_{j}^{(P)} =& y_{j}^{(P)} - \abs{\Signal}^{-1}\sum_{j': \delta(j')>0}y_{j'}^{(P)}\\
    a(\mu_g) =& \bm{\mu}_{\text{int}_2} + \bm{\Sigma}_{\text{int}_{12}}/\bm{\Sigma}_{\text{int}_{11}}(\mu_g - \bm{\mu}_{\text{int}_1}), \quad b = \bm{\Sigma}_{\text{int}_{22}} - \bm{\Sigma}_{\text{int}_{12}}^2/\bm{\Sigma}_{\text{int}_{11}},
\end{align*}
where $\mathcal{X}(a;b)$ is the density at $a$ of a $\chi_{b}^2$. We first note that for fixed $\mu_g$, $h_{\noise}(\mu_g)$ is trivial and $h_{\signal}(\mu_g)$ can easily be approximated with a Laplace approximation. However, since $\log\{h_{\noise}(\mu_g)\}$ is a seventh degree polynomial and $h_{\signal}(\mu_g)$ does not have an analytic form, it is difficult to quickly calculate or approximate $x^{(\text{int})}$ with standard methods. To circumvent this, we note that due to the large number of noise peaks in MS/MS spectra and because $\mu_g$ is the mean log-intensity of noise peaks, $\bar{y}_{\noise}^{(S_g)} = \abs{\Noise}^{-1}\sum_{i \in \Noise}y_i^{(S_g)} \mid \mu_g \approx N(\mu_g, \hat{s}^2/\abs{\Noise})$ for $\hat{s}^2$ the sample variance of $\{ y_i^{(S_g)} \}_{i \in \Noise}$. For $\tilde{\mu}_g = \E\{\mu_g \mid \bar{y}^{(S_g)}_{\noise}\}$ assuming $\bar{y}^{(S_g)}_{\noise}$ is normally distributed, we therefore approximate $x^{(\text{int})}$ as $h_{\signal}(\tilde{\mu}_g)h_{\noise}(\tilde{\mu}_g)$. Lastly, we use a similar technique to approximate $z^{(\text{int})}$ as
\begin{align*}
    z^{(\text{int})} =& \int \prod_{i \in n^{(S_g)}} D_{\noise}^{\IntSuper}\{y_i^{(S_g)} - \mu_g\}\mathcal{N}(\mu_g; \bm{\mu}_{\text{int}_1}, \bm{\Sigma}_{\text{int}_{11}})\text{d}\mu_g \approx \prod_{i \in n^{(S_g)}} D_{\noise}^{\IntSuper}\{y_i^{(S_g)} - \tilde{\mu}_g\}\\
    \tilde{\mu}_g =& \E\{ \mu_g \mid \bar{y}^{(S_g)}_{\noise} \}, \quad \bar{y}^{(S_g)}_{\noise} = \{n^{(S_g)}\}^{-1}\sum_{i \in [n^{(S_g)}]} y_i^{(S_g)}\\
    \bar{y}^{(S_g)}_{\noise} \mid \mu_g \sim & N(\mu_g,\hat{s}^2/n^{(S_g)}), \quad  \hat{s}^2 = \text{Sample\_variance}\left[\{ y_i^{(S_g)} \}_{i \in [n^{(S_g)}]}\right].
\end{align*}

\subsection{Estimating signal hyperparameters}
\label{supp:subsection:SignalHyp}
\begin{figure}[t!]
    \centering
    \includegraphics[width=0.8\textwidth]{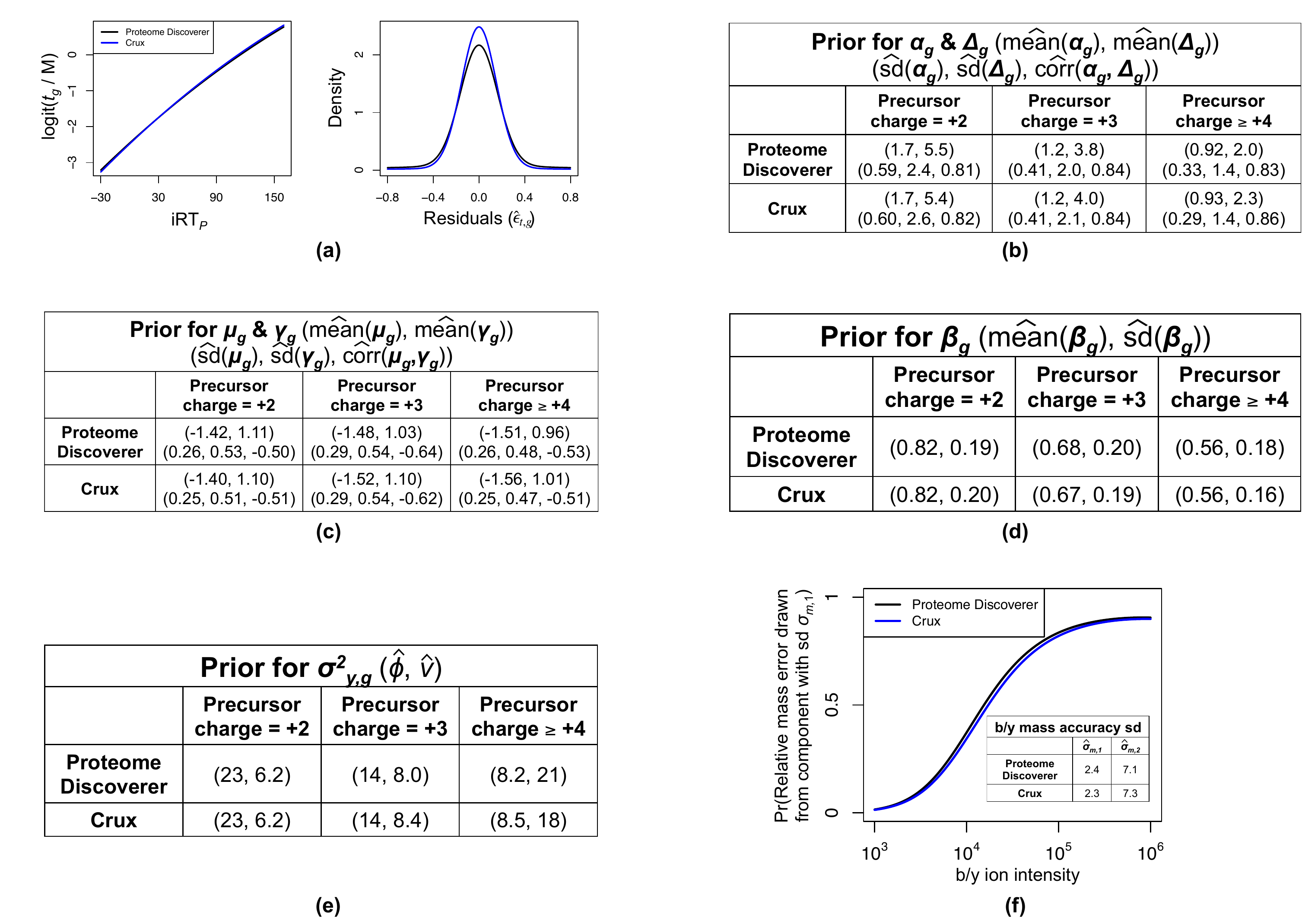}
    \caption{A comparison of the signal hyperparameter estimates derived from Proteome Discoverer and Crux, a commonly used open-source method to score peptide-spectrum matches \citep{Crux_Pvalues}. We used Proteome Discoverer to derive hyperparameters in Figure~\ref{Figure:TrainSignal}, Section~\ref{section:Simulations}, and Section~\ref{section:Results}. (\textbf{a}): Retention time defined in \eqref{equation:RTModel}. Curves indicate the estimates for $f^{(\text{t})}$ (left) and the density of $\epsilon_{t,g}$ (right). (\textbf{b}): Prior for $\alpha_g$ and $\Delta_g$ from \eqref{equation:Obs}. (\textbf{c}): Prior for $(\mu_g,\gamma_g)^T$ from step (ii) of Algorithm~\ref{algorithm:SpectrumGen}. (\textbf{d}): Prior for $\beta_g$ from \eqref{equation:ySignal:prior}. (\textbf{e}): Prior for $\sigma_{y,g}^2$ in \eqref{equation:ySignal:prior}. (\textbf{f}): Prior for relative part per million mass accuracy $r_{\delta(j)}$ defined in \eqref{equation:mSignal}. Curves are estimates for $\pi_{\signal}(x)$.}\label{supp:figure:Crux}
\end{figure}

Here we describe our estimates for the signal hyperparameters $\Theta^{(\text{rt})}=\{ f_{\signal}^{(\text{t})}, \sigma_{t,1}^2,\sigma_{t,2}^2,\pi_{t} \}$, $\{\bm{\mu}_{\Gen},\bm{\Sigma}_{\Gen}\} \subset \Theta^{(\text{gen})}$, $\{\mu_{\beta},\sigma_{\beta}^2,\phi,v, \bm{\mu}_{\Int},\bm{\Sigma}_{\Int}\} \subset \Theta^{(\text{int})}$, and $\{f_{\signal}^{\Mass},\sigma_{m,1}^2,\sigma_{m,2}^2\} \subset \Theta^{(\text{ma})}$. We first note that the estimates for $\Theta^{(\text{rt})}$ and $\{f_{\signal}^{\Mass},\sigma_{m,1}^2,\sigma_{m,2}^2\} \subset \Theta^{(\text{ma})}$ were the same for all precursor charges, whereas the remaining estimates depend on precursor charge, where we partition precursor charge as $+2$, $+3$, and $\geq +4$. We apply the same partition in Section~\ref{supp:subsection:NoiseHyp}. For both this set of hyperparameters and the noise hyperparameters estimated in Section~\ref{supp:subsection:NoiseHyp}, we first curate a set of high-confidence peptide-spectrum matches (PSMs) using an existing search engine. For the estimates shown in Figure~\ref{Figure:TrainSignal} and used in Sections~\ref{section:Simulations} and \ref{section:Results}, we used the commercial software Proteome Discoverer \citep{ProteomeDiscoverer} licensed by Thermo Fisher Scientific to identify PSMs at a 1\% decoy-determined false discovery rate (FDR) with normalized delta scores $>0.05$. The second criterion ensures the selected PSMs have unambiguous score ordering, which, as shown in Figure~\ref{Figure:Database}, is important in nanobody proteomes. As indicated in Figure~\ref{supp:figure:Crux}, we find that the search engine has little to no effect on hyperparameter estimates.\par 
\indent For the $\Theta^{(\text{rt})}$, we fit the quadratic function $f_{\signal}^{(\text{t})}$ by regressing $\logit\{t_g/M\}$ onto spectrum $g$'s generator's indexed retention time, $\irt_P$, using Huber's loss to account for any potential incorrectly assigned PSMs. We then estimate $\sigma_{t,1}^2$, $\sigma_{t,2}^2$, and $\pi_{t}$ using an EM algorithm. To estimate the remaining hyperparameters, we map predicted peaks $j$ to observed peaks $i$ for each PSM using the procedure outlined in Remark~\ref{remark:delta}. We then estimate $\bm{\mu}_{\Gen}$ and $\bm{\Sigma}_{\Gen}$ as
\begin{align*}
    \{\hat{\bm{\mu}}_{\Gen}, \hat{\bm{\Sigma}}_{\Gen}\} =& \argmax_{\bm{\mu},\bm{\Sigma}} \prod_{g=1}^q \int \left[ \prod_{j:\delta(j)>0} \expit\{ \alpha_g y_j^{(P)} + \Delta_g \} \right] \left( \prod_{j:\delta(j)=0} [ 1-\expit\{ \alpha_g y_j^{(P)} + \Delta_g \} ] \right)\\
    &\times \mathcal{N}\{(\alpha_g,\Delta_g)^T; \bm{\mu},\bm{\Sigma}\}\text{d}\alpha_g\text{d}\Delta_g,
\end{align*}
where $q$ is the number of spectra in the training set. We note that the map $\delta$ and matched peptide $P$ implicitly depend on the spectrum $g$.\par 
\indent For $\mu_{\beta}$, $\sigma_{\beta}^2$, $\phi$, and $v$, we use ordinary least squares (OLS) to regress $\{y_{\delta(j)}^{(S_g)}\}_{j \in \delta^{-1}(\Signal)}$ onto $\{y_{j}^{(P)}\}_{j \in \delta^{-1}(\Signal)}$, which gives us an estimate for $\beta_g$, $\hat{\beta}_g$, an estimate for $\V(\hat{\beta}_g \mid \beta_g)$, $\hat{s}_g^2$, and an estimate for $\sigma_{y,g}^2$, $\hat{\sigma}_{y,g}^2$, where we note that $\Signal$ implicitly depends on $g$. We then let $\hat{\mu}_{\beta}=q^{-1}\sum\limits_{g=1}^q \hat{\beta}_g$ and estimated $\sigma_{\beta}^2$ as the method of moments estimator $\hat{\sigma}_{\beta}^2 = (q-1)^{-1}\sum\limits_{g=1}^q \{(\hat{\beta}_g - \hat{\mu}_{\beta})^2 - \hat{s}_g^2\}$. Finally, we estimate $\phi$ and $v$ using maximum likelihood assuming $\hat{\sigma}_{y,g}^2 \mid \sigma_{y,g}^2 \sim \sigma_{y,g}^2 (d_g^{-1}\chi_{d_g}^2)$ and $\sigma_{y,g}^{-2} \sim (\phi/v)\chi_v^2$, where $d_g = \abs{I_{\signal}} - 2$.\par \indent For $\bm{\mu}_{\Int}$ and $\bm{\Sigma}_{\Int}$, we let $\hat{\bm{\mu}}_{\Int} = q^{-1}\sum\limits_{g=1}^q( \bar{y}_{\noise}^{(S_g)}, \bar{y}_{\signal}^{(S_g)}-\bar{y}_{\noise}^{(S_g)} )^T$ for $\bar{y}_{\noise}^{(S_g)} = \abs{\Noise}^{-1}\sum\limits_{i \in \Noise} y_{i}^{(S_g)}$, where we note that both $\Signal$ and $\Noise$ implicitly depend on $g$. We then let $\hat{\bm{\Sigma}}_{\Int}$ be the method of moments estimator
\begin{align*}
    \hat{\bm{\Sigma}}_{\Int} = q^{-1}\sum\limits_{g=1}^q \left\{\bm{x}_g\bm{x}_g^T - \begin{pmatrix} \hat{a}_g^2 & -\hat{a}_g^2\\-\hat{a}_g^2 & \hat{a}_g^2 + \hat{b}_g^2 \end{pmatrix}\right\}, \quad \bm{x}_g = ( \bar{y}_{\noise}^{(S_g)}, \bar{y}_{\signal}^{(S_g)}-\bar{y}_{\noise}^{(S_g)} )^T - \hat{\bm{\mu}}_{\Int},
\end{align*}
where $\hat{a}_g^2$ is the sample variance of $\{y_{i}^{(S_g)}\}_{i \in \Noise}$ and $\hat{b}_g^2$ is the estimate for the variance of the OLS estimate for the intercept in the regression described in the previous paragraph.\par
\indent Lastly, for $f_{\signal}^{\Mass}$, $\sigma_{m,1}^2$, and $\sigma_{m,2}^2$, let $r_{\delta(j)}$, $j \in \delta^{-1}(\abs{\Signal})$, be as defined in step (iv) of Algorithm~\ref{algorithm:SpectrumGen}, where we assume $\{r_{\delta(j)}\}_{j \in [n^{(P)}]; g \in [q]}$ are independent and identically distributed. Note that $\delta$ implicitly depends on spectrum $g$. Assuming $f_{\signal}^{\Mass}$ is a constant, we first use an EM algorithm to estimate $\sigma_{m,1}^2$, and $\sigma_{m,2}^2$. We next estimate $f_{\signal}^{\Mass}$ using a second EM algorithm by fixing the estimates for $\sigma_{m,1}^2$, and $\sigma_{m,2}^2$. The estimates changed only slightly upon iterating this procedure.

\begin{remark}
\label{supp:remark:Heterogeneity}
The estimates for the priors of $(\alpha_g,\delta_g)$, $(\mu_g,\gamma_g)$, $\beta_g$, and $\sigma_{y,g}^2$ in Figure~\ref{supp:figure:Crux} demonstrate the inter-spectra heterogeneity. For example, the coefficient of variations for $\alpha_g$ and $\sigma_{y,g}^2$, defined in \eqref{equation:Obs} and \eqref{equation:ySignal}, ranges from 0.34 to 0.36 and 0.15 to 0.23, respectively.
\end{remark}

\subsection{Estimating noise hyperparameters}
\label{supp:subsection:NoiseHyp}
It remains to define estimates for the noise hyperparameters $D_{\noise}^{\IntSuper}(y)$, $\lambda_{\noise,g}^{\Mass}(m)$, and $d_{\noise}^{\Mass}(r \mid m; y)$, where the first two were found to depend on precursor charge, and the third showed little dependence on precursor charge. We therefore return three estimates for the first two (one for each precursor charge group defined in Section~\ref{supp:subsection:SignalHyp} above) and one for the third. For $D_{\noise}^{\IntSuper}(y)$, we plot a histogram of $\{y_i^{(S_g)}-\bar{y}_{\noise}^{(S_g)}\}_{i \in \Noise, g \in [q]}$, where $q$ is the number of training spectra in the precursor charge bin, $\Noise$ implicitly depends on $g$, and $\bar{y}_{\noise}^{(S_g)}=\abs{\Noise}^{-1}\sum_{i \in \Noise} y_i^{(S_g)}$. We then use the technique outlined in \citet{PoiDens} to regress bin counts $N_s$ onto their midpoints $m_s$ assuming $N_s\mid m_s \sim  \text{Poi}\{ N\Delta_m\exp(\sum_{k=0}^7 b_k m_s^k) \}$, where $N$ is the total number of points, $\Delta_m$ is the bin width, and $b_k$ are the regression coefficients. The density estimate is $\hat{D}_{\noise}^{\IntSuper}(y) \propto \exp( \sum_{k=1}^7 \hat{b}_k y^k )$, and we normalize $\hat{D}_{\noise}^{\IntSuper}(y)$ so that it integrates to 1. The choice to make the log density a seventh degree polynomial was to ensure estimates were unbiased and consistent across different training sets, although estimates were similar when we let the degree be as low as five. Histogram bin width $\Delta_m$ had a negligible effect on estimates.\par 
\indent We next present our estimator for $d_{\noise}^{\Mass}(r \mid m; y)$ in Algorithm~\ref{supp:algorithm:NoiseMA} below, which utilizes the estimate for $d_{\signal}^{\Mass}(r \mid y)$, $\hat{d}_{\signal}^{\Mass}(r \mid y)$, defined using the hyperparameter estimates for $\{ f_{\signal}^{\Mass}, \sigma_{m,1}^2, \sigma_{m,2}^2\} \subset \Theta^{(\text{ma})}$ obtained in Section~\ref{supp:subsection:SignalHyp}, and is therefore uniquely determined by the estimate for $f_{\noise}^{\Mass}$ defined in \eqref{equation:dnoise:ma}. We let $\precmass_g$ and $z_g$ be spectrum $g$'s precursor's monoisotopic mass and charge for the remainder of Section~\ref{supp:subsection:NoiseHyp}.
\begin{myalgorithm}[Estimate for $f_{\noise}^{\Mass}$]
\label{supp:algorithm:NoiseMA}
\textbf{Input}: The estimate $\hat{d}_{\signal}^{\Mass}$, threshold $\Delta>0$, a target database $\Targ$, and training spectra $S_g$ whose matched signal peaks have been removed.\\
\textbf{Output}: Estimates for $a_0,a_1 \in \mathbb{R}$ used to define the function $f_{\noise}^{\Mass}(m) = a_0 + a_1 \log(m)$.
\begin{enumerate}[label=(\roman*)]
\item Fix $g$ and let $\mathcal{N}_g$ be sampled uniformly at random from $\mathcal{S}_g=\{\{$b/y ion m/z's with charges 1 through $\min(3,z_g)$ from a randomly permuted peptide in $\Targ$ with monoisotopic mass $\precmass\}: \abs{\precmass_g-\precmass} \leq z_g\Delta\}$.\label{supp:item:Pg}
\item Use the procedure outlined in Remark~\ref{remark:delta} to map predicted m/z's $m \in \mathcal{N}_g$ to noise m/z's $\hat{m}$ in $S_g$, where $m$ matches $\hat{m}$ if $r(\hat{m},m) = 10^6\abs{\hat{m}/m-1} \leq \omega$. Let $\mathcal{M}_g = \{(m,r(\hat{m},m),\hat{y})\}_{\text{$m$ and $\hat{m}$ match}}$, where $\hat{y}$ is the log-intensity of the noise peak with m/z $\hat{m}$.\label{supp:item:Mg}
\item Repeat steps \ref{supp:item:Pg} and \ref{supp:item:Mg} for all spectra $g \in [q]$, where $q$ is the number of training spectra.
\item Assuming $r(\hat{m},m) \mid m,\hat{y} \sim \pi_{\noise}(m)\mathcal{U}_{\pm \omega}\{r(\hat{m},m)\} + \{1-\pi_{\noise}(m)\}\hat{d}_{\signal}^{\Mass}\{ r(\hat{m},m) \mid \hat{y} \}$ for $\mathcal{U}_{\pm \omega}(r)$ the density at $r$ of a $U[-\omega,\omega]$ and $\pi_{\noise}(m) = \expit\{ a_0 + a_1\log(m) \}$, estimate $a_0$ and $a_1$ via maximum likelihood using $\{\mathcal{M}_g\}_{g \in [q]}$ and an EM algorithm.
\end{enumerate}
\end{myalgorithm}
\noindent To make Algorithm~\ref{supp:algorithm:NoiseMA} as principled as possible, we let $\Delta$ be the mass spectrometer's precursor isolation window. However, we note that varying $\Delta$ or defining $\mathcal{S}_g$ from step~\ref{supp:item:Pg} in terms of the precursor's relative part per million mass accuracy $10^6\abs{\precmass_g/\precmass-1}\leq \Delta$ resulted in virtually identical estimates for $a_0$ and $a_1$. We discuss our choice for $\mathcal{N}_g$ in Remark~\ref{supp:remark:Pg} after the presentation of Algorithm~\ref{supp:algorithm:lambda} below.\par
\indent We lastly describe our estimate for $\lambda_{\noise,g}^{\Mass}(m)$. For each precursor charge group defined in Section~\ref{supp:subsection:SignalHyp}, we partition spectra by precursor monoisotopic mass $\precmass_g$ using the below partition, where $\lambda_{\noise,g}^{\Mass}(m) = \lambda_{\noise,h}^{\Mass}(m)$ if spectra $g$ and $h$ lie in the same group.
\begin{enumerate}[label=(\alph*)]
\item $z_g=+2$: $\precmass_g \in (0,\infty)$
\item $z_g=+3$: $\precmass_g \in (0,m_3)$ or $\precmass_g \in [m_3,\infty)$
\item $z_g\geq +4$: $\precmass_g \in (0,m_4)$ or $\precmass_g \in [m_4,\infty)$
\end{enumerate}
where $m_3$ and $m_4$ are the median $\precmass_g$ for training spectra with $z_g=+3$ and $z_g\geq +4$, respectively. This partition, consisting of five groups, was chosen because $\lambda_{\noise,g}^{\Mass}(m)$ appeared consistent within each group. For the purposes of Algorithm~\ref{supp:algorithm:lambda} below, we let $\mathcal{J}: \mathbb{N}\times \mathbb{R} \to [5]$ define the above partition (i.e. $\mathcal{J}(z_g,\precmass_g)=j$ if $g$'s precursor charge-mass pair falls in group $j \in [5]$) and $\lambda_{\noise,j}^{\Mass}(m)$ to be such that $\lambda_{\noise,g}^{\Mass}(m)=\lambda_{\noise,j}^{\Mass}(m)$ if $\mathcal{J}(z_g,\precmass_g)=j$ for group $j \in [5]$.
\begin{myalgorithm}[Estimate for $\lambda_{\noise,g}^{\Mass}$]
\label{supp:algorithm:lambda}
\textbf{Input}: A target database $\Targ$, non-overlapping bins $\{\mathcal{B}_s\}_s$ such that $\cup_s \mathcal{B}_s = (0,\infty)$, thresholds $\Delta,t > 0$, group $j \in [5]$, and training peptide-spectrum match pairs $(P_g,S_g)$ such that $\mathcal{J}(z_g,\precmass_g)=j$.\\
\textbf{Output}: The estimator $\hat{\lambda}_{\noise,j}^{\Mass}(m) = I(m\leq t)\sum_s \hat{b}_{js} I(m \in \mathcal{B}_s) + I(m > t)\hat{g}_j(m)$, where $\hat{b}_{js},\hat{g}_j(m) \geq 0$.
\begin{enumerate}[label=(\roman*)]
\item Given $\Delta$, let $\mathcal{N}_g$ be as defined step~\ref{supp:item:Pg} of Algorithm~\ref{supp:algorithm:NoiseMA}, and let $\tilde{\mathcal{N}}_g = \{m \in \mathcal{N}_g:m$ is not the m/z of a b1, b1$^c$, y1, or y1$^c$ ion in $P_g$'s predicted spectrum$\}$, where the superscript ``$c$'' denotes the ion's complement.\label{supp:item:Ptildeg} 
\item For $\mathcal{M}_g$ as defined in step~\ref{supp:item:Mg} of Algorithm~\ref{supp:algorithm:NoiseMA}, let $\tilde{\mathcal{M}}_g = \{ m > 0:m \in \tilde{\mathcal{N}}_g$ and $(m,r(m,\hat{m}),\hat{y}) \in \mathcal{M}_g\}$.\label{supp:item:Mtildeg}
\item Repeat steps \ref{supp:item:Ptildeg} and \ref{supp:item:Mtildeg} for all training spectra in group $j$. For $n_{\noise}^{(S_g)}$ the number of peaks in $S_g$ after removing its matched signal peaks, define
\begin{align*}
    \hat{p}_{gjs} =& \abs*{ \mathcal{B}_s \cap \tilde{\mathcal{M}}_g }/\{ n_{\noise}^{(S_g)}\abs*{ \mathcal{B}_s \cap \tilde{\mathcal{N}}_g } \}, \quad g \in  \{\text{group $j$ training spectra $g'$}: \mathcal{B}_s \cap \tilde{\mathcal{N}}_{g'} \neq \{\emptyset\}\}\\
    \hat{b}_{js} =& \frac{ \sum_g I(\mathcal{B}_s \cap \tilde{\mathcal{N}}_g \neq \{\emptyset\})\hat{p}_{gjs}}{ \sum_g I(\mathcal{B}_s \cap \tilde{\mathcal{N}}_g \neq \{\emptyset\}) },
\end{align*}
where $\abs*{E}$ is the number of elements in the set $E$.
\item Let $m_s$ be the midpoint of bin $\mathcal{B}_s$. Then $\log\{\hat{g}_j(m)\}$ is the estimator obtained by regressing $\log(\hat{b}_{js})$ onto $m_s$ for $m_s>t$ using B-splines of degree three.
\end{enumerate}
\end{myalgorithm}

\begin{remark}
\label{supp:remark:Pg}
Like step~\ref{supp:item:Pg} of Algorithm~\ref{supp:algorithm:NoiseMA}, $\mathcal{N}_g$ in step~\ref{supp:item:Ptildeg} of Algorithm~\ref{supp:algorithm:lambda} is meant to contain elements from $\RandMZ_g$ defined in both Theorem~\ref{theorem:noise} and Assumption~\ref{supp:assumption:Mg}. We define $\mathcal{N}_g$ in terms of the fragments of a potentially observable peptide to sample the space of peptide fragment m/z's that may generate a noise peak in $S_g$, which circumvents having to enumerate $\RandMZ_g$. We further restrict $\mathcal{N}_g$ to be the elements in $\tilde{\mathcal{N}}_g$ in Algorithm~\ref{supp:algorithm:lambda} because noise peaks cannot exist in the regions surrounding $P_g$'s predicted m/z's after $S_g$'s matched signal m/z's have been removed. We only remove $P_g$'s b1, y1, and their compliment ions to simplify the implementation of Algorithm~\ref{supp:algorithm:lambda}, but note that including additional ions in this exclusion set had a negligible effect on the estimator $\hat{\lambda}_{\noise,j}^{\Mass}(m)$.
\end{remark}

\begin{remark}
\label{supp:remark:LambdaForm}
Since Theorem~\ref{theorem:noise} states that $\lambda_{\noise,g}^{\Mass}(m)$ can be approximated with a continuous function, we assume $\lambda_{\noise,g}^{\Mass}(m)$ is locally a constant and let the estimator $\hat{\lambda}_{\noise,j}^{\Mass}(m)$ be a smoothed version of the locally constant step function $\sum_s \hat{b}_{js} I(m \in \mathcal{B}_s)$. In practice, we only smooth for $m>t=350\text{m/z}$, since smoothing the entire function tended to underestimate $\lambda_{\noise,g}^{\Mass}(m)$ at small $m$.
\end{remark}

\begin{remark}
\label{supp:remark:pgjs}
If $m \in \mathcal{B}_s$ is the predicted m/z of a peptide fragment, the estimator $\hat{p}_{gjs}$ in step (iv) of Algorithm~\ref{supp:algorithm:lambda} is the method of moments estimator for the probability a noise peak in spectrum $g$ is within $\omega$ parts per million (ppm) of $m$. Our data processing procedure outlined in Section~\ref{section:Data} ensures only one noise peak can lie within $\omega$ ppm of any peptide fragment m/z in practice.
\end{remark}

\subsection{A justification for the noise intensity model}
\label{supp:subsection:NoiseInt}
We assume that noise log-intensities are drawn from a location family of distributions. That is, $y_i^{(S_g)} \sim \mu_g + Y_{gi}$ for $i \in \Noise$, where $\mu_g$ is defined in step (ii) of Algorithm~\ref{algorithm:SpectrumGen} and $Y_{gi}$ is a mean zero random variable with density $D_{\noise}^{\IntSuper}(\cdot)$. This differs from previous work that assumes $\{y_i^{(S_g)}\}_{i\in\Noise,g \in[q]}$ are independent and identically distributed \citep{UniformNoise,LikelihoodScoring,MSGFplus}, or simply normally distributed \citep{StatNormal}. Figure~\ref{supp:figure:Crux}(c) suggests the variation in $\mu_g$ is non-trivial ($\text{CV}(\mu_g)$, its estimated coefficient of variation, is around 0.2), indicating $\E\{y_i^{(S_g)}\}$ is spectrum-specific. To explore how other aspects of $y_i^{(S_g)}$'s distribution vary across spectra $g$, we examined the variation in $y_i^{(S_g)}$'s scale, skew, and tails across spectra using Tukey's g-and-h family of distributions \citep{TukeyGandH}. Briefly, we used quantile least squares to fit the following model for each training spectrum $g$ in the dataset:
\begin{align*}
    y_i^{(S_g)} = \begin{cases}
    a_g + b_g \frac{\exp(\text{g}_{g}Z_{gi}) - 1}{\text{g}_{g}}\exp(\text{h}_g Z_{gi}^2/2) & \text{if $\text{g}_{g}\neq 0$}\\
    a_g + b_g Z_{gi} \exp(\text{h}_g Z_{gi}^2/2) & \text{if $\text{g}_{g}=0$}
    \end{cases}, \quad Z_{gi} \stackrel{i.i.d}{\sim} N(0,1), \quad i \in \Noise,
\end{align*}
where we abuse notation and let the subscript ``$g$'' index spectrum and ``$\text{g}$'' be a parameter. Here, $a_g=\text{median}\{y_i^{(S_g)}\}$, $b_g$ gives the scale, $\text{g}_{g}$ specifies the skew, and $\text{h}_g \geq 0$ determines the size of $y_i^{(S_g)}$'s tails. Since nearly all estimates for $\text{h}_g$ were exactly 0, we set $\text{h}_g=0$ for all spectra $g$. We then used the asymptotic results derived in \citet{QLS_Tukey} in conjunction with ashr \citep{Ashr} to derive empirical Bayes prior distributions $b_g\sim \Pi_{b}(\cdot)$ and $\text{g}_g\sim \Pi_{\text{g}}(\cdot)$. We used these priors to estimate that $\text{CV}(b_g),\text{CV}(\text{g}_g) < 0.1$, which suggests that in comparison to the variation in noise location, the variation in scale, skew, and tail size is trivial, and was therefore assumed to be zero.

\section{Additional simulation details and results}
\label{supp:section:simulations}
\subsection{Uniform noise model}
\label{supp:subsection:simulationDetails}
The uniform noise model was chosen to match that used in \citet{LikelihoodScoring,UniformNoise}, and was defined to be $D_{\noise,g}^{\Mass}(m \mid y) \propto I\{m \in [ m_{\min}^{(S_g)}, m_{\max}^{(S_g)} ]\}$, where $m_{\min}^{(S_g)} = \min\left( \{m_i^{(S_g)}\}_{i \in [n^{(S_g)}]} \right)$ and $m_{\max}^{(S_g)} = \max\left( \{m_i^{(S_g)}\}_{i \in [n^{(S_g)}]} \right)$ for observed spectrum $S_g$. As a consequence, the probability a noise peak $i \in \Noise$ gets mapped to predicted peak $j$ is
\begin{align*}
    \Prob\{ m_i^{(S_g)} \in m_j^{(P)}(1 \pm 10^{-6}\omega) \} = 2 \times 10^{-6}\omega\frac{ m_j^{(P)} }{ m_{\max}^{(S_g)} - m_{\min}^{(S_g)} } = O(10^{-6}\omega),
\end{align*}
which follows from the fact that $\frac{ m_j^{(P)} }{ m_{\max}^{(S_g)} - m_{\min}^{(S_g)} } \lesssim 1$ because the range $m_{\max}^{(S_g)} - m_{\min}^{(S_g)}$ is typically comparable with $S_g$'s precursor's mass. Since $\omega = 20$ in our application, this probability is $O(10^{-5})$.\par
\indent We compared MSeQUiP's model for noise m/z's to the uniform model in Section~\ref{section:Simulations} by replacing $\lambda_{\noise,g}^{\Mass}$ and $d_{\noise}^{\Mass}$ in the expressions for $BF_g^{(\Gen)}(P)$ and $BF_g^{(\text{ma})}(P)$ in Section~\ref{subsection:GlobalHyp} by those implied under the uniform noise model. Specifically,
\begin{align*}
    &\lambda_{\noise,g}^{\Mass}\{m_j^{(P)}\} = 2 \times 10^{-6}\omega\frac{ m_j^{(P)} }{ m_{\max}^{(S_g)} - m_{\min}^{(S_g)} }, \quad d_{\noise}^{\Mass}\{r \mid m_j^{(P)}; y\} = (2\omega)^{-1}.
\end{align*}

\subsection{Simulation results}
\label{supp:subsection:simulationResults}
\begin{figure}
    \centering
     \includegraphics[width=0.4\textwidth]{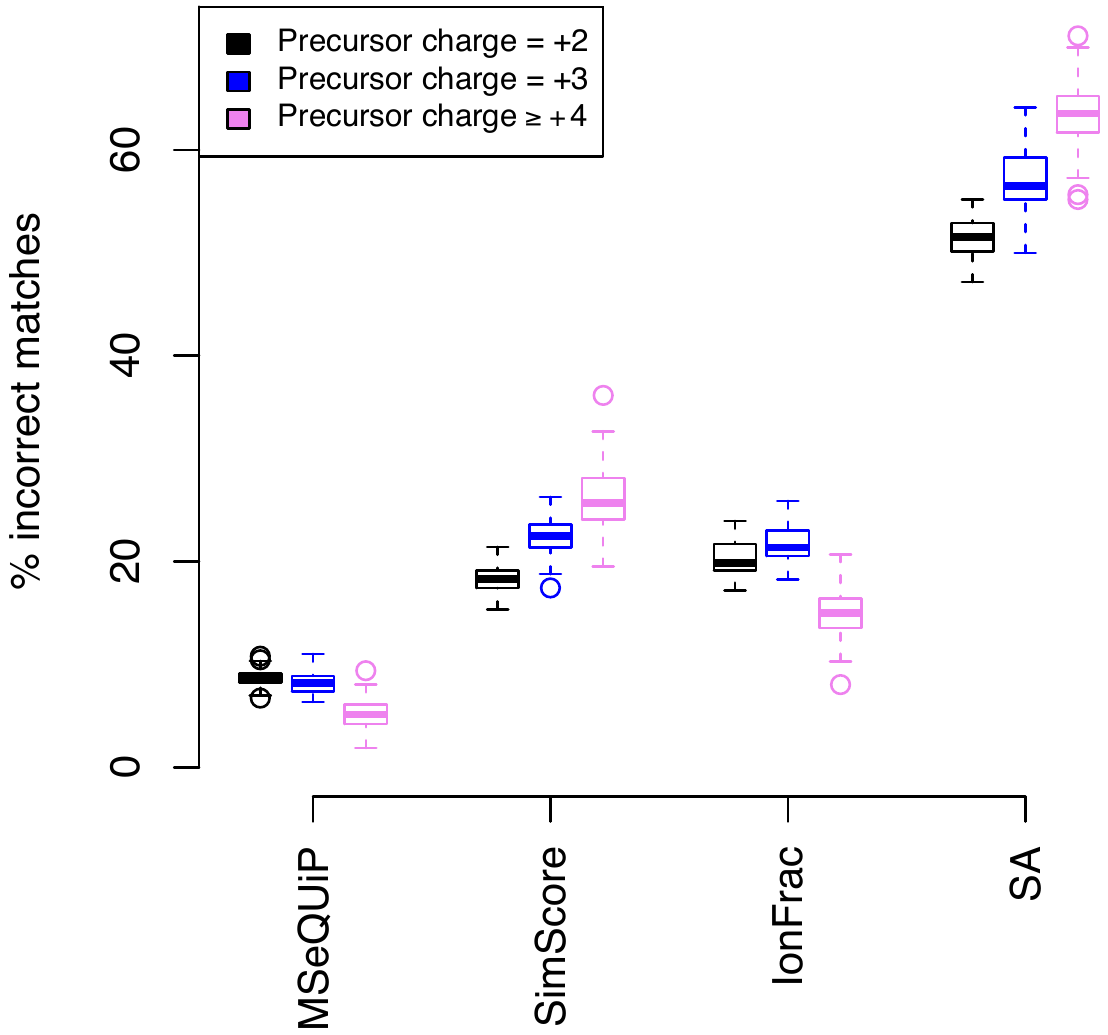}
    \caption{Simulation results comparing MSeQUiP to other scoring functions.}\label{supp:Figure:SimResults}
\end{figure}

Here we include additional simulation results comparing inference with MSeQUiP and existing scoring functions that incorporate relative intensity information. The latter include the SimScore \citep{SimScore}, ion fraction \citep{NN_pred}, and spectral angle (SA) \citep{NN_pred}, and, for $I_p$ and $I_o$ the predicted and observed relative intensities, are defined as
\begin{align*}
    \text{SimScore}&=\frac{\sum (I_p I_o)^{1/2}}{\sum I_p I_o}, \quad \text{IonFrac} = \frac{\text{\#observed ions}}{\text{\#predicted ions}}\, \quad \text{SA} = f\{\text{Corr}(I_p,I_o)\},
\end{align*}
where $f$ is a strictly decreasing function. We note that SA is used by \citet{NN_pred} as the loss function to train their spectrum prediction machine learner, and both SA and IonFrac are used by Prosit \citep{NN_pred} to rank peptides. We used each scoring function to score potential generators, and defined each method's inferred generator to be the peptide that gave the maximum (SimScore, IonFrac) or minimum (SA) score. The results are given in Figure~\ref{supp:Figure:SimResults}.

\section{Additional real data analysis results}
\label{supp:section:Analysis}
\subsection{PSM-fdr calibration}
\label{supp:subsection:Calibration}
\begin{figure}
    \centering
    \includegraphics[width=0.4\textwidth]{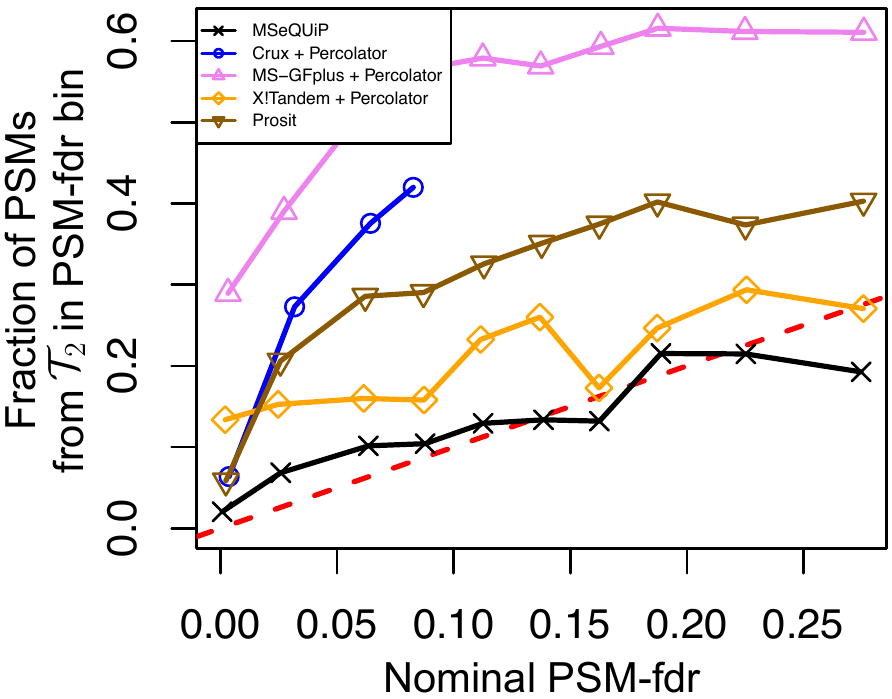}
    \caption{The same as Figure~\ref{Figure:CDR3_FDR}(a), except for all PSMs, not just CDR3 PSMs.}\label{supp:Figure:FullCal}
\end{figure}

We first provide $\PSMfdr$ calibration results for all group 2 PSMs, where group 2 contains the three datasets analyzed in Section~\ref{section:Results}, in Figure~\ref{supp:Figure:FullCal}. While MSeQUiP performs markedly better than all competing methods, we hypothesize its slight inflation is due to the incompleteness of $\Targ_{\text{a}}$ and cross-antigen degeneracies in nanobody framework, CDR1, and CDR2 regions \citep{xu2000diversity}, which, as discussed in Section~\ref{subsection:Calibration}, could imply $\Entrap$ contains non-CDR3 generating peptides.

\begin{figure}
    \centering
    \includegraphics[width=0.4\textwidth]{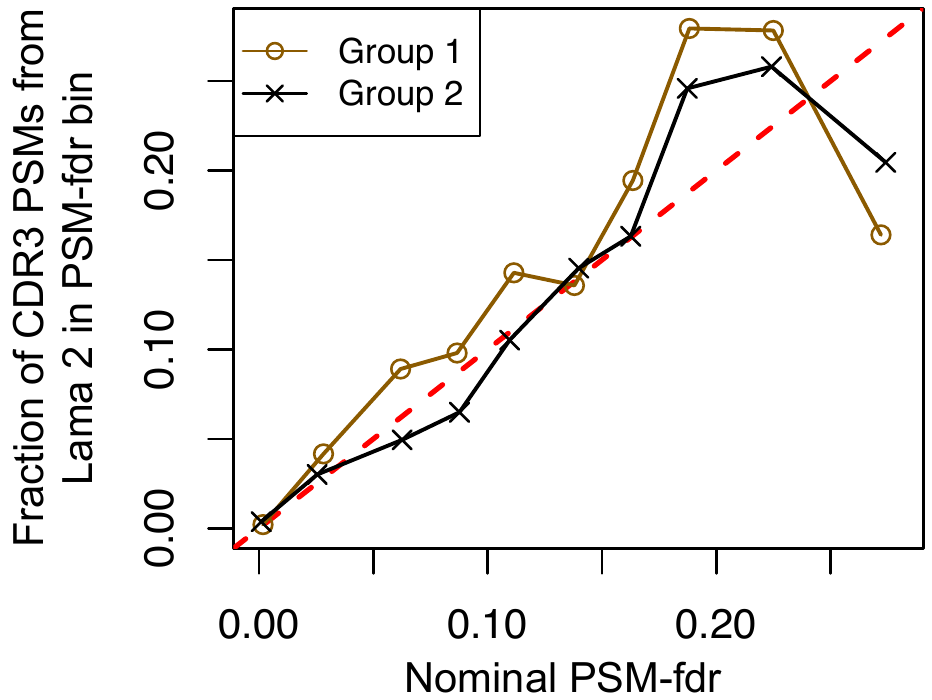}
    \caption{PSM-fdr calibration results for the group 1 and group 2 datasets. The black curve is the exact same as that plotted in Figure~\ref{Figure:CDR3_FDR}(a).}\label{supp:Figure:HighAndLowPEP}
\end{figure}

\indent To further demonstrate the calibration of MSeQUiP's $\PSMfdr$, we evaluated MSeQUiP's performance on the three group 1 datasets by repeating the analysis used to generate Figure~\ref{Figure:CDR3_FDR}(a). Figure~\ref{supp:Figure:HighAndLowPEP} contains the results. Note that group 1 data were used to train Algorithm~\ref{algorithm:SpectrumGen}'s hyperparameters in both the group 1 and group 2 analyses.

\subsection{A justification for the entrapment FDP estimate}
\label{supp:subsection:EntrapmentEst}
Here we argue that the entrapment false discovery proportion (FDP) used in Section~\ref{subsection:Calibration} of the main text to evaluate the fidelity of MSeQUiP's and existing methods' $\PSMfdr$'s is justified. The entrapment FDP is the fraction of the number of CDR3 PSMs that match to $\Entrap$, where $\abs*{\Entrap}/\abs*{\Targ_{\text{a}}\cup \Targ_{\text{b}}} = 0.48$ and 41\% of CDR3 PSMs matched to a peptide from $\Entrap$. Since some CDR3 PSMs from $\Targ_{\text{a}}$ may also be incorrect, the entrapment FDP is a lower bound on the fraction of incorrect PSMs.\par
\begin{figure}
\centering
\includegraphics[width=0.7\textwidth]{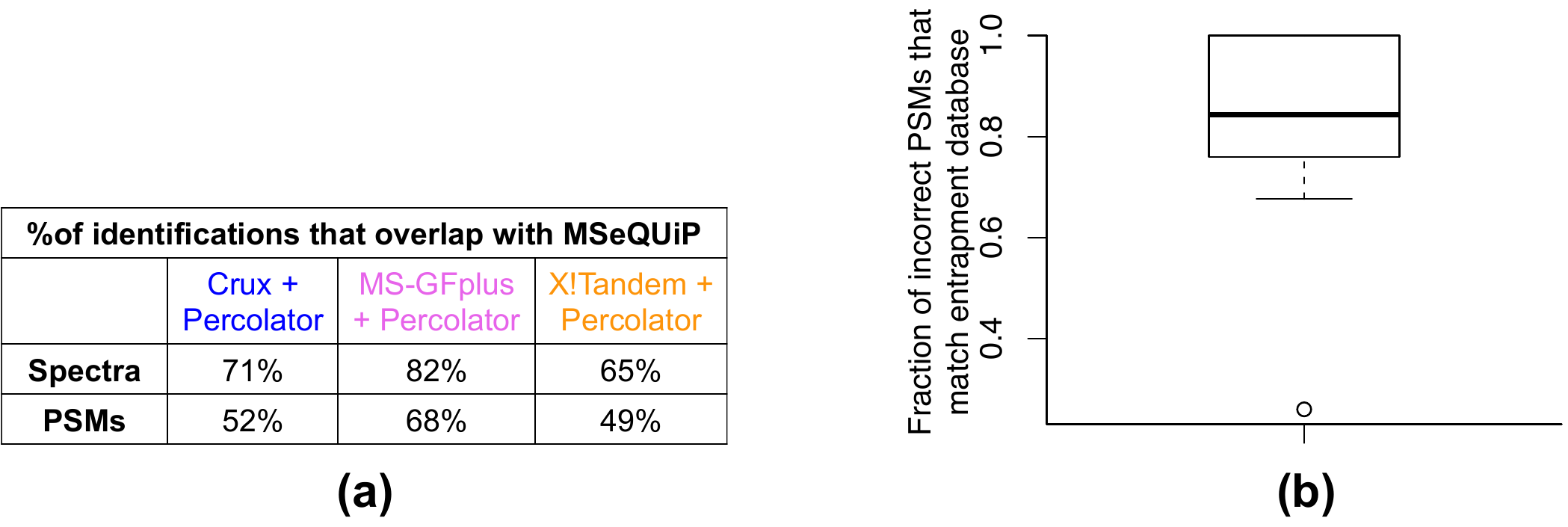}
\caption{(a): Overlap with MSeQUiP at a 1\% $\PSMfdr$-derived global false discovery rate. A spectrum (PSM) overlapped if both methods identified the spectrum (PSM) as significant. MS-GFplus returned fewer and Crux and X!Tandem returned more significant PSMs that MSeQUiP. The small overlap with X!Tandem is likely because their software default includes several post-translational modifications not considered by MSeQUiP (see Section~\ref{supp:subsection:search}). (b): A boxplot of $\hat{\Prob}\{D_g^{(\text{b})} \mid J_g\}$ for simulated spectra $g$ with $\PSMfdr_g \leq 0.2$, where each point represents a simulation and $\hat{\Prob}\{D_g^{(\text{b})} \mid J_g\}=1$ in 30\% of simulations.}\label{supp:figure:DB2Assumption}
\end{figure}

\indent To elaborate on this, let $J_g=\{$PSM $g$ is incorrect$\}$ and $D_g^{(j)}=\{$PSM $g$ matched to $\Targ_j\}$ for $j=\text{a},\text{b}$. Then because $\Prob\{J_g \cap D_g^{(\text{b})}\} = \Prob\{D_g^{(\text{b})}\}$,
\begin{align*}
    \Prob(J_g) = \sum_{j\in\{\text{a},\text{b}\}}\Prob\{J_g \cap D_g^{(j)}\} = \Prob\{D_g^{(\text{b})}\}\left[ 1+\frac{1-\Prob\{D_g^{(\text{b})} \mid J_g\}}{\Prob\{D_g^{(\text{b})} \mid J_g\}} \right].
\end{align*}
Therefore, $\Prob(J_g) \approx \Prob\{D_g^{(\text{b})}\}$ if $\Prob\{D_g^{(\text{b})} \mid J_g\}\approx 1$. To show this is the case for MSeQUiP, we note that because of the sharp contrast from other work arguing existing methods control database incompleteness errors and perform well in standard proteomes whose PSMs are not prone to score-ordering errors \citep{Entrapment1}, as well as their small delta scores for known incorrect CDR3 PSMs (Figure~\ref{Figure:CDR3_FDR}(c)), we hypothesize that the majority of incorrect CDR3 PSMs are the result of score-ordering error, as opposed to database incompleteness errors. This is supported by the fact that even though existing methods inflate $\PSMfdr$'s (see Figure~\ref{Figure:CDR3_FDR}), there was substantial overlap between the spectra MSeQUiP and existing methods identify as significant (Figure~\ref{supp:figure:DB2Assumption}(a)). For example, since MS-GFplus returns egregiously inflated $\PSMfdr$'s, only 68\% of its significant PSMs overlap with MSeQUiP's. However, over 80\% of their significant spectra overlapped, suggesting that the ambiguity was not so much whether the spectrum was generated by a peptide in the target database, but rather the spectrum's correct match. Therefore, we argue $\Prob\{D_g^{(\text{b})} \mid J_g\}\approx 1$ by showing $\Prob\{D_g^{(\text{b})} \mid J_g,H_g=1\}\approx 1$ for $H_g$ as defined in Section~\ref{section:Model}.

\indent We used simulated data to show this. Briefly, we re-analyzed the simulated data from Section~\ref{section:Simulations} by matching spectra to all peptides from $\Targ_{\text{a}}\cup \Entrap$, where since simulated spectra where all generated by peptides in $\Targ_{\text{a}}$, matches to $\Entrap$ are known to be incorrect. For each of the 50 datasets we simulated, we defined $\hat{\Prob}\{D_g^{(\text{b})} \mid J_g\}$ to be the fraction of incorrect PSMs that matched to $D_g^{(\text{b})}$, where Figure~\ref{supp:figure:DB2Assumption}(b) contains results for PSMs with $\PSMfdr_g \leq 0.2$. We chose this $\PSMfdr$ threshold because all but 5\% of MSeQUiP-derived PSMs in Figures~\ref{Figure:CDR3_FDR}(a) and \ref{supp:Figure:HighAndLowPEP}, and all but 20\% in Figure~\ref{Figure:CDR3_FDR}(b), have score-ordering error rate bounded above by 0.2. These results suggest $\Prob\{D_g^{(\text{b})} \mid J_g,H_g=1\}$, and therefore $\Prob\{D_g^{(\text{b})} \mid J_g\}$, is approximately 1, which indicate the y-axes in Figures~\ref{Figure:CDR3_FDR}(a), \ref{Figure:CDR3_FDR}(b), and \ref{supp:Figure:HighAndLowPEP} mirror MSeQUiP's true false discovery proportion.\par
\indent We hypothesize this result suggesting incorrect, but nominally correct, matches are more likely to match $\Entrap$ than $\Targ_{\text{a}}$ is due to us only considering PSMs with small $\PSMfdr$'s. To see this, suppose a spectrum was generated by a peptide $P \in \Targ_{\text{a}}$ but the match $P'$, which may be in $\Targ_{\text{a}}$ or $\Entrap$, is different from $P$. If $P'\in \Targ_{\text{a}}$, then since CDR3 peptides in $\Targ_{\text{a}}$ bind to $\Anto$ and therefore have similar sequences, $P$ will likely be similar to $P'$. As such, even though $BF_g(P')$ may be large, $BF_g(P') \approx BF_g(P)$, implying the $\PSMfdr$ will be large, and therefore is unlikely to be considered by MSeQUiP. However, if $P' \in \Entrap$, then because of the dissimilarity between $\Antt$ and $\Anto$, $P'$ will not necessarily be similar to $P \in \Targ_{\text{a}}$. Therefore, given that we are only considering relatively small $\PSMfdr$'s, these should be dominated by PSMs that match to $\Entrap$. To provide additional support for our hypothesis, we also considered PSMs with $\PSMfdr$'s between 0.4 and 0.6, as these imply $BF_g(P') \approx BF_g(P)$, and consequently suggests $P'$ is similar to $P$. If our reasoning is valid, incorrect PSMs from this set should therefore be dominated by $\Targ_{\text{a}}$ matches. We observed this to be the case, as nearly 80\% of these incorrect PSMs had peptide matches from $\Targ_{\text{a}}$.


\section{Proofs of all theory presented in Sections~\ref{subsection:lfdr} and \ref{supp:section:theory}}
\label{supp:section:Proofs}
\subsection{Proof of Proposition~\ref{Proposition:Pvalues}}
\label{supp:subsection:Prop}
We first prove Proposition~\ref{Proposition:Pvalues} from the main text.
\begin{proof}[Proof of Proposition~\ref{Proposition:Pvalues}]
We can re-write $p_g$ as
\begin{align*}
    p_g = I\{z_g^{(\mathcal{T})} \geq z_g^{(\mathcal{D})}\}/(q+1) + [1+\sum\limits_{h \neq g}^q I\{z_g^{(\mathcal{T})} \geq z_h^{(\mathcal{D})}\}]/(q+1),
\end{align*}
where $I\{z_g^{(\mathcal{T})} \geq z_g^{(\mathcal{D})}\}/(q+1) = O(q^{-1})$. The result then follows because the elements of $\{z_g^{(\mathcal{T})}$,$\{z_h^{(\mathcal{D})}\}_{h \in [q]\setminus \{g\}}\}$ are independent and identically distributed with continuous distribution functions (i.e. no ties in the order statistics) conditional on $H_g=0$.
\end{proof}

\subsection{Proofs of theoretical statements made in Section~\ref{supp:section:theory}}
\label{supp:subsection:ProofTheorems}

\begin{lemma}
\label{supp:lemma:logHolder}
Suppose $\log\{f(x)\},\log\{g(x)\}$ are $\alpha$-H\"{o}lder continuous on the compact interval $\mathcal{I}$. Then $\log\{f(x) + g(x)\}$ is $\alpha$-H\"{o}lder continuous on $\mathcal{I}$.
\end{lemma}
\begin{proof}
$\log\{f(x)\}$ is $\alpha$-H\"{o}lder continuous on $\mathcal{I}$ if and only if $\abs*{f(x_1) - f(x_2)} \leq f(x_2) c \abs*{x_1 - x_2}^{\alpha}$ for some constant $c > 0$ and all $x_1,x_2 \in \mathcal{I}$. We see that 
\begin{align*}
    &\text{$\log\{f(x) + g(x)\}$ is $\alpha$-H\"{o}lder continuous}\\
    &\text{$\Leftrightarrow$ $\abs{f(x_1) - f(x_2) + g(x_1) - g(x_2)} \leq \{f(x_2)+g(x_2)\}c \abs{x_1-x_2}^{\alpha}$ for some $c>0$}\\
    &\text{$\Leftarrow$ $\abs{f(x_1) - f(x_2)} + \abs{g(x_1) - g(x_2)} \leq \{f(x_2)+g(x_2)\}c \abs{x_1-x_2}^{\alpha}$ for some $c>0$}\\
    &\text{$\Leftarrow$ $\abs{f(x_1) - f(x_2)} \leq f(x_2)\frac{c}{2} \abs{x_1-x_2}^{\alpha}, \abs{g(x_1) - g(x_2)} \leq g(x_2)\frac{c}{2} \abs{x_1-x_2}^{\alpha}$ for some $c>0$},
\end{align*}
where the latter holds by the assumptions on $f(x),g(x)$.
\end{proof}

\begin{lemma}
\label{supp:lemma:subexp}
Let $p:\mathbb{R} \to [0,1]$ and $\lambda: \mathbb{R}\to [0,\infty)$ be continuous functions, where $p$'s maximum is 1, and for $\RandMZ \subset \mathbb{R}$ a Poisson point process with intensity $\lambda$, define $D = \sum_{m \in \RandMZ} p(m)$. Then the following hold:
\begin{enumerate}[label=(\roman*)]
\item $M=\E(D) = \smallint p(x)\lambda(x)\text{d}x$ and $V=\V(D) = \smallint \{p(x)\}^2\lambda(x)\text{d}x$.\label{supp:item:MeanVar}
\item $\Prob( \abs*{ D-M }/M \geq z ) \leq \begin{cases}
    -\frac{M^2}{4V}z^2 & \text{if $z \leq \frac{2tV}{M}$}\\
    -\frac{tM}{2}z & \text{if $z > \frac{2tV}{M}$}
    \end{cases}$ for some universal constant $t>0$ that does not depend on $p$ or $\lambda$.\label{supp:item:SubExp}
\end{enumerate}
\end{lemma}
\begin{proof}
The expressions in \ref{supp:item:MeanVar} are straightforward. For \ref{supp:item:SubExp}, we see that
\begin{align*}
    K(t)=\log\{ \E(\exp[t\{D - \E(D)\}]) \} =  \smallint \lambda(x)\{ e^{tp(x)} - tp(x) - 1 \}\text{d}x.
\end{align*}
Let $C = V^{1/2}/M$. Since $p(x)\geq 0$ has maximum 1, there exists a $t^*>0$ that does not depend on $p$ or $\lambda$ such that $K(t) \leq t^2 V$ for all $t \in [0,t^*]$, meaning
\begin{align*}
    \log\{ \Prob( \abs*{ D-M }/M \geq z ) \} \leq \begin{cases}
    -\frac{1}{4C^2}z^2 & \text{if $z \leq \frac{2t^* V^{1/2}}{C}$}\\
    -\frac{t^*M}{2}z & \text{if $z > \frac{2t^* V^{1/2}}{C}$}
    \end{cases}.
\end{align*}
This completes the proof.
\end{proof}

For the remainder of the section, we define
\begin{align*}
    \tilde{d}_{\signal}^{\Mass}(u \mid y; m) = \frac{10^6}{m} d_{\signal}^{\Mass}\{ 10^{6}(u/m-1) \mid y \}
\end{align*}
to be the density of $M_g \mid (Y_g=y, M_g \in F_m)$ for $F_m = [m(1-\eta),m(1+\eta)]$ and $\eta=10^{-6}\omega$.

\begin{proof}[Proof of Theorem \ref{supp:theorem:CondSmall}]
Let $u_m \in F_m$ be such that $L(F_m)B(u_m) = \int I(u \in F_m)B(u)\text{d}u$. Then for all $u \in F_m$,
\begin{align*}
    &L(F_m)\Prob(M_g = u \mid Y_g=y, M_g \in F_m, \RandMZ_g) = \frac{a_g(m)\{B(u)/B(u_m)\} + b_g(m)\{d_{\signal}^{\Mass}(u \mid y; m)L(F_m)\}}{a_g(m) + b_g(m)}\\
    &a_g(m) = \pi_0 \alpha_g B(u_m), \quad b_g(m) = (1-\pi_0)\tilde{p}_g(m)/L(F_m)
\end{align*}
Since $B(m)$ and $\tilde{p}_g(m)$ are log $\alpha$-H\"{o}lder continuous, this completes the proof.
\end{proof}


\begin{proof}[Proof of Theorem \ref{supp:theorem:dModerate}]
Let $G_m = [m\frac{1-10^{-6}\omega}{1+10^{-6}\omega}, m\frac{1+10^{-6}\omega}{1-10^{-6}\omega}]$ and define the random sets $R_1 = G_m \cap (\RandMZ_g \setminus \{m\})$ and $R_2 = G_m^c \cap (\RandMZ_g \setminus \{m\})$. Note that $\tilde{d}_{\signal}^{\Mass}(u \mid y; m') = 0$ if $u \in F_m$ and $m' \in G_m^c$. Conditional on $m \in \RandMZ_g$, the random normalizing constant is $\beta_g^{-1} = \sum_{m' \in R_1}p(m') + \sum_{m' \in R_2}p(m') + p(m) = P_1 + P_2 + p(m)$, where $P_1$ and $P_2$ are independent with distributions that implicitly depend on $g$. Then
\begin{align*}
    &\Prob(M_g=u \mid Y_g=y,m \in \RandMZ_g) = \pi_0 \alpha_g B(m)(1+O[\{L(F_m)\}^{\alpha}])\\
    +& (1-\pi_0)\underbrace{\E\left\{ \frac{p(m)\tilde{d}_{\signal}^{\Mass}(u \mid y; m)+ \sum\limits_{m' \in R_1}p(m')\tilde{d}_{\signal}^{\Mass}(u \mid y; m') }{P_1 + P_2 + p(m)} \mid m \in \RandMZ_g\right\}}_{F_g(u)},
\end{align*}
where for $\tau_g(m) = \E[\{P_1/p(m) + P_2/p(m) + 1\}^{-1}]$,
\begin{align*}
    F_g(u) = \tau_g(m) \tilde{d}_{\signal}^{\Mass}(u \mid y; m) + \underbrace{\E\left\{ \frac{\sum\limits_{m' \in R_1} \frac{p(m')}{p(m)}\tilde{d}_{\signal}^{\Mass}(u \mid y; m') }{P_1/p(m) + P_2/p(m) + 1} \right\}}_{\tilde{F}_g(u)}.
\end{align*}
Let $N_1$ be the number of elements in $R_1$, and for ease of notation, set $\delta_m = L(F_m)$. Then $P_1/p(m) = N_1\{1 + O(\delta_m^{\alpha})\}$ and
\begin{align*}
    \tilde{F}_g(u) &= \E \left[  \frac{\sum\limits_{i=1}^{N_1}\E\left\{ A\frac{p(x_i)}{p(m)}\tilde{d}_{\signal}^{\Mass}(u \mid y; x_i) \mid N_1,P_2 \right\}}{N_1 + P_2/p(m) + 1} \right]\\
    A &= \frac{N_1 + P_2/p(m) + 1}{P_1/p(m) + P_2/p(m) + 1} = \frac{1 + \{P_2/p(m)\}/(N_1+1)}{1+O(\delta_m^{\alpha}) + \{P_2/p(m)\}/(N_1+1)} = 1 + O(\delta_m^{\alpha})
\end{align*}
where the $x_1,\ldots,x_{N_1}$ are independent and identically distributed with density $\tilde{\lambda}(x) = \lambda(x)I(x \in G_m)/\{\smallint_{G_m}\lambda(z)\text{d}z\}$ conditional on $N_1$ and $P_2$. Since $p(x)/p(m),\lambda(x)/\lambda(m) = 1 + O(\delta_m^{\alpha})$ for all $x_i \in G_m$,
\begin{align*}
    \E\left\{ A\frac{p(x_i)}{p(m)}\tilde{d}_{\signal}^{\Mass}(u \mid y; x_i) \mid N_1,P_2 \right\} =& \{1 + O(\delta_m^{\alpha})\} \{L(G_m)\}^{-1}\int_{G_m} \tilde{d}_{\signal}^{\Mass}(u \mid y; x) \text{d}x\\
    =& \{1 + O(\delta_m^{\alpha})\} \{L(G_m)\}^{-1},
\end{align*}
where $L(G_m)$ is the Lebesgue measure of $G_m$ and the last equality follows from the proof of Theorem \ref{supp:theorem:Condlarge} below. Putting this all together gives us
\begin{align*}
    \tilde{F}_g(u) = \{1 + O(\delta_m^{\alpha})\} \{L(G_m)\}^{-1}\E[ N_1\{N_1 + P_2/p(m)+1\}^{-1} ], \quad u \in F_m,
\end{align*}
where the error $O(\delta_m^{\alpha})$ is uniform over $g \in [q]$ and $u \in F_m$, and $v \in \mathbb{R}$. The proves the results in \eqref{supp:equation:dmoderate}.\par 
\indent It remains to show that $\log\{\tilde{\pi}_g(m)\}$ is $\alpha$-H\"{o}lder continuous, where $\tilde{\pi}_g(m)$ is defined in \eqref{supp:equation:dmoderate}. By Lemma \ref{supp:lemma:logHolder}, the amounts to showing that $\log[\E\{\tilde{N}_m\tilde{p}_g(m)\}]$ and $\log[\E\{(\tilde{N}_m+1)\tilde{p}_g(m)\}]$ are $\alpha$-H\"{o}lder continuous. Let $m_2>m_1 > T(1-\eta)$ and define 
\begin{align*}
    &S_1 = G_{m_1} \cap G_{m_2}^c, \quad R_1 = \sum\limits_{m' \in \RandMZ_g} p(m')I(m' \in S_1)\\
    &S_2 = G_{m_1}^c \cap G_{m_2}, \quad R_2 = \sum\limits_{m' \in \RandMZ_g} p(m')I(m' \in S_2)\\
    &S_3 = G_{m_1} \cap G_{m_2}, \quad R_3 = \sum\limits_{m' \in \RandMZ_g} p(m')I(m' \in S_3)\\
    &S_4 = G_{m_1}^c \cap G_{m_2}^c, \quad R_4 = \sum\limits_{m' \in \RandMZ_g} p(m')I(m' \in S_4).
\end{align*}
Then
\begin{align*}
    \tilde{N}_{m_2}\tilde{p}_g(m_2) &=  \frac{\tilde{N}_{m_2}}{1 + \frac{R_1+R_2+R_3}{p(m_1)} \frac{p(m_1)}{p(m_2)} + \frac{R_4}{p(m_1)}\frac{p(m_1)}{p(m_2)}}\\
    & = \tilde{N}_{m_2}\tilde{p}_g(m_1) \{1 + c(R_1,R_2,R_3,R_4,m_1,m_2)\abs*{m_1 - m_2}^{\alpha}\},
\end{align*}
where $\abs*{c(R_1,R_2,R_3,R_4,m_1,m_2)} \leq M$ for some constant $M$ that does no depend on $R_1$, $R_2$, $R_3$, $R_4$, $m_1$, $m_2$ or $g$. Therefore,
\begin{align*}
    \abs{\frac{\E\{ \tilde{N}_{m_2}\tilde{p}_g(m_2) \}}{\E\{ \tilde{N}_{m_2}\tilde{p}_g(m_1) \}} - 1} \leq k_1\abs*{m_1 - m_2}^{\alpha}
\end{align*}
for some universal constant $k>0$. It is easy to see that these same techniques can be used to show that $\log[\E\{\tilde{p}_g(m)\}]$ is $\alpha$-H\"{o}lder continuous. Therefore, we need only show that
\begin{align*}
    \abs{\frac{\E\{ \tilde{N}_{m_2}\tilde{p}_g(m_1) \}}{\E\{ \tilde{N}_{m_1}\tilde{p}_g(m_1) \}} - 1} \leq k_2\abs*{m_1 - m_2}^{\alpha}
\end{align*}
for some constant $k_2>0$. To do this, we see that
\begin{align*}
    \frac{\E\{\tilde{N}_{m_2}\tilde{p}_g(m_1)\}}{\E\{\tilde{N}_{m_1}\tilde{p}_g(m_1)\}} - 1 =& 
    \frac{ \E\{\tilde{M}_{1}\tilde{p}_g(m_1)\} }{\E\{\tilde{M}_{1}\tilde{p}_g(m_1)\} + \E\{\tilde{M}_{2}\tilde{p}_g(m_1)\} + \E\{\tilde{M}_{3}\tilde{p}_g(m_1)\}}\\
    &- \frac{ \E\{\tilde{M}_{2}\tilde{p}_g(m_1)\} }{\E\{\tilde{M}_{1}\tilde{p}_g(m_1)\} + \E\{\tilde{M}_{2}\tilde{p}_g(m_1)\} + \E\{\tilde{M}_{3}\tilde{p}_g(m_1)\}}\\
    \tilde{M}_{j} =& \frac{1-\eta^2}{2}\sum\limits_{m' \in \RandMZ_g} I(m' \in S_j) = \frac{1-\eta^2}{2} M_j, \quad j =1,2,3
\end{align*}
where for $\delta = \Delta(m_2)+\abs{m_1-m_2}$,
\begin{align*}
    \E\{\tilde{M}_{j}\tilde{p}_g(m_1)\} &= \frac{1-\eta^2}{2}\E\left\lbrace \frac{M_j}{ 1 + \frac{R_1}{p(m_1)} + \frac{R_2}{p(m_1)} +\frac{R_{3}}{p(m_1)} + \frac{R_{4}}{p(m_1)} } \right\rbrace\\
    & = \frac{1-\eta^2}{2} \E\left\lbrace \frac{M_j}{1 + M_1 + M_2 + M_3 +\frac{R_{4}}{p(m_1)}} \right\rbrace \{ 1+O( \delta^{\alpha} ) \}, \quad j=1,2,3.
\end{align*}
We then see that,
\begin{align*}
    &\E\left\lbrace \frac{M_j}{1 + M_1 + M_2 + M_3 +\frac{R_{4}}{p(m_1)}} \right\rbrace\\
    =& \E\left\lbrace \frac{M_1 + M_2 + M_3}{1 + M_1 + M_2 + M_3 + \frac{R_{o}}{p(m_1)}} \E\left( \frac{M_j}{M_1 + M_2 + M_3} \mid M_1+M_2+M_3,R_4 \right) \right\rbrace\\
    =& \frac{\lambda(x_j)L(M_j)}{\lambda(x_1)L(M_1)+\lambda(x_2)L(M_2)+\lambda(x_3)L(M_3)} \E\left\lbrace \frac{M_1 + M_2 + M_3}{1 + M_1 + M_2 + M_3 + \frac{R_{o}}{p(m_1)}} \right\rbrace, \quad j=1,2,3\\
    &x_1 \in G_{m_1} \cap G_{m_2}^c, \quad x_2 \in G_{m_1}^c \cap G_{m_2}, \quad x_3 \in G_{m_1} \cap G_{m_2}
\end{align*}
where $L(M_1),L(M_2) = O[\min\{ \abs*{m_1-m_2},\Delta(m_2) \}]$, $L(M_3) = L(G_{m_1}\cup G_{m_2}) - \{L(M_1)+L(M_2)\}$, $L(G_{m_1}\cup G_{m_2}) = O\{ \Delta(m_1)+\Delta(m_2) \}$, and $\lambda(x_r)/\lambda(x_s) = 1+O(\abs*{m_1-m_2}^{\alpha})$ for $r,s \in \{1,2,3\}$. This shows that $\log[\E\{ \tilde{N}_m\tilde{p}_g(m) \}]$ is $\alpha$-H\"{o}lder continuous, and completes the proof.
\end{proof}

\begin{proof}[Proof of Corollary \ref{supp:corollary:dModAvg}]
We implicitly condition on the event
\begin{align*}
    \{ m_1,\ldots,m_K \in \RandMZ_g \} \cap \{\text{all $K$ intensities$=y$}\}
\end{align*}
in all probability statements in the proof. Let $\mathcal{F} = \{ M_{g1} \in F_{m_1}, \ldots, M_{gK} \in F_{m_K} \}$ for $M_{gk}$ the $k$th (random) observed noise m/z, where $M_{gk}$ is identically distributed to $M_g$. Define
\begin{align*}
    f_k(r \mid \RandMZ_g) =& d_{\noise,g}^{\Mass}(r \mid y; m_k)\\
    =& 10^{-6}m_k\Prob\{ M_{gk}=m_k(10^{-6}r+1) \mid M_{gk} \in F_{m_k}, \RandMZ_g \}\\
    f_k(r) =& 10^{-6}m_k\Prob\{ M_{gk}=m_k(10^{-6}r+1) \mid M_{gk} \in F_{m_k}\}.
\end{align*}
We break this proof into two smaller lemmas. First, we show that $\E\{ f_k(r \mid \RandMZ_g) \mid \mathcal{F} \} \approx f_k(r)$. Second, we prove that $\V\{\sum_{k=1}^K f_k(r \mid \RandMZ_g)\}$ is small.
\begin{lemma}
\label{supp:lemma:Expf}
Suppose the assumptions of Corollary \ref{supp:corollary:dModAvg} hold and let $E = \E\{ \sum_{m \in \RandMZ_g}p(m)I(m \notin \{m_1,\ldots,m_K\}) \}$. Then
\begin{align*}
    \E\{ f_k(r \mid \RandMZ_g) \mid \mathcal{F} \} = f_k(r)\{1 + O(E^{-\alpha+2\delta})\}, \quad k \in [K],
\end{align*}
where the error is uniform across $r \in [-\omega,\omega]$.
\end{lemma}

\begin{proof}
Note that $M_{g1},\ldots,M_{gK} \mid \RandMZ_g$ are independent. Define the function $R(m_1,m_2) = 10^6(m_2/m_1 - 1)$. Then 
\begin{align}
\label{supp:equation:fk}
    \E\{ f_k(r \mid \RandMZ_g) \mid \mathcal{F} \} = \frac{ \E[ \Prob\{ R(M_{gk},m_k)=r \mid \RandMZ_g \} \prod\limits_{j \neq k}^K \Prob( M_{gj} \in F_{m_j} \mid \RandMZ_g ) ] }{ \E[ \Prob( M_{gk} \in F_{m_k} \mid \RandMZ_g ) \prod\limits_{j \neq k}^K \Prob( M_{gj} \in F_{m_j} \mid \RandMZ_g ) ] },
\end{align}
where for $P_k = 1 + \sum_{j\neq k}^K p(m_j)/p(m_k)=K\{ 1+O[\{\Delta(m_K)\}^{\alpha}\}]$, $D = \sum_{m \in \RandMZ_g}p(m)I(m \notin \{m_1,\ldots,m_K\})$, and $\tilde{G}_{m_k} = G_{m_k}\setminus \{m_k\}$,
\begin{align*}
    &\Prob\{ R(M_{gk},m_k)=r \mid \RandMZ_g \} = a_k + \frac{M_k}{P_k + D/p(m_k)}, \quad a_k=\pi_0 \alpha_g B(u_k)(10^{-6}m_k)\\
    &\Prob( M_{gk} \in F_k \mid \RandMZ_g ) = b_k + \frac{\tilde{M}_k}{P_k + D/p(m_k)}, \quad b_k=\pi_0 \alpha_g \smallint_{F_{m_k}} B(x)\text{d}x\\
    &M_k = (1-\pi_0)[d_{\signal}(r \mid y) + \sum\limits_{m \in \tilde{G}_{m_k}} \frac{p(m)}{p(m_k)}\{10^{-6}m \tilde{d}_{\signal}(u_k \mid y;m)\}]\\
    &\tilde{M}_k = (1-\pi_0)\{ 1 + \sum\limits_{m \in \tilde{G}_{m_k}} \frac{p(m)}{p(m_k)} z_k(m) \}\\
    &z_k(m)=\begin{cases} \int_{ 10^6\{ \frac{m_k}{m}(1-10^{-6}\omega) - 1 \} }^{\omega}d^{\Mass}(r\mid y)\text{d}r & \text{if $m\leq m_k$}\\
    \int_{-\omega}^{10^6\{1-\frac{m_k}{m}(1+10^{-6}\omega)\}}d_{\signal}^{\Mass}(r\mid y)\text{d}r  & \text{if $m> m_k$}
    \end{cases},
\end{align*}
where the terms in the summand in the expressions for $M_k$ and $\tilde{M}_k$ are uniformly bounded from above. We can therefore write the denominator of \eqref{supp:equation:fk} as
\begin{align}
\label{supp:equation:Numerator}
    \E\left\{ \left(b_k + \frac{\tilde{M}_k}{P_k + D/p(m_k)}\right) \prod\limits_{j \neq k} \left( b_j + \frac{\tilde{M}_j}{P_j + D/p(m_j)} \right) \right\},
\end{align}
where $(M_1,\tilde{M}_1),\ldots,(M_K,\tilde{M}_K)$ are independent and $a_1,b_1,P_1,\ldots,a_K,b_K,P_K$ are deterministic. When factored, the expression inside the expectation will be the sum of $2^K$ terms. We first study last term, which can be expressed as
\begin{align*}
    \E\left( \prod\limits_{j=1}^K \frac{\tilde{M}_j}{P_j + D/p(m_j)} \right) = \E\left( \prod\limits_{j=1}^K \frac{P_j + E/p(m_j)}{P_j + D/p(m_j)} \prod\limits_{j=1}^K \frac{\tilde{M}_j}{P_j + E/p(m_j)} \right),
\end{align*}
where our analysis can then be easily extended to the remaining $2^K-1$ terms. This term can be re-written as
\begin{align}
\label{supp:equation:Dtimes}
    \prod_{j=1}^K \E\left( \frac{\tilde{M}_j}{P_j + E/p(m_j)} \right) + \E\left\lbrace \left(1 - \prod\limits_{j=1}^K \frac{P_j + E/p(m_j)}{P_j + D/p(m_j)}\right)\prod\limits_{j=1}^K \frac{\tilde{M}_j}{P_j + E/p(m_j)} \right\rbrace.
\end{align}
Let $A=\prod_{j=1}^K \E\left( \frac{\tilde{M}_j}{P_j + E/p(m_j)} \right)$, $Z = \prod\limits_{j=1}^K \frac{\tilde{M}_j}{P_j + E/p(m_j)}$, $S = \prod\limits_{j=1}^K \frac{P_j + E/p(m_j)}{P_j + D/p(m_j)}$, and $H$ be the second term in the above expression. Then if $\abs{D-E}/E \leq z<1$, $S \in [1/(1-z)^K,1/(1+z)^K]$, meaning
\begin{align*}
    &\abs{H} \leq c_z A + \abs*{\E\{ I(\abs{D-E}/E>z) (1-S)Z \}}\\
    &c_z = \max[\{ (1+z)^K-1\}(1+z)^{-K}, \{ 1-(1-z)^K \}(1-z)^{-K}],
\end{align*}
where $c_z  \leq cKz$ for some universal constant $c>1$ and all $z>0$ sufficiently small. We next see that
\begin{align*}
    \abs*{\E\{ I(\abs{D-E}/E>z) (1-S)Z \}} \leq \{\E(Z^2)\}^{1/2} [\E\{ I(\abs{D-E}/E>z)(1-S)^2  \}]^{1/2},
\end{align*}
where since $\tilde{M}_k \leq \tilde{c}N_k$ for some universal constant $\tilde{c}>0$ and $N_k \sim 1+\text{Poi}\{\smallint_{G_{m_k}} \lambda(x)\text{d}x\}$, $\{\E(Z^2)\}^{1/2} \leq \bar{c}A$ for some universal constant $\bar{c}>0$. Next, let $a = \{\sum_{k=1}^K p(m_k)\}^{-1}$ and $\tilde{a}=aE/(1+aE)$. Then $S = 1/\{1+\tilde{a}(D/E - 1)\}^K$ and
\begin{align*}
    \E\{I(\abs{D/E - 1}>z)(1-S)^2\} =& \Prob(\abs{D/E - 1}>z) - 2\E\{I(\abs{D/E - 1}>z)S\}\\
    &+ \E\{I(\abs{D/E - 1}>z)S^2\}.
\end{align*}
Since identical bounds exist for the last two terms, we bound the second term in the above expression below. By Lemma \ref{supp:lemma:subexp} and for $\bar{t}=2tV/M, V=\V(D)$ (for $t>0$ as defined in Lemma \ref{supp:lemma:subexp}),
\begin{align*}
    \E\{I(\abs{D/E - 1}>z)S\} \leq& \Prob(\abs{D/E - 1}>z) + c\exp\left\{-\frac{tE}{2}z + K\log(aE)\right\}\\
    &+ I(z<\bar{t})c\exp\left\lbrace-\frac{E^2}{4V}z^2 + K\log(aE)\right\rbrace
\end{align*}
for some universal constant $c>0$. Putting this all together,
\begin{align*}
    &\E\left\lbrace \prod_{j=1}^K \frac{\tilde{M}_j}{P_j + D/p(m_j)} \right\rbrace = \prod_{j=1}^K \frac{\E(\tilde{M}_j)}{P_j + E/p(m_j)} \{ 1 + O(K z) \}\\
    &z = c_1E^{-\alpha+\delta}, \quad K \leq c_2 E^{\delta}
\end{align*}
for some universal constants $c_1,c_2>0$ and $\delta \in (0,\alpha/2)$. Since the error is multiplicative and only depends on $K$, the number of times $D$ appears in \eqref{supp:equation:Dtimes}, \eqref{supp:equation:Numerator} can be written as
\begin{align*}
    \left\lbrace b_k + \frac{\E(\tilde{M}_k)}{P_k + E/p(m_k)}\right\rbrace \prod_{j \neq k}\left\lbrace b_j + \frac{\E(\tilde{M}_j)}{P_j + E/p(m_j)}  \right\rbrace \{1+O(Kz)\}
\end{align*}
and
\begin{align*}
    \Prob(M_{gj} \in F_{m_j})=&\E\{\Prob(M_{gj} \in F_{m_j} \mid \RandMZ_g)\}= b_k + \E\left\lbrace \frac{\tilde{M}_j}{P_j + D/p(m_j)} \right\rbrace = b_k\\
    &+ \E\left\lbrace \frac{\tilde{M}_j}{P_j + E/p(m_j)} \frac{P_j + E/p(m_j)}{P_j + D/p(m_j)} \right\rbrace\\
    =& \left[b_k+\frac{\E(\tilde{M}_j)}{P_j + E/p(m_j)}\right]\{1+O(z)\}.
\end{align*}
Therefore, \eqref{supp:equation:Numerator} can be expressed as
\begin{align*}
    \Prob(M_{gk} \in F_{m_k})\prod_{j \neq k} \Prob(M_{gj} \in F_{m_j})\{1+O(Kz)\}.
\end{align*}
An identical analysis can be used to show the numerator of \eqref{supp:equation:fk} can be written as
\begin{align*}
    \Prob\{R(M_{gk},m_k)=r\}\prod_{j \neq k} \Prob(M_{gj} \in F_{m_j})\{1+O(Kz)\},
\end{align*}
which completes the proof.
\end{proof}

\begin{lemma}
\label{supp:lemma:Varf}
Suppose the assumptions of Corollary \ref{supp:corollary:dModAvg} hold and let $E = \E\{ \sum_{m \in \RandMZ_g}p(m)I(m \notin \{m_1,\ldots,m_K\}) \}$. Then
\begin{align*}
    \V\left\{K^{-1}\sum\limits_{k=1}^K f_{k}(r \mid \RandMZ_g) \mid \mathcal{F}\right\} = O\left( K^{-1} + E^{-\alpha+2\delta} \right),
\end{align*}
where the error is uniform across $r \in [-\omega,\omega]$.
\end{lemma}
\begin{proof}
Let $S(r \mid \RandMZ_g) = K^{-1}\sum\limits_{k=1}^K f_{k}(r \mid \RandMZ_g)$. Then
\begin{align*}
    \V\{S(r \mid \RandMZ_g) \mid \mathcal{F}\} =& K^{-2}\sum\limits_{k=1}^K \V\{f_{k}(r \mid \RandMZ_g) \mid \mathcal{F}\}\\
    &+ 2 K^{-2}\sum\limits_{k_1<k_2} \C\{ f_{k_1}(r \mid \RandMZ_g), f_{k_2}(r \mid \RandMZ_g) \mid \mathcal{F}\}\\
    =& O(K^{-1}) + 2 K^{-2}\sum\limits_{k_1<k_2} \C\{ f_{k_1}(r \mid \RandMZ_g), f_{k_2}(r \mid \RandMZ_g) \mid \mathcal{F}\},
\end{align*}
where the second equality follows because $f_{k}(r \mid \RandMZ_g)$ is uniformly bounded from above. Next, for $k_1 \neq k_2 \in [K]$,
\begin{align*}
    &\C\{ f_{k_1}(r \mid \RandMZ_g), f_{k_2}(r \mid \RandMZ_g) \mid \mathcal{F}\}\\
    =& \frac{ \E[ \Pr\{R(M_{g k_1},m_{k_1})=r \mid \RandMZ_g\}\Pr\{R(M_{gk_2},m_{k_2})=r \mid \RandMZ_g\} \prod\limits_{j \notin \{k_1,k_2\} }\Pr(M_{gj} \in F_{m_j} \mid \RandMZ_g) ] }{ \E\{ \Pr(M_{g k_1} \in F_{m_{k_1}} \mid \RandMZ_g)\Pr(M_{gk_2} \in F_{m_{k_2}} \mid \RandMZ_g) \prod\limits_{j \notin \{k_1,k_2\} } \Pr(M_{gj} \in F_{m_j} \mid \RandMZ_g) \} }\\
    &- \E\{ f_{k_1}(r \mid \RandMZ_g)\mid \mathcal{F} \}\E\{ f_{k_2}(r \mid \RandMZ_g)\mid \mathcal{F} \}.
\end{align*}
Identical techniques used in the proof of Lemma \ref{supp:lemma:Expf} can be used to show that this is $O(E^{-\alpha + 2\delta})$, where the error is uniform across $k_1,k_2$ and $r$. This completes the proof.
\end{proof}
Lemmas \ref{supp:lemma:Expf} and \ref{supp:lemma:Varf} shows that
\begin{align*}
    K^{-1}\sum\limits_{k=1}^K f_{k}(r \mid \RandMZ_g) = K^{-1}\sum\limits_{k=1}^K f_k(r) + O_P(K^{-1/2} + E^{-\alpha/2+\delta})
\end{align*}
for $E$ defined in the statements of Lemmas \ref{supp:lemma:Expf} and \ref{supp:lemma:Varf}. The results then follow by Theorem \ref{supp:theorem:dModerate}.
\end{proof}

\begin{proof}[Proof of Corollary \ref{supp:corollary:dModerate}]
To prove the result, it suffices to study $\E\{\tilde{N}_m\tilde{p}_g(m)\}/\E\{\tilde{p}_g(m)\}$. Let $\delta_m = \{\Delta(m)\}^{\alpha}$. Using the proof of Theorem \ref{supp:theorem:dModerate}, we see that
\begin{align*}
    \E\{\tilde{N}_m\tilde{p}_g(m)\} &= \frac{1-\eta^2}{2}\E\left\{ \frac{M_1}{1 + \frac{R_1}{p(m)} + \frac{R_2}{p(m)}} \right\} = \frac{1-\eta^2}{2}\E\left\{ \frac{M_1}{1 + M_1 + \frac{R_2}{p(m)}} \right\}\{1 + O(\delta_m)\}\\
    \E\{\tilde{p}_g(m)\} &= \frac{1-\eta^2}{2} \E\left\{ \frac{1}{1 + M_1 + \frac{R_2}{p(m)}} \right\}\{1 + O(\delta_m)\}\\
    M_1 &= \sum\limits_{m' \in \RandMZ_g\setminus \{m\}} I(m' \in G_m)\\
    R_1 &= \sum\limits_{m' \in \RandMZ_g\setminus \{m\}} p(m')I(m' \in G_m), \quad R_2 = \sum\limits_{m' \in \RandMZ_g\setminus \{m\}} p(m')I(m' \in G_m^c).
\end{align*}
Let $\mu_m=\int_{G_m}\lambda(x)dx$. Then there exists a constant $c_1>0$ that does not depend on $m$ such that for all $t \in (0,c_1\mu_m^{1/2})$,
\begin{align*}
    \E\left\{ \frac{M_1}{1 + M_1 + \frac{R_2}{p(m)}} \mid R_2 \right\} \geq & \frac{k_{m,t}}{1 + k_{m,t} + \frac{R_2}{p(m)}}\{1-\exp(-c_2 t^2)\}\\
    \E\left\{ \frac{1}{1 + M_1 + \frac{R_2}{p(m)}} \mid R_2 \right\} \leq & \frac{1}{1 + k_{m,t} + \frac{R_2}{p(m)}}\{1-\exp(-c_2 t^2)\}\\
    &+ \frac{1}{1 + k_{m,t} + \frac{R_2}{p(m)}}\left\lbrace 1 + \frac{k_{m,t}}{1 + \frac{R_2}{p(m)}}\right\rbrace\exp(-c_2 t^2)\\
    k_{m,t} =& \mu_m(1-t\mu_m^{-1/2})
\end{align*}
for some universal constant $c_2 > 0$, meaning
\begin{align*}
    \frac{\E\left\{ \frac{M_1}{1 + M_1 + \frac{R_2}{p(m)}}\right\}}{ \E\left\{ \frac{1}{1 + M_1 + \frac{R_2}{p(m)}}\right\} } \leq \frac{ k_{m,t} }{ 1 + (1+k_{m,t})\frac{\exp(-c_2 t^2)}{\{1-\exp(-c_2 t^2)\}} }.
\end{align*}
Setting $t = \mu_m^{\epsilon}$ for $\epsilon>0$ small enough completes the proof.
\end{proof}

\begin{proof}[Proof of Theorem \ref{supp:theorem:Condlarge}]
Similar to the proof of Theorem \ref{supp:theorem:CondSmall}, for all $u \in F_m$,
\begin{align*}
    &\Prob(M_g=u \mid M_g \in F_m, Y_g=y, \RandMZ_g) = \frac{a_g(m)\{L(F_m)^{-1} B(u)/B(u_m)\} + b_g(m,v) \frac{\sum\limits_{m' \in \RandMZ_g} p(m')\tilde{d}_{\signal}^{\Mass}(u \mid y; m')}{k(v,m)} }{a_{g}(m) + b_g(m,v) }\\
    &a_{g}(m) = \pi_0 \alpha_g B(u_m), \quad b_g(m,v) = (1-\pi_0) k(y,m)\\
    &k(y,m) = \sum_{m' \in \RandMZ_g}p(m') \int I(u \in F_m) \tilde{d}_{\signal}^{\Mass}(u \mid y; m')
\end{align*}
for $u_m \in F_m$ defined in the proof of Theorem \ref{supp:theorem:CondSmall}. Assumption \ref{supp:assumption:dnoise}\ref{supp:item:C} implies
\begin{align*}
    L(F_m)^{-1}B(u)/B(u_m) = L(F_m)^{-1}I(u \in F_m)[ 1+O\{L(F_m)^{\alpha_c}\} ].
\end{align*}
Therefore, wee need only understand how $\{k(y,m)\}^{-1}\sum\limits_{m' \in \RandMZ_g} p(m')\tilde{d}_{\signal}^{\Mass}(u \mid y; m')$ behaves. Define $\delta=10^{-6}\omega$. We see that for $G_u = [u/(1+\delta),u/(1-\delta)]$
\begin{align*}
 f(u;y) = \sum\limits_{m' \in \RandMZ_g} p(m')\tilde{d}_{\signal}^{\Mass}(u \mid y; m') = \sum\limits_{m' \in G_u \cap \RandMZ_g} p(m')\tilde{d}_{\signal}^{\Mass}(u \mid y; m').
\end{align*}
Let $A\subseteq F_m$ be Lebesgue-measurable and define $G_A = \cup_{x \in A} (x/(1+\delta),x/(1-\delta))$. Then
\begin{align*}
    \E\{f(u; y)\} &= \smallint_{x \in G_u} p(x)\tilde{d}_{\signal}^{\Mass}(u \mid y; x)\lambda(x)\text{d}x\text{d}u\\
    \E\{\smallint_{u\in A} f(u; y)\text{d}u\} &= \smallint_{u \in A}\smallint_{x \in G_u} p(x)\tilde{d}_{\signal}^{\Mass}(u \mid y; x)\lambda(x)\text{d}x\text{d}u\\
    \V\{ f(u; y)\text{d}u\} &= \smallint_{x \in G_u} \{p(x) \tilde{d}_{\signal}^{\Mass}(u \mid y; x)\}^2 \lambda(x)\text{d}x\\
    \V\{\smallint_{u\in A} f(u; y)\text{d}u\} &= \smallint_{x \in G_A} \{p(x)\}^2 \{ \smallint_{u \in A} \tilde{d}_{\signal}^{\Mass}(u \mid y; x) \text{d}u\}^2\lambda(x)\text{d}x.
\end{align*}
Let $\phi_{\mu,\sigma}(x)$ be the pdf of a normal with mean $\mu$ and variance $\sigma^2$. Then for $\sigma_{j,x} = 10^{-6}\sigma_{m,j}x$ for $\sigma_{m,j}$, $j=1,2$, defined in \eqref{equation:mSignal},
\begin{align*}
    &\smallint_{x \in G_u} \tilde{d}_{\signal}^{\Mass}(u \mid y; x) \text{d}x = \sum_{j=1}^2 \pi_{\signal,j}(y)k_j \smallint_{x \in G_u} \phi_{u,\sigma_{j,x}}(x) \text{d}x\\
    &\pi_{\signal,1}(y) = \pi_{\signal}(y), \quad \pi_{\signal,2}(y) = 1-\pi_{\signal}(y), \quad k_j^{-1} = \smallint_{-\omega/\sigma_{m,j}}^{\omega/\sigma_{m,j}} \phi_{0,1}(z)\text{d}z.
\end{align*}
We first see that
\begin{align*}
    \phi_{u,\sigma_{j,x}}(x) = \phi_{u,\sigma_{j,u}}(x) \left[ \frac{u}{x}\exp\left\{ -1/2 \left( \frac{x-u}{\sigma_{j,u}} \right)^2\left(u^2/x^2-1\right) \right\} \right], \quad u/x \in [1-\delta,1+\delta].
\end{align*}
Therefore,
\begin{align*}
    \phi_{u,\sigma_{j,x}}(x)/\phi_{u,\sigma_{j,u}}(x) \in [1-c_1\delta,1+c_1\delta]
\end{align*}
for some constant $c_1$ that does not depend on $x$ or $u$. We then see that
\begin{align*}
    k_j \int_{x \in G_u} \phi_{u,\sigma_{j,u}}(x) \text{d}x = k_j \int_{-\omega/\sigma_{m,j}\{1 + \delta/(1+\delta) \}}^{\omega/\sigma_{m,j}\{1+\delta/(1-\delta)\}} \phi_{0,1}(x)\text{d}x = 1+c_2\delta
\end{align*}
for some constant $c_2>0$ that does not depend on $x$ or $u$. Let
\begin{align*}
    x_{+} = \argmax_{x \in [m(1-\delta)/(1+\delta),m(1+\delta)/(1-\delta)]} p(x)\lambda(x), \quad x_{-} = \argmin_{x \in [m(1-\delta)/(1+\delta),m(1+\delta)/(1-\delta)]} p(x)\lambda(x).
\end{align*}
We then get that for some universal constant $c_3>0$ and all Lebesgue-measurable $A \subseteq F_m$,
\begin{align*}
    &\E\{ f(u; y) \} \in \left[ p(x_-)\lambda(x_-)(1-c_3\delta), p(x_+)\lambda(x_+)(1+c_3\delta) \right]\\
    &\E\{\smallint_{u\in A} f(u; y)\} \in \left[ p(x_-)\lambda(x_-)(1-c_3\delta)L(A), p(x_+)\lambda(x_+)(1+c_3\delta)L(A) \right]
\end{align*}
where $L(A)$ is the Lebesgue measure of $A$. Therefore,
\begin{align*}
    \E\{k(m;y)\} &= p(x_+)\lambda(x_+)L(F_m)\{1 + O(\delta^{\alpha_p})\}\\
    \E\{ f(u; y) \}/\E\{k(m;y)\} &= \{L(F_m)\}^{-1}\{ 1+O(\delta^{\alpha_p}) \}, \quad u \in F_m,
\end{align*}
where the error $O(\delta^{\alpha_p})$ does not depend on $u$, $y$, or $m$. For the variance, we first see that
\begin{align*}
    \V\{k(m;y)\}&=\V\{\smallint_{u\in F_m} f(u; y)\text{d}u\} \leq \int_{x \in G_{F_m}} \{p(x)\}^2 \lambda(x)\text{d}x\\
    &\leq c_4 p(\tilde{x}) \{p(x_+)\lambda(x_+)L(F_m)\}\\
    \tilde{x}&= \argmax_{x \in [m(1-\delta)/(1+\delta),m(1+\delta)/(1-\delta)]} p(x).
\end{align*}
for some universal constant $c_4>0$. Therefore,
\begin{align}
\label{supp:equation:kmv}
    k(m;y) = \E\{k(m;y)\}(1+O_P[ \{\lambda_- L(F_m)\}^{-1/2}] ).
\end{align}
Using similar techniques, we also have
\begin{align}
\label{supp:equation:var}
    \V[f(u; y)/\E\{k(m;y)\}] \leq c_5 \{L(F_m)\}^{-2}\{ \lambda(x_+)L(F_m) \}^{-1}
\end{align}
for some universal constant $c_5>0$. Lastly, let $\mathcal{B}(a,\Delta) \subseteq F_m$ be a closed ball centered at $a$ with radius $\Delta>0$, and for $u_1,u_2 \in \mathcal{B}(a,\Delta)$, define $\mathcal{I} = \RandMZ_g \cap (G_{u_1} \cap G_{u_2} )$ and $\mathcal{R} = \RandMZ_g \cap \{(G_{u_1} \cup G_{u_2} )\setminus (G_{u_1} \cap G_{u_2})\}$. Then
\begin{align*}
    \abs{f(u_1;y) - f(u_2;y)} \leq & \sum\limits_{m' \in \mathcal{I}} p(m')\abs*{ \tilde{d}_{\signal}^{\Mass}(u_1 \mid y; m')-\tilde{d}_{\signal}^{\Mass}(u_2 \mid y; m')}\\
    &+ \sum\limits_{m' \in \mathcal{R}} p(m') \abs*{\tilde{d}_{\signal}^{\Mass}(u_1 \mid y; m')-\tilde{d}_{\signal}^{\Mass}(u_2 \mid y; m') }\\
    \leq & c_6 \Delta/\{L(F_m)\}^2 \sum\limits_{m' \in \mathcal{I}}p(m') + c_6/L(F_m) \sum\limits_{m' \in \mathcal{R}}p(m'),
\end{align*}
meaning for $\mathcal{S}_{\Delta} = \{\mathcal{B}(a,r)\subseteq F_m: a \in F_m,r=\Delta \}$,
\begin{align}
\label{supp:equation:sup}
    \E\{ \sup_{\mathcal{B}(a,\Delta) \in \mathcal{S}_{\Delta}} \sup_{u_1,u_2 \in \mathcal{B}(a,\Delta)} \abs{f(u_1;y) - f(u_2;y)} \} \leq c_6 p_+\lambda_+\Delta/L(F_m).
\end{align}
for some constant $c_6>0$. A standard stochastic equicontinuity argument using \eqref{supp:equation:var} and \eqref{supp:equation:sup} completes the proof.
\end{proof}

\begin{proof}[Proof of Theorem \ref{supp:theorem:lambda}]
Statement \ref{supp:item:lambdaOnem} follows directly from the assumptions on $C$ in Assumption \ref{supp:assumption:dnoise}. Statements \ref{supp:item:lambdaNmGen} and \ref{supp:item:lambdaNm} follow from the proof of Theorem \ref{supp:theorem:dModerate} and the properties of $k(m;v)$ in \eqref{supp:equation:kmv} in the proof of Theorem \ref{supp:theorem:Condlarge}, respectively. Lastly, \eqref{supp:equation:ContFunctions} follow by the proofs of Theorems \ref{supp:theorem:CondSmall} and \ref{supp:theorem:dModerate}.
\end{proof}

\begin{proof}[Proof of Corollary \ref{supp:corollary:lambda}]
The proof of Corollary \ref{supp:corollary:lambda} is a simple extension of the proof of Corollary \ref{supp:corollary:dModAvg}, and has been omitted.
\end{proof}

\newpage

\end{document}